\documentclass[useAMS,usenatbib] {mn2e}
\usepackage{graphicx}
\usepackage{float}
\usepackage{times}
\usepackage{natbib}

\newcommand{\enzo}{{ENZO\ }}
\newcommand{\visit}{{VisIt\ }}

\begin{document}

\title[Filaments in Cosmological Simulations]{ 
Properties of Cosmological Filaments extracted from Eulerian Simulations}
\author[C. Gheller, F. Vazza, J. Favre. M. Br\"{u}ggen]{C. Gheller$^{1}$, F. Vazza$^{2,3}$, J. Favre$^{1}$, M. Br\"{u}ggen$^{2}$\\
$^{1}$ ETHZ-CSCS, Via Trevano 131, Lugano, Switzerland \\
$^{2}$ Hamburger Sternwarte, Gojenbergsweg 112, 21029 Hamburg, Germany\\
$^{3}$ INAF-Istituto di Radio Astronomia, Via Gobetti 101, Bologna, Italy}

\date{Accepted ???. Received ???; in original form ???}
\maketitle

\begin{abstract}

Using a new parallel algorithm implemented within the \visit framework, we analysed large
 cosmological grid simulations to study the properties of baryons in filaments. 
The procedure allows us to build large catalogues with up to $\sim 3 \cdot 10^4$ filaments per 
simulated volume
and to investigate the properties of cosmic filaments for very large volumes at high resolution (up to $300^3 ~\rm Mpc^3$ simulated with $2048^3$ cells). 
We determined scaling relations for the mass, volume, length and temperature of filaments and
compared them to those of galaxy clusters. The longest filaments 
have a total length of about $200 ~\rm Mpc$ with a mass of several $10^{15} M_{\odot}$. We also investigated the effects of different gas physics. Radiative cooling significantly modifies the thermal properties of the warm-hot-intergalactic
medium of filaments, mainly by lowering their mean temperature via line cooling. On the other hand, powerful 
feedback from active galactic nuclei in surrounding halos can heat up
the gas in filaments. The impact of shock-accelerated cosmic rays from diffusive shock acceleration on filaments is small and the ratio of between cosmic ray and gas pressure within filaments is of the order of $\sim 10-20$ percent. 

\end{abstract}


\label{firstpage} 

\vskip 0.4cm \fontsize{11pt}{11pt} \selectfont  

\section{Introduction}
\label{sec:intro}

Simulations of structure formation predict that
about 50\% of the baryons reside in a filamentary web of tenuous matter
at temperatures f $10^5-10^7$K connecting already virialized structures
\citep[e.g.][]{1999ApJ...514....1C,2001ApJ...552..473D}.
This plasma is referred to as ``Warm-Hot Intergalactic Medium'' (WHIM).

The total matter (mostly Dark Matter, DM) of the cosmic web is traced by galaxies that populate filamentary and wall--like structures, the largest of which have sizes in excess of $100 ~\rm Mpc/h$ \citep[e.g.][]{1982Natur.300..407Z,1984MNRAS.206..529E, 1989Sci...246..897G,2005ApJ...624..463G}.
A first glimpse of this cosmic web in the local
Universe was offered by the first CfA redshift slice \citep[e.g.][]{1986ApJ...302L...1D}.
In recent years, this view has been expanded by the 2dFGRS \citep[e.g.][]{2003yCat.7226....0C}, SDSS \citep[e.g.][]{2004ApJ...606..702T} and
IRAS \citep[e.g.][]{2013AJ....146...69C} galaxy redshift surveys. 

A direct detection of the gas in the cosmic web is more challenging because the low
densities and temperatures are unfavourable, and so far only very little evidence has been produced.
Baryons in filaments may be revealed in the soft X--ray band but the few reported detections are still controversial \citep[e.g.][]{2003A&A...410..777F,2008A&A...482L..29W,2010ApJ...715..854N,2013ApJ...769...90N}.
In  the radio band, only very few possible detections have been reported \citep[][]{2002NewA....7..249B,2007ApJ...659..267K,2010A&A...511L...5G,2013ApJ...779..189F}.
These sources are more likely to be related to merger shocks than to the accretion shocks 
described by numerical simulation. 

Recently, the Sunyaev-Zeldovich (SZ) effect has been used to probe filamentary gas connections 
between  galaxy clusters, and a first indication for the detection of a filament between the cluster pair A399-A401 has been reported by  \citet[][]{2013A&A...550A.134P}.
They observed a significant thermal SZ  signal in the regions beyond the virial radii. A joint X-ray SZ analysis constrained the temperature of the filamentary region to
kT = $7.1 \pm 0.9$ keV and the baryon density to $3.7 \pm 0.2 \cdot 10^{-4} \rm cm^{-3}$. However, this should represent only the dense and small-size version of much larger  ($\sim 10-100$ Mpc long) objects that the cosmic volume can contain. 

The evolution of the cosmic web can be explained through the
interaction of the initial pattern of density waves at different scales,
with random and uncorrelated spatial phases. 
The main skeleton of the cosmic web is determined by the initial gravitational potential
field \citep[][]{2011A&A...531A..75E}.  Recently, \citet{2013ApJ...762..115O} studied in detail 
the emergence of the filamentary structure of the cosmic web from the phase information embedded 
in the pattern of initial cosmological fluctuations.
During their evolution initial fluctuations interact in a non-linear way,
with the generation of non-random and correlated phases,  
which lead to the spatial pattern of the present cosmic web \citep[][]{1996Natur.380..603B}. 
The non-linear evolution can be described by N-body simulations, that provide an accurate picture 
of the evolution of the gravitational potential and of the Dark Matter halos.

With the growing capabilities of N-body cosmological simulations, increasing effort has been devoted to
the implementation of reliable methodologies for the identification of complex structures 
in the matter distribution, aiming at the accurate segmentation of the cosmic web into clusters, filaments, 
walls and voids \citep[e.g.][and references therein for a recent review]{2014MNRAS.441.2923C}. 
The list of attempted methods is too long to enable a complete summary here. The algorithms can be broadly 
grouped into: a) geometrical and tessellation methods, based on the topological analysis of the density field by means of sophisticated mathematical approaches  \citep[e.g.][]{2005A&A...434..423S, 2008MNRAS.383.1655S, 2010MNRAS.408.2163A, 2010MNRAS.407.1449G, 2014arXiv1412.0510B, 2015arXiv150105303C}, b) morphological methods, that classify the 3D distribution of matter based on the density Hessian and/or the tidal or velocity shear fields \citep[e.g.][]{2007MNRAS.381...41H, 2007A&A...474..315A, 2014MNRAS.441.2923C}.
Recently, \citet{2014MNRAS.441.2923C} compared the results of different filament detection methods with the {\it NEXUS} algorithm, which is an advanced multiscale analysis tool to identify the different morphologies of cosmic structures, based on the analysis of density, tidal field, velocity divergence and velocity shear as tracers of the Cosmic Web. They concluded that most methods agree on the largest filaments, but parameters such as the diameter of filaments depend on the resolution and the method. Most of the differences are found in the most rarefied environments, where density contrasts are very small and methods working on shears or local differences are less sensitive. As a result, different algorithms agree well on the total mass in filaments ($\sim 90$ percent of the total mass is captured in all methods) and worse on the total volume (only a $\sim 60$ percent of the volume is equally identified).\\

So far, the study of the properties of the baryons in the low-density components of the cosmic web  (such as in filaments) have been investigated less systematically. 
Low-density environments outside of galaxy clusters have been studied with cosmological hydrodynamical simulations \citep[e.g.][and references therein]{do08, 2001ApJ...552..473D,2005MNRAS.360.1110V,2006MNRAS.370..656D}. The effect
of energy feedback from galactic activity on the observable properties of the WHIM has been analysed by \citet[e.g][]{2006ApJ...650..560C,2006MNRAS.368...74R,ka07,2007ApJ...671...27H,2012MNRAS.424.1012R, 2014MNRAS.438.2499B,2014Natur.509..177V}.
Hydrodynamical simulations suggest that regions of moderate overdensity such as filaments, host a significant fraction of the WHIM \citep[e.g.][]{1999ApJ...514....1C,2001ApJ...552..473D}, and they are surrounded by strong stationary accretion
shocks, where the cosmic gas is first shock heated to temperatures $\geq 10^{4} \rm ~K$ 
\citep[e.g.][]{mi00,ry03,pf06}. Using a more idealised setup, \citet{2012MNRAS.423..304K} simulated the impact of different physical processes as well as of the scale dependencies in the formation of $\sim 5$ Mpc filaments starting from a single-scale perturbation. The simulated filaments exhibit an isothermal core, whose temperature is balanced by radiative cooling and heating due to the UV background, yielding a multiphase medium.
The WHIM is also expected to host supersonic turbulent motions \citep[][]{iapichino11,va14mhd}, and the decay of such
motions can cause the amplification of weak primordial magnetic fields up to the $\sim \rm nG$ level \citep[][]{2008Sci...320..909R,va14mhd}.\\

In this work, we combined state-of-the art cosmological hydrodynamic simulations
performed using the \enzo code (Section \ref{sec:enzo}) with a novel methodology 
for the identification of the cosmic web, based on the gas matter distribution.
The gas distribution can be accurately described by \enzo's Eulerian hydrodynamic solver irrespectively
of the mass density. Using the gas mass density, we have developed an {\it Isovolume}
based technique to identify the cosmic web structures. 
Isovolumes represent a class of algorithms that tackle the problem of  
separating regions of space with distinct physical properties, in our case 
with density above and below a given threshold. 
We tuned the selection in gas overdensity specifically to extract the mass distribution
associated with the filamentary structure of the cosmic web. However, the same approach can also be used to identify
other structures, such as voids, sheets and halos.

The resulting methodology is described in detail in Section \ref{sec:filaments}. The corresponding performance and
scaling properties on high-performance computing systems are presented in Appendix A, together with a discussion on the influence on the results
of the parameters characterizing our approach.

The methodology is simpler than the aforementioned methods while avoiding the drawbacks of algorithms that rely on DM particles.
In regions where the matter density is comparable to the mean cosmic density, 
as in the WHIM, the numerical noise arising from the graininess of the DM particles 
distribution can affect the estimates of the properties of the cosmic web.
Further advantages of using the gas component are that it is immediately related to observations,
and that the effect of additional physical processes (e.g. cooling and feedback from galactic activity)
influencing the gas properties, can be studied in detail.
Using the gas component as a tracer of the gravitational potential, the method is
insensitive to cosmic shear fields and to the local dynamics of DM. 

We have analysed the statistical and thermodynamic properties of filaments identified through our procedure, and investigated their dependence on the spatial 
resolution and the assumed physics, in particular the cooling and heating processes.
The results are presented in Sections \ref{sec:tuning} and \ref{sec:prop} and discussed in Section 
\ref{sec:discussion}. 
Our finally summary is given in Section \ref{sec:conclusions} where we discuss possible future numerical developments that will allow  a more quantitative prediction of the observable properties of the WHIM through the inclusion of mechanisms such as chemical enrichment, metal-dependent cooling and magnetic fields.

\section{Simulations}
\label{sec:enzo}

Our simulations are run with
a customised version of the grid code {\enzo} \citep[][]{enzo13}, presented in  \citet{va14curie}.
{\enzo} is a parallel code for cosmological (magneto-)hydro-dynamics, which uses a particle-mesh N-body method (PM) to follow
the dynamics of the DM and (in this case) the Piecewise Parabolic Method (PPM, \citealt{cw84}) to evolve the gas component.  
On the basis of the public version of {\enzo}, we have implemented specific methods to describe the evolution and feedback of cosmic-ray (CR) particles  \citep[][]{scienzo}, as well as our implementation of energy release from AGN \citep[][]{va13feedback} and supernovae. The suite of simulations analysed in this work has been designed to study the properties of
CRs, the acceleration of relativistic hadrons at shocks and their energy
feedback on the baryon gas \citep[][]{va14curie}.\\
The simulations presented in this paper have been 
run with two flavours of CR-injection, based on a high efficiency 
model by \citet{kj07} (model ``0") and on a lower efficiency model by \citet{kr13} (model ``1"). 

Radiative cooling  (``c" in the name descriptor of Table~\ref{tab:tab2}) has been modelled assuming a primordial composition of a fully ionized H-He plasma
with a uniform metallicity of $Z=0.3 ~Z_{\odot}$ (where $Z_{\odot}$ is the solar metallicity), using the APEC emission model \citep[e.g.][]{2001ApJ...556L..91S}.
For the cold gas, with temperature $T \leq 10^4$ K, we use the cooling curve of \citet{2011ApJ...731....6S}, which is derived from a complete set of of metals (up to an atomic number 30), obtained with the chemical network of the photo-ionization software {\it Cloudy} \citep[][]{1998PASP..110..761F}.
In order to model the UV re-ionization background  \citep[][]{hm96}, we enforced a temperature floor for the gas in the redshift range $4 \leq z \leq 7$, as discussed in  \citet{va10kp}. 

All radiative simulations presented here also include the effect of feedback from AGN, which are placed in high-density peaks ($n \geq n_{\rm AGN}=10^{-2} \rm cm^{-3}$). This method by-passes, both, the
problem of monitoring the mass accretion rate onto the central black hole within each
galaxy, and the complex small-scale physical processes which
couple the energy from the black hole to the surrounding gas (i.e. the launching of strong winds due to the radiation pressure of photons from the accretion disc). This is unavoidable, given that our best resolution is orders of magnitudes larger than the accretion disc region, let alone the difficulty of modelling the radiative transfer of photons from the accretion region.
At each feedback event we add an energy of $E_{\rm AGN} \sim 10^{59} \rm erg$ during a single time step, which typically raises the temperature
in the cell to $\sim 5 \cdot 10^{7}-5 \cdot 10^{8} \rm K$ at the injection burst. We model the feedback as a ``bipolar thermal outflow'', i.e.
the thermal energy is released into the ICM along two cells on opposite sides of the AGN cell, and
the direction of the two jets is randomly selected along one of the
three coordinate axes of the simulation.
In the runs analysed for this work we include two slightly different implementations of AGN feedback in simulations with radiative cooling, which produce rather different cluster scaling relations at $z=0$. In the first run (``c1''), we trigger thermal feedback whenever the  local maximum density exceeds the (comoving) gas density threshold, $n_{\rm AGN}$, starting from $z=1$. In the second run, the AGN feedback is started at $z \leq 4$ (run ``c2''). While the first approach is found to quench the radiative cooling
inside most of halos but fails to produce cluster scaling relations that match observations (i.e. the $M-T$ and $L_{\rm X}-T$ relations are much flatter than observations), the second produces a tilt in scaling relations that agrees well with observations, and is thus our preferred AGN-feedback method.  

At all redshifts the heating by the reionizing background largely exceeds the heating contribution from CRs, which lose energy via interactions with thermal particles of the ICM \citep[][]{guo08}. This energy exchange is modelled
at run-time in our method for CRs, but it becomes important only for high gas 
densities ($\rho/(\mu m_{\rm p})>10^{-2} \rm cm^{-3}$) that are typical of galactic cool cores \citep[][]{va14curie}. 

The runs analysed here assume a WMAP 7-year cosmology with
$\Omega_0 = 1.0$, $\Omega_{B} = 0.0455$, $\Omega_{DM} =
0.2265$, $\Omega_{\Lambda} = 0.728$, Hubble parameter $h = 0.702$, a normalisation for the primordial density power
spectrum $\sigma_{8} = 0.81$ and a spectral index of $n_s=0.961$ for the primordial spectrum of initial matter
fluctuations, starting the runs at $z_{\rm in}=30$ \citep[][]{2011ApJS..192...18K}.

Table~\ref{tab:tab2} lists the simulations used for this work, showing the different choices for the physical mechanisms, as well as the grid size.  We simulated three independent cosmological volumes with boxes of sides $216 ~ \rm Mpc/h$,  $108 ~\rm Mpc/h$ and $54 ~\rm Mpc/h$, respectively.  In order to monitor resolution effects for each case, we produced re-simulations of each of these with a different number of cells (from $2048^3$ to $256^3$) and DM particles.
In general, the missing effects of non-gravitational physics, such as radiative gas cooling and AGN feedback, are much more evident at high resolutions, where gas density and hence radiative losses are higher. For this reason, our study of the impact of non-gravitational physics is based largely
on the smallest box (with side $54 ~\rm Mpc/h$). On the other hand, the impact of different CR injection efficiencies on the outer cluster profiles is also captured at a lower
spatial resolution, and therefore all our boxes include a study using different
acceleration recipes.

\begin{table*}
\caption{List of the simulations run for this project. Column 1: box size. Column 2: number of grid cells. Column 3: spatial resolution. Column 4: physical implementation. Column 5: identification name of each run. The ID must be read as follows: the first number refers to the box size, 1 referring to 216 Mpc/h, 2 to 108 Mpc/h, 3 to 54 Mpc/h. The second symbol refers to the physical model. Finally, the third number indicates the 1D computational mesh size in cells. These IDs will be adopted throughout the paper.}
\centering \tabcolsep 5pt
\begin{tabular}{c|c|c|c|c}
  $L_{\rm box}$ [Mpc/h]& $N_{\rm grid}$ & $\Delta x$[kpc/h] & physics & ID \\  \hline
 216 & $2048^3$ & $105$ & non-rad.+CR(KR13) & 1-1\_2048 \\
  216 & $1024^3$ & $210$ & non-rad.+CR(KR13)& 1-1\_1024\\
  216 & $1024^3$ & $210$ & non-rad.+CR(KJ07)& 1-0\_1024\\
  108 & $1024^3$ & $105$ & non-rad.+CR(KR13) & 2-1\_1024 \\
  108 & $1024^3$ & $105$ & non-rad.+CR(KJ07) & 2-0\_1024 \\
  108 & $1024^3$ & $105$ & cool1.+AGN feedback+CR(KR13) & 2-c1\_1024 \\
 108 & $1024^3$ & $105$ & cool2.+AGN feedback+CR(KR13) & 2-c2\_1024\\
  54 & $1024^3$ & $52$ &  non-rad.+CR(KR13) & 3-1\_1024\\
   54 & $1024^3$ & $52$ &  non-rad.+CR(KJ07) & 3-0\_1024\\
   54 & $1024^3$ & $52$ &  cool1.+AGN feedback+CR(KR13) & 3-c1\_1024\\
   54 & $512^3$ & $105$ & non-rad.+CR(KR13) & 3-1\_512 \\
   54 & $512^3$ & $105$ & non-rad.+CR(KJ07) & 3-0\_512 \\
   54 & $512^3$ & $105$ & cool1.+AGN feedback+CR(KR13)+AGN & 3-c2\_512 \\
   54 & $256^3$ & $210$ & non-rad.+CR(KR13) & 3-1\_256 \\
\end{tabular}
\label{tab:tab2}
\end{table*}

\section{Identification of Filaments}
\label{sec:filaments}

Our procedure to identify filaments uses the gas mass density to
separate over- from under-dense regions.
With this criterion, a single physical parameter determines the identification
procedure, with no further arbitrary settings of other physical quantities. The selection is further refined
by considering the numerical
resolution of the simulations and the expected geometry of a filament. These additional
criteria mainly affect small objects (of typical size less than 1 Mpc$^3$).
The resulting methodology is implemented based 
on a combination of different algorithms developed and optimised in the
context of material interface reconstruction.  
The accurate identification of the interface
between two or more materials represents a major issue in these applications since 
a single computational cell can be composed of pieces of different materials.
Various solutions have been developed in order to reconstruct 
interfaces, both for simulations and for visualisation and
data analysis. Examples are provided by \citet{Meredith:2010:VAR:2421836.2421898}, 
\citet{conf/visualization/FujishiroMS95}, \citet{conf/egpgv/HarrisonCG11}, \citet{10.1109/TVCG.2010.17}
and references therein.

We can identify two distinct material phases: the warm/hot gas which fills collapsed cosmological structures on one side, 
and the cold under-dense phase typical of voids on the other side. The 
accurate representation of the boundaries between these two phases is the key 
to define the properties of the identified objects. 
In this work, we have adopted an {\it Isovolume}-based approach,  
described in \citet{Meredith2004}. 
A quantitative evaluation of this approach and of its performance, compared to
other methodologies, can be found in \citet{Meredith:2010:VAR:2421836.2421898}. 
It has been further evaluated for 
our specific needs. The resulting accuracy and performance are
presented in Section \ref{sec:validation} and Appendix A 
respectively. 

In order to implement a computationally efficient, flexible and extensible 
filament reconstruction procedure,
we have exploited the \visit data visualisation and
analysis framework \citep{Childs11visit:an}. 
\visit is designed for the efficient processing of large
datasets through a combination of optimized algorithms, the
exploitation of high-performance computing architectures, in particular
through parallel processing, based on the MPI standard,
and the support of client-server capabilities,
allowing efficient remote visualization. It can be used interactively, through
a graphical user interface, and it can be scripted
using the Python programming language in order to automate the data processing.
Finally, it natively supports a number of file format adopted by popular astrophysical
simulation codes, such as {Cosmos}, {FLASH}, {Gadget}, {Chombo} (besides, of course, \enzo)
and in general all files adopting the VTK format.
\footnote{http://www.visitusers.org/index.php?title=Detailed\_list\_of\_file\_formats\_VisIt\_supports}
We refer to the official documentation
\footnote{https://wci.llnl.gov/codes/visit/manuals.html}
for all the technical details on the software.

\begin{figure*}
  \includegraphics[width=0.7\textwidth]{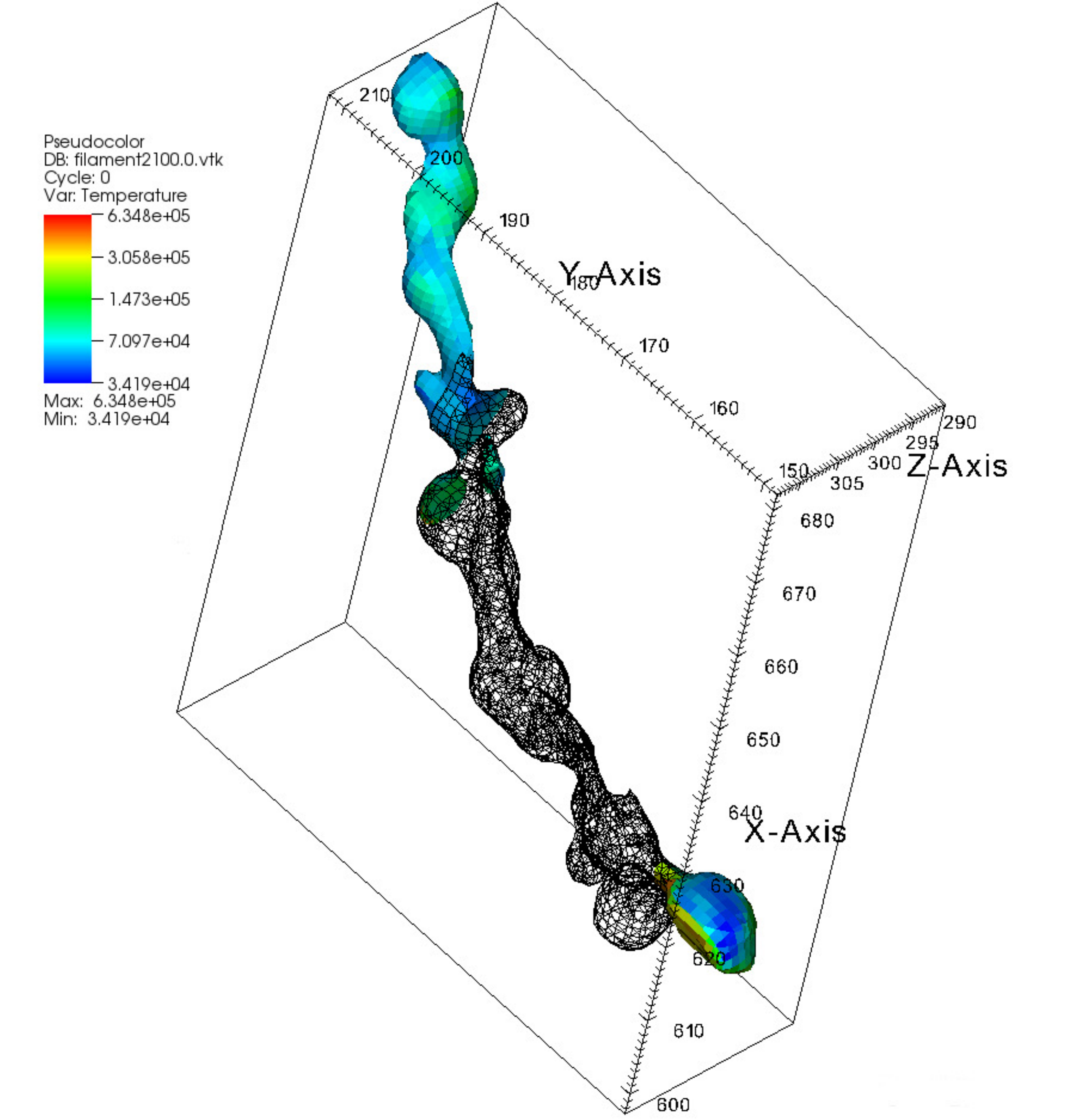}
  \caption{
Example of a filament identified by the Isovolume method. Part of the object is
outlined by the reconstructed surface mesh. The rest of the object is coloured
with the temperature.
}
  \label{fig:mesh}
\end{figure*}

\visit includes the aforementioned Isovolume algorithm, which, combined with 
the {\it Connected Components} filter (also provided by the software, \citealt{conf/egpgv/HarrisonCG11}), allows 
the segmentation of the data into distinct objects. 
An example of a filament identified and reconstructed by our procedure is shown by Figure \ref{fig:mesh}.

\subsection{Filament identification procedure}
\label{sec:identification}

The main steps performed by our implemented identification procedure can be summarised as follows: 

\begin{enumerate}

\item {\it Identification of large clumps.} 
In a first step we separate filaments from galaxy clusters. Clusters can be described as high-density clumps in the matter distribution.
Such clumps can be identified first by selecting isovolumes with

\begin{equation}
{\varrho_{\rm BM}\over\varrho_0} \ge a_{\rm cl}, \label{eq:BM}
\end{equation}
where $\varrho_{\rm BM}$ is the cell's baryon mass density, $a_{\rm cl}$ is a proper threshold  
and $\varrho_0$ is the critical density at
present time. For each clump, \visit ~ returns the volume $V_{\rm cl}$ and the coordinates of its centre.
At this point, assuming spherical symmetry, we calculate the radius of each cluster as:

\begin{equation}
R = \beta\left({3 V_{\rm cl}\over 4\pi}\right)^{1/3},
\end{equation}
where $\beta$ is a multiplicative factor needed to rescale the radius to values of the order of 1 Mpc (expected for galaxy clusters).
Setting $a_{\rm cl} = 100$ and $\beta = 10$ proved to be effective for a proper cluster characterization.

Clusters with radius $R \ge 1$ Mpc are discarded by removing all cells 
falling within spheres with centre in the cluster centroid and radius $R$.

\item {\it Filament identification.} 
The Isovolume algorithm is used once more on the residual cells 
to identify the volumes that obey Equation \ref{eq:BM}, only with a new threshold $a_{\rm fil}$. The 
exact results of the identification procedure depends critically on this parameter, 
whose setting will be discussed in Section \ref{sec:tuning} and Appendix A.
The connected components algorithm is then used to combine cells belonging to distinct (i.e. non 
intersecting) filaments, assigning to each filament an Id (an integer number) and marking each cell 
belonging to a filament with that Id. 
For each filament, the volume ($V_{\rm fil}$), the mass ($M_{\rm fil}$) and 
the average temperature ($T_{\rm fil}$), are calculated. 

\item  {\it Small clumps removal. } 
The next step consists of removing small objects with volumes
$V_{\rm fil} < V_{\rm res}$. 
In order to have a unique volume cutoff for all simulations, the parameter $V_{\rm res}$ is set equal to the lowest
spatial resolution of our runs, $V_{\rm res} = $(0.3 Mpc)$^3$.
This ensures that a filament is always larger than a single cell.

\item {\it Shape selection.}
Although the previous steps are designed to eliminate small clumps and large galaxy clusters,
nonetheless a few outsiders can still be present at this stage. 
Hence, we use the following two-stage cleaning procedure to remove all remaining round-shaped, isolated
structures. First, we calculate the bounding box enclosing each identified object and accept as filaments
all those objects with:
\begin{equation}
{\rm MAX}(r_{xy}, r_{xz}, r_{yz}) > \alpha ,
\end{equation}
where $r_{ab}$ is the ratio between axes $a$ and $b$ of the bounding box. A value of $\alpha = 2$
ensures that the selected object stretches in a specific direction without being too restrictive.
The second stage is intended to recover all those objects which do not match criterion purely based on axis ratio. 
This happens, for instance, to filaments laying along the bounding box diagonal. 
The second shape selection consists of calculating the filling factor $F_V$ of each remaining object, defined as:
\begin{equation}
F_V = {V_{fil}\over V_{box}},
\end{equation}
where $V_{box}$ is the bounding box volume. All objects with $F_V < \varphi$ are classified as filaments.
The value of $\varphi$ is set according to the ratio between the volume 
of a cubic bounding box and that of a cylindrical 
filament lying on its diagonal, with a base radius set as a fraction, namely 1/5, of the box side. 
The values of the parameters $\alpha$ and $\varphi$  are discussed in Appendix A.

\item {\it Data export.}  The properties of the cells belonging to each filament, 
such as the mass density, temperature, velocity, geometric and topological information,
are exported to output files using the VTK format\footnote{http://www.vtk.org/VTK/img/file-formats.pdf}.
Since the number of cells identified as filaments is some fraction of the total 
computational mesh, the file size is reduced correspondingly, making further data 
management and processing significantly easier and faster, which is very convenient for the large runs we use in this work.

\end{enumerate}

\subsection{Validation}
\label{sec:validation}

We validate the Isovolume algorithm using spherically symmetric mass distributions modelled by
Gaussian functions, representing idealised clusters and clumps, and cylindrical shapes 
to mimic filaments. For the filaments, the mass distribution orthogonal to 
the main axis is again modelled as Gaussian. 
We have first evaluated the accuracy of surface areas, volumes and 
density estimates as a function of the mesh resolution for single objects. 
Table \ref{tab:single} shows the error, calculated as
the fractional difference between the calculated and the exact value of each of the three
quantities, as a function of the radius, expressed in computational cells. 
For both spheres and cylinders, the density estimate is always accurate to $\leq 1$\%. 
Volumes and surface areas are more sensitive to the resolution 
(see also \citealt{Meredith:2010:VAR:2421836.2421898}). For radii smaller
than 4 cells, the error approaches (or is even larger, as in the case of the 
sphere's area) 10\%. 

\begin{figure*}
  \includegraphics[width=1.0\textwidth]{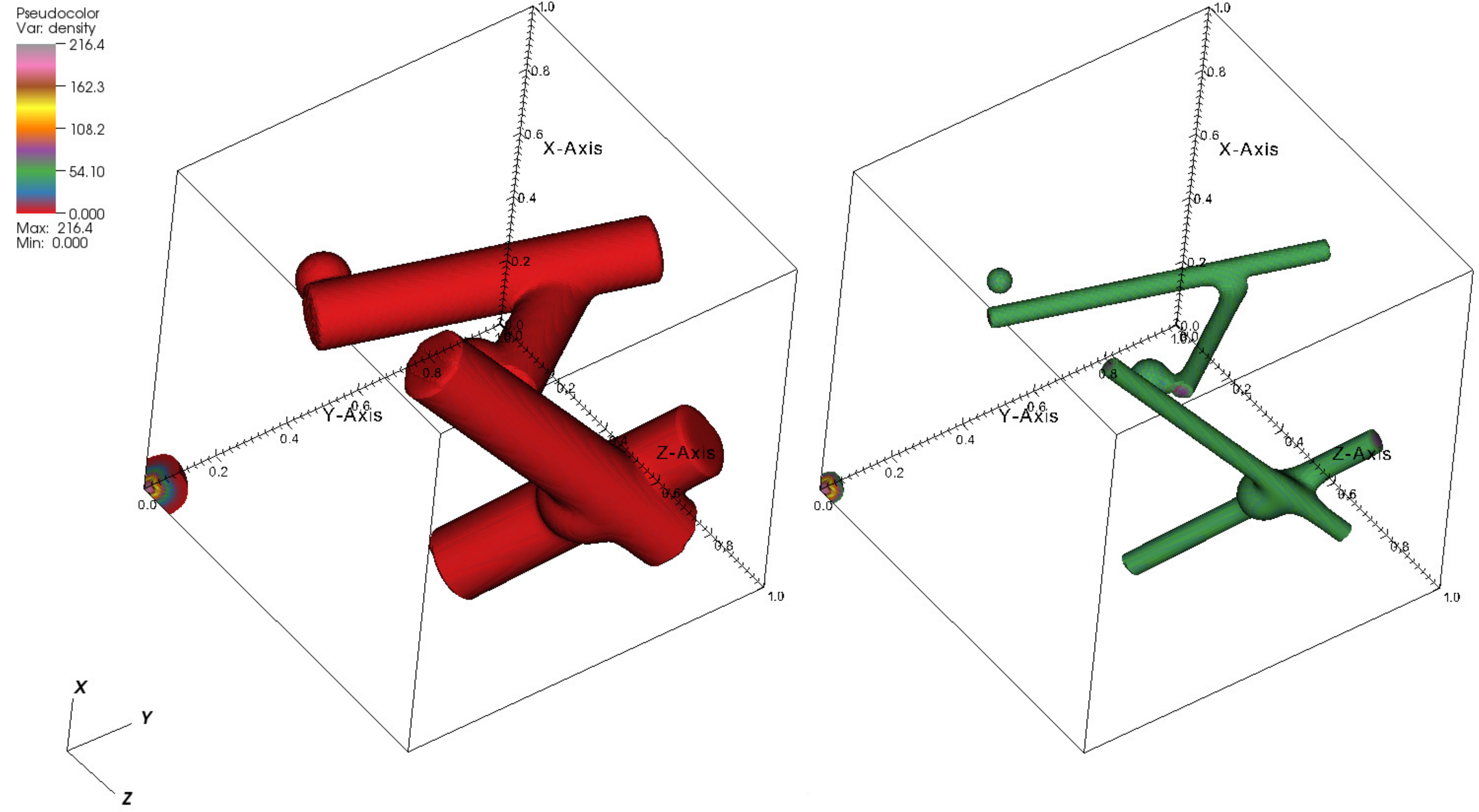}
  \caption{Isovolumes at $a_{\rm fil}=1$ (left) and $a_{\rm fil}=50$ (right) extracted from a test dataset 
combining spheres and cylinders with different mass distribution}
  \label{fig:shapes}
\end{figure*}

Datasets with various combinations of spheres and cylinders
have been used to model the  
matter distribution from a cosmological simulation. For all tests, a computational mesh
of $128^3$ cells is used. An example is presented in Figure \ref{fig:shapes}, 
where isovolumes at $a_{\rm fil}=1$ and $a_{\rm fil}=50$ are shown in 
the left and right panels, respectively.
The accuracies of the volume and density estimates have been calculated as a function of 
$a_{\rm fil}$, obtaining the results
presented in Table \ref{tab:multi}. Volume and density estimates are compared to those
obtained by a simple algorithm which sums the contributions of all the cells above the given
threshold, run at much higher resolution ($1024^3$).   
In all tests, the number of identified objects at each density threshold 
is correct. Only objects 
crossing the faces of the periodic computational box
could not be counted properly because our procedure does not yet support periodic boundary 
conditions. In this case, a single object 
is split into multiple components, each treated as a separate filament. This leads
also to underestimate the length of any filament crossing one of the box's side, as well as to the artificial
increase of small-scale objects because of this artificial fragmentation. However, this problem is statistically not very relevant, as our volumes are typically much larger than the largest objects in the volume. 

The differences in the volume estimates are
between the 1.6 and 4.7\%. The maximum difference is at the highest density thresholds where the size of the
various objects approaches the mesh resolution. The minimum is for $a_{\rm fil} = 5.0$ 
with objects that are still extended 
and have a simple geometry (disjoint spheres or cylinders). Density estimates show
differences of the order of or less than 2\%. 

\begin{table}
\caption{Fractional error in the calculation of surface areas (columns 2 and 5), volumes (column 3 and 6) and densities (column 4 and 7) for spheres and cylinders as a function of the distance (in cells) from the centre/axis (column 1) of the sphere/cylinder.}
\centering \tabcolsep 5pt
\begin{tabular}{c|c|c|c|c|c|c}
  $ R $   & $\delta A/A$ & $\delta V/V$ & $\delta\varrho/\varrho$ & $\delta A/A$ & $\delta V/V$ & $\delta\varrho/\varrho$\\  \hline
  & Sphere & & & Cylinder & & \\ \hline
  3.84  & 0.0762 & 0.1203 & 0.0024 & 0.0616 & 0.0836 & 0.0010\\
  6.40  & 0.0265 & 0.0421 & 0.0011 & 0.0423 & 0.0453 & 0.0007\\
  12.80 & 0.0049 & 0.0079 & 0.0002 & 0.0306 & 0.0183 & 0.0010\\ 
  25.60 & 0.0003 & 0.0004 & 0.0002 & 0.0200 & 0.0048 & 0.0069\\
  38.40 & 0.0014 & 0.0020 & 0.0021 & 0.0067 & 0.0207 & 0.0193\\
\end{tabular}
\label{tab:single}
\end{table}

\begin{table}
\caption{Fractional error in the calculation of volumes (column 2) and densities (column 3) for combination of spheres and cylinders as a function of $a_{\rm fil}$ (column 1).}
\centering \tabcolsep 5pt
\begin{tabular}{c|c|c}
  $ a_{\rm fil} $ & $\delta V/V$ & $\delta\varrho/\varrho$ \\  \hline
  0.25  & 0.026 & 0.010 \\
  0.50  & 0.025 & 0.011 \\
  0.75  & 0.024 & 0.012 \\
  1.0   & 0.023 & 0.013 \\
  2.0   & 0.020 & 0.018 \\
  5.0   & 0.016 & 0.022 \\
  50.0  & 0.047 & 0.004 \\
\end{tabular}
\label{tab:multi}
\end{table}

\begin{figure*}
  \includegraphics[width=1.0\textwidth]{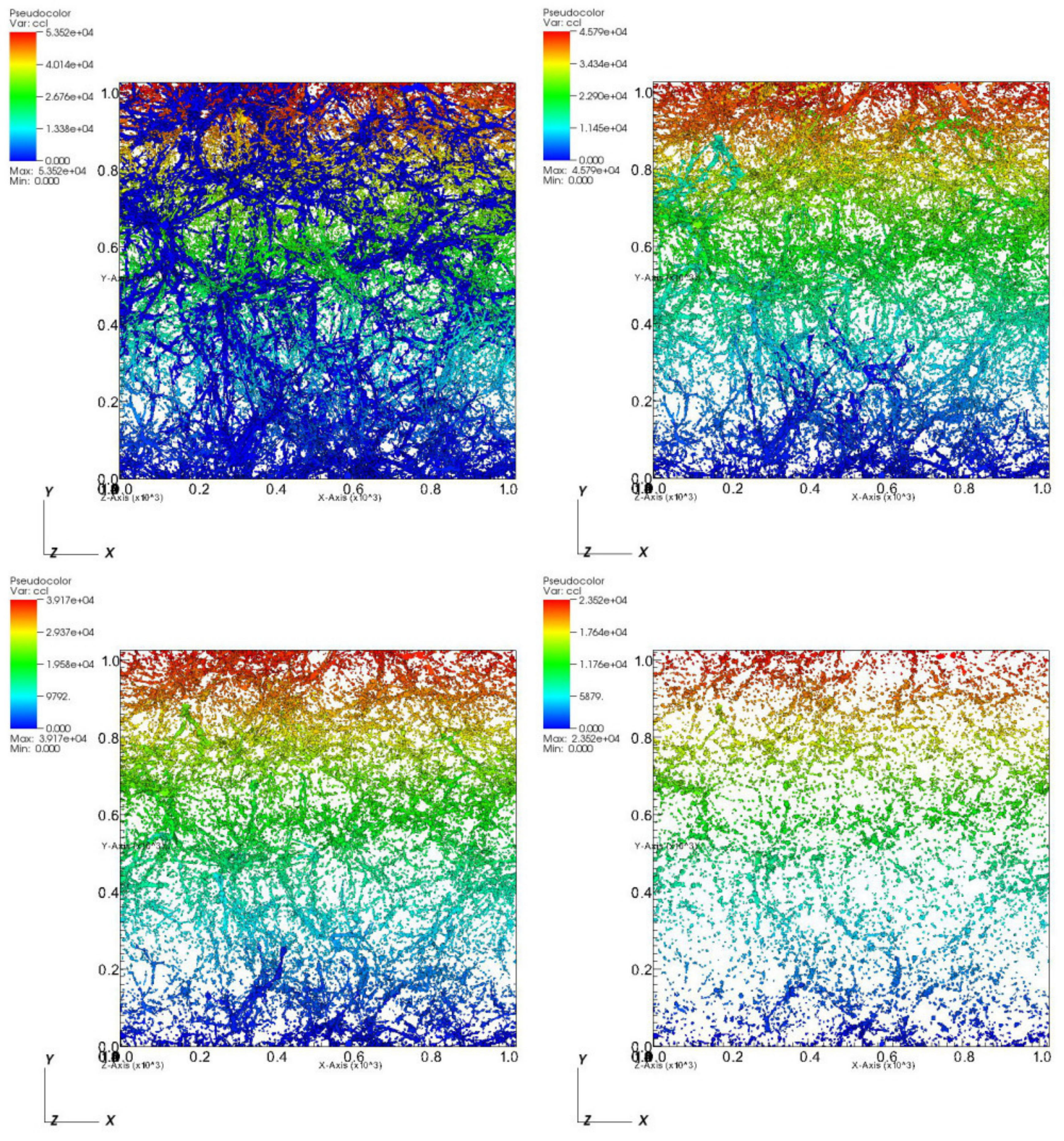}
  \caption{Mass distribution obtained using $a_{\rm fil}$ = 0.5 (top left panel),
   0.75 (top right), 1.0 (bottom left) and 2.0 (bottom right), for the model 2-1\_1024. Each different object identified by our algorithm has a different colour. }
  \label{fig:visit}
\end{figure*}

\subsection{Density threshold and spatial resolution}
\label{sec:tuning}

The parameter $a_{\rm fil}$ is critical for the characterization of the filament 
distribution. 
Its value is discussed below with additional considerations  presented in Section \ref{subsec:singleprop} and in Appendix A.
This parameter determines, together with the 
numerical spatial distribution, the features and the connectivity of the detected structures. 
Figure \ref{fig:visit} shows the mass distribution obtained using four different values 
of $a_{\rm fil}$, ranging from 0.5 to 2 for the model 2-1\_1024,
keeping all else constant. 
Similar results are obtained for the other models (not shown).
Values of the mass density within this range are those expected for filaments \cite[see e.g.][]{1999ApJ...514....1C,2001ApJ...552..473D}.

A qualitative visual inspection of the distributions suggests the adoption 
of a value of $a_{\rm fil}$ in the range 0.75-1.0, smaller values leading to 
structures percolated across the whole computational box, 
and higher values generating clumpy distributions with limited (or even no) 
connectivity. 

\begin{figure*}
  \includegraphics[width=0.7\textwidth]{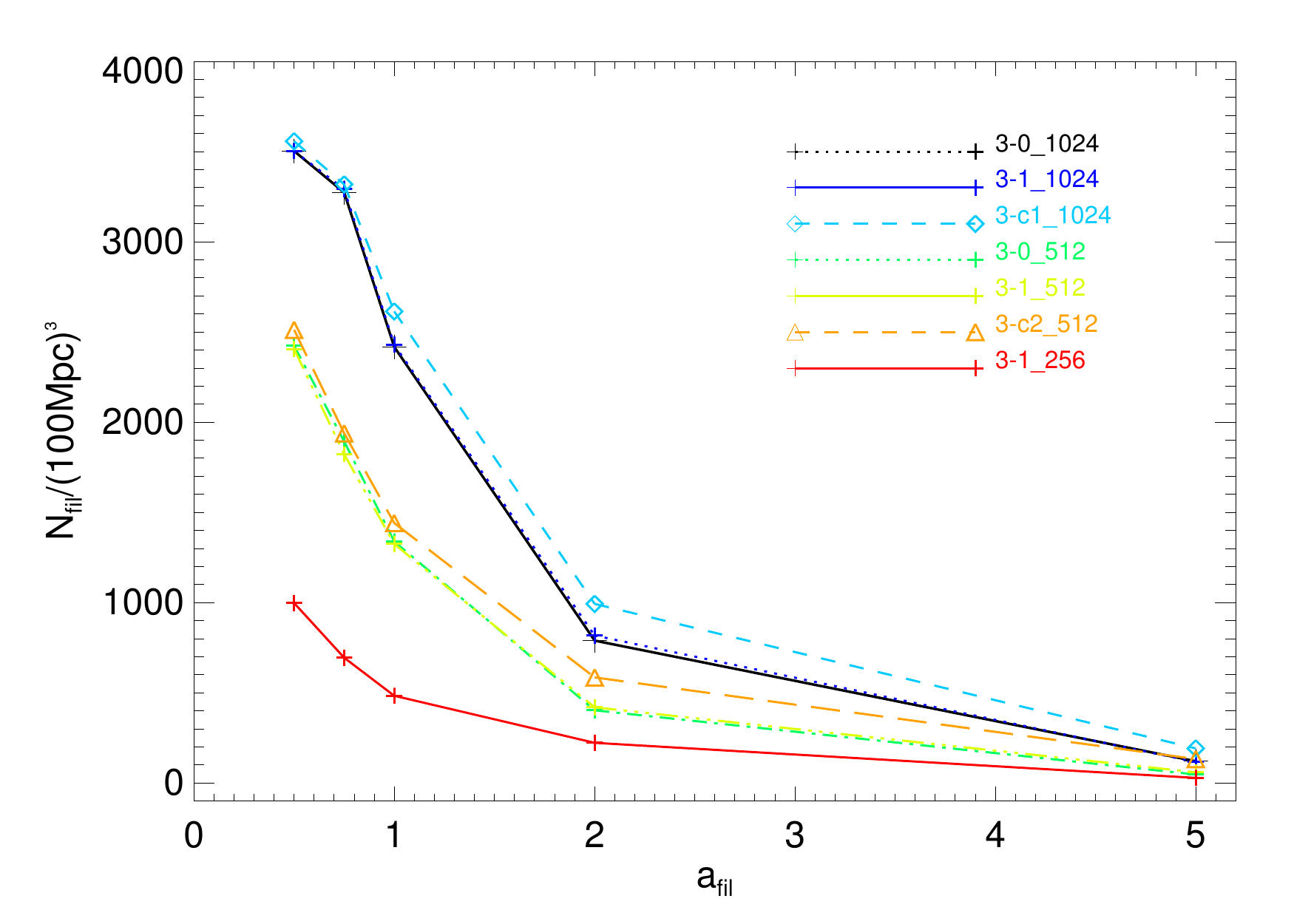}
  \caption{Number of filaments per 100 cubic Mpc as a function of $a_{\rm fil}$ }
  \label{fig:numbers}
\end{figure*}

Figure \ref{fig:numbers} shows the number of objects extracted from the 
simulations with box size of $(50 \rm Mpc/h)^3$ as a function of $a_{\rm fil}$. Results obtained with computational meshes
of $1024^3$, $512^3$ and $256^3$ cells are presented, in order to investigate 
also the impact of spatial resolution on the results. 
Spatial resolution directly affects the number of objects identified by our algorithm, the differences decreasing
at large values of the threshold parameter $a_{\rm fil}$.
The number of filaments increases by a factor of $\sim 3.5$ for $a_{\rm fil}=0.5$ going from the $256^3$ to the $1024^3$ run, while the increase
is only by a factor $\sim 2$ for the largest threshold, $a_{\rm fil}=5$.   
This shows that resolution strongly affects the identification of 
low-density filamentary structures below a given spatial scale, identifiable at about 0.2 Mpc,
due to the larger diffusion of the gas at that resolution, which  smoothens
the mass density in the outer parts of the filaments, connecting structures otherwise distinct and
leading to an overall decrease in the number of detected objects.
As a consequence, among our datasets (Table~1), we expect the resolution to affect mainly the
coarsest runs, with spatial resolution of $210 ~\rm kpc/h$ 
(1-0\_1024, 1-1\_1024 and 3-1\_256).

It is worth noting that for values of $a_{\rm fil}$ between 0.5 and 1, the various
physical processes acting in the different simulations have the smallest influence on the statistics. 
Furthermore, at the highest resolution ($1024^3$ cells) 
the value $a_{\rm fil} = 1$ represents an inflection point for all the runs, suggesting 
a change in the properties of the distributions under investigation.

\begin{figure*}
\centering
\begin{minipage}[c]{0.48\linewidth}
  \includegraphics[width=1.1\textwidth]{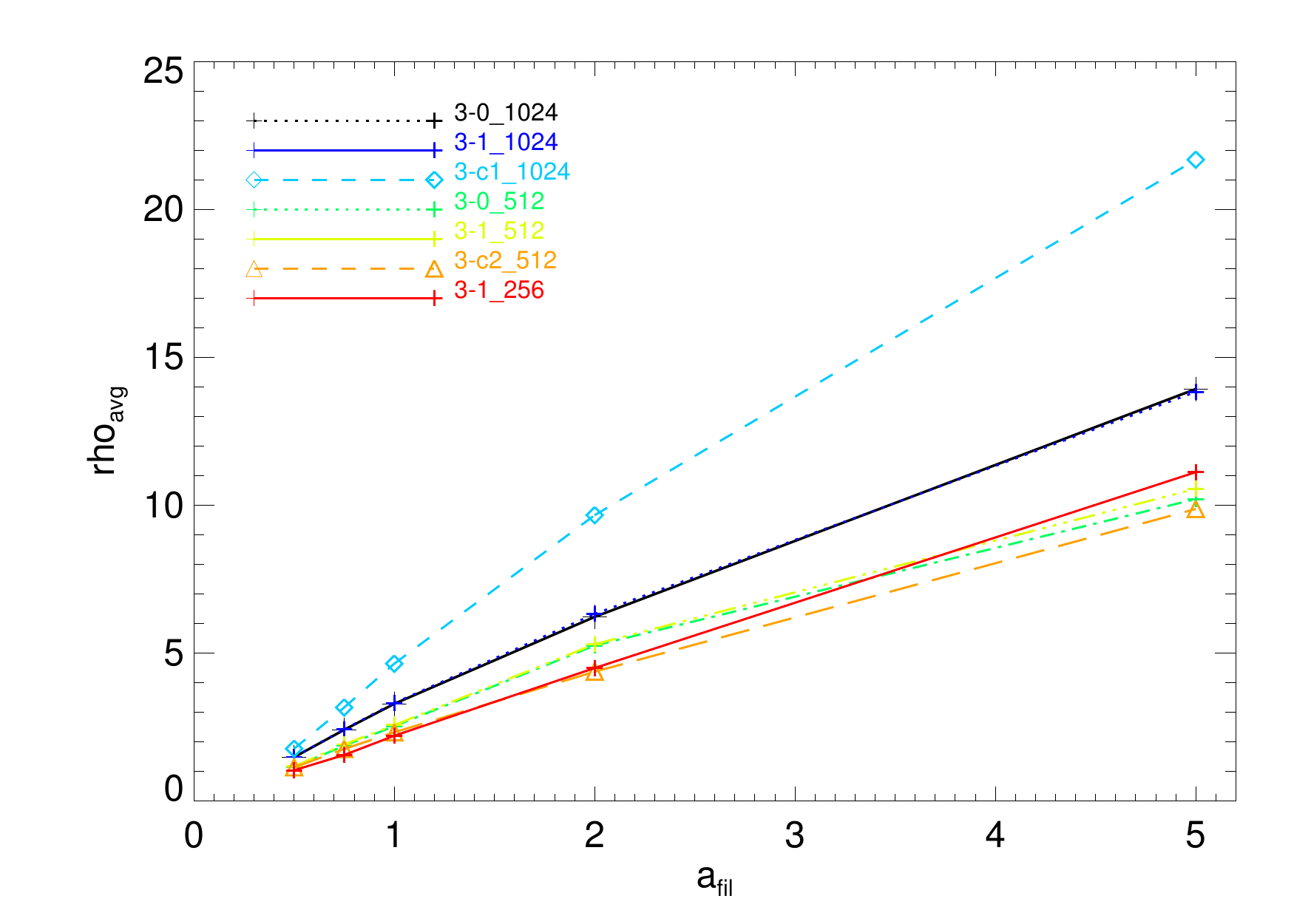}
\end{minipage}
\quad
\begin{minipage}[c]{0.48\linewidth}
  \includegraphics[width=1.1\textwidth]{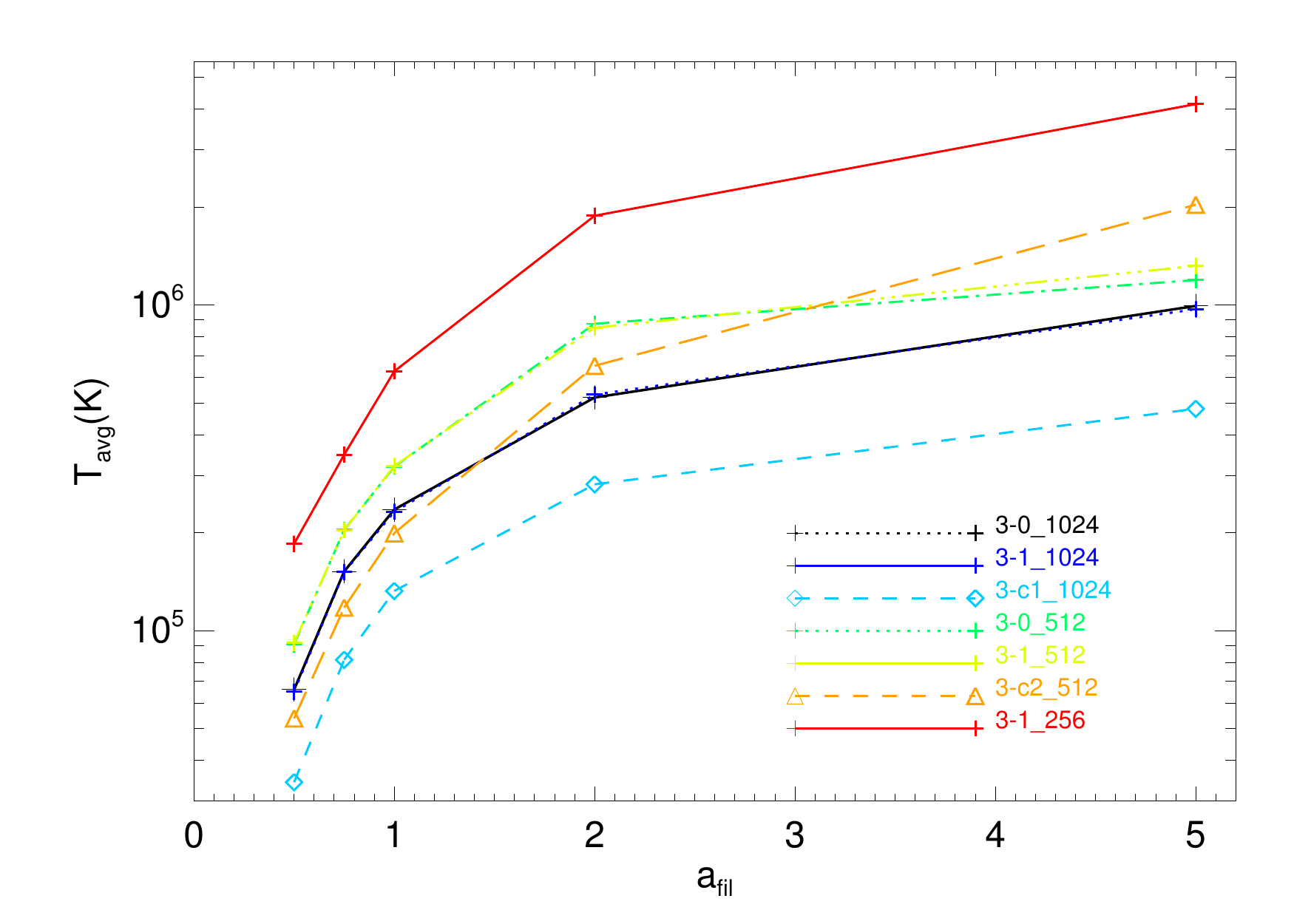}
\end{minipage}
\caption{Average baryonic mass density (left panel) and temperature (right panel)
in filaments as a function of $a_{\rm fil}$.}
\label{fig:denstemp}
\end{figure*}

Figure \ref{fig:denstemp} (left panel), shows the dependence of the average mass density of a filament,
$\varrho_{\rm avg}$, on the parameter $a_{\rm fil}$. As expected, the quantity $\varrho_{\rm avg}$
grows with increasing $a_{\rm fil}$ since lower-density regions are progressively
cut out.
The growth is almost linear and the slope higher for the ``c1'' model,
due to the cooling processes which dominate the AGN feedback. In model ``c2'',
that has effective AGN feedback, the effect of cooling is completely compensated
by the energy injection, which raises the pressure and stops the infall of matter. The runs
with CRs, ``0'' and ``1'', give similar results at all resolutions. This indicates
that CRs only have a small effect on the dynamics of the gas within filaments.
The model ``c1'' has densities higher than the other models, due to the much lower 
thermal support within the collapsing structures, leading to highly compressed matter
distributions. 
The trend with resolution of the mean gas density shows that for most of the investigated
threshold values $a_{\rm fil}$ (with the exception of $a_{\rm fil}>4$ extractions, which should be dominated by low-number statistics) the mean density increases with resolution, due to the higher achievable compression and the
increased presence of substructures within filaments.

The right panel of Figure \ref{fig:denstemp} shows the average temperature
of gas in filaments, $T_{\rm avg}$, as a function of $a_{\rm fil}$.
Also the temperature increases with the density threshold parameter, faster
for $a_{\rm fil} < 1.0$. At higher values, the temperature
grows slowly with density. Inner parts of the filaments are thermalized at an almost constant temperature
by shock waves propagating from the centre outwards. The outer part of the filaments, captured
only at low $a_{\rm fil}$, can comprise non-shocked cells that lower the average
temperature. The most striking feature of the graph is the trend of temperature with resolution.
The lowest resolution run ($256^3$) presents the largest average temperature, at all investigated $a_{\rm fil}$, and further increase in resolution is followed by a decreased mean temperature. This trend is caused by the weakening of outer accretion shocks (yielding a
lower thermalisation efficiency in filaments) when resolution is increased, an effect already mentioned in \citet{va14curie}.
On the opposite end, the model 3-c1\_1024 has the lowest average temperatures
at all values of $\varrho_{\rm avg}$, as a result of the strong effect of cooling. 
The ``c2'' model has a peculiar behaviour. At low densities, cooling tends to dominate AGN feedback and the resulting
average temperatures are clearly below all the other models (all but ``c1''). However, when higher-density
objects are selected, AGN heating dominates, leading to the highest temperatures
(neglecting the $256^3$ case).
The analysis of these results indicates that overall a value $a_{\rm fil} = 1$ is a 
suitable choice to identify filamentary structures. This choice of parameter is further supported by
the analysis of the density and mass profiles discussed in Section \ref{subsec:singleprop}. Hence, this will be the fiducial value for the rest of the paper.

\begin{figure*}
  \includegraphics[width=0.99\textwidth]{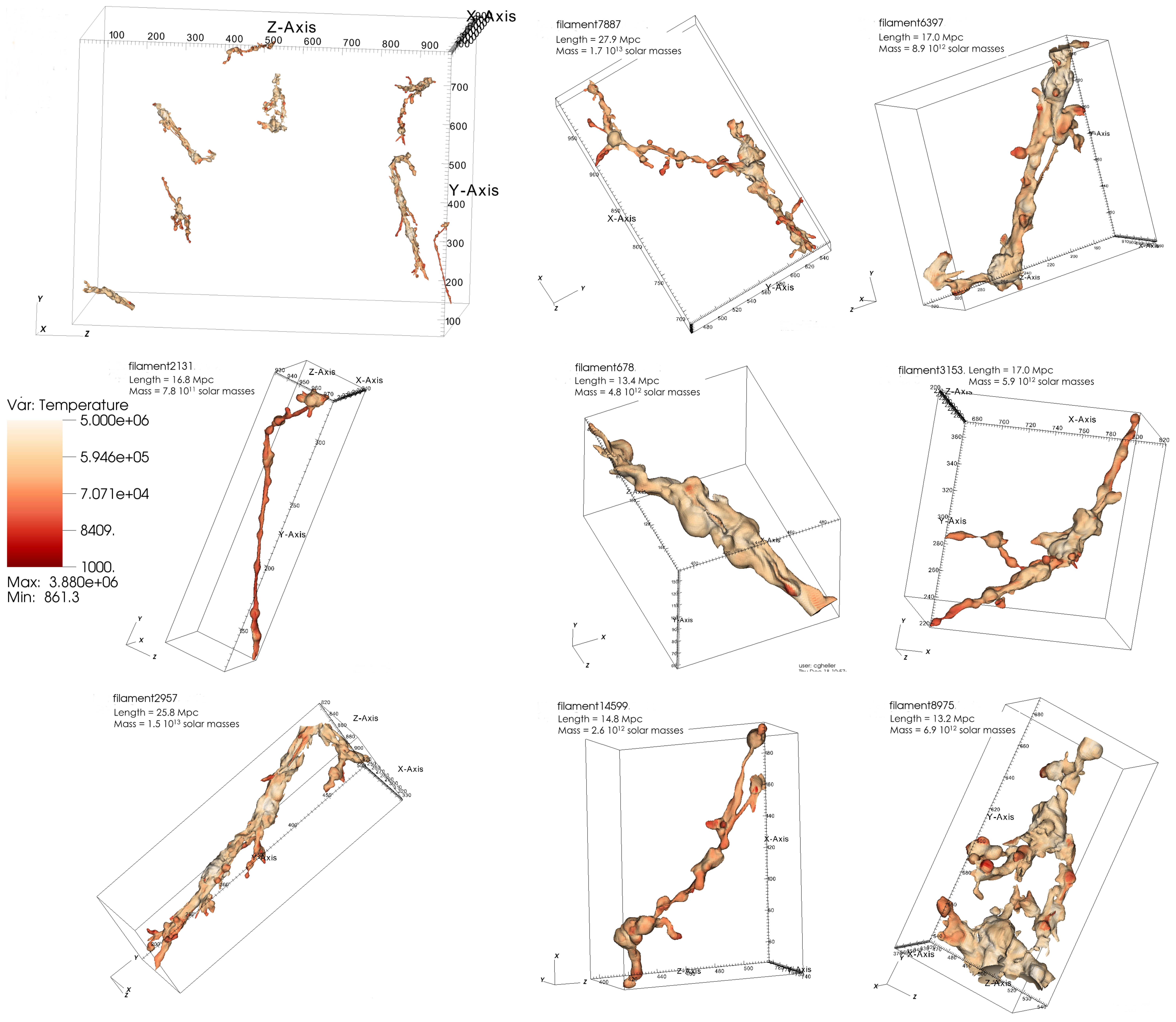}
  \caption{Eight filaments selected from simulation 3-1\_1024. Each object is identified
by a different ID (an integer number). The top left
panel shows the selected filaments altogether in the simulation box, for comparison.
The other panels are close-ups on each single filament. Colours represent the gas temperature in the outer surface of each object.}
  \label{fig:filaments}
\end{figure*}

\begin{figure*}
  \includegraphics[width=0.7\textwidth]{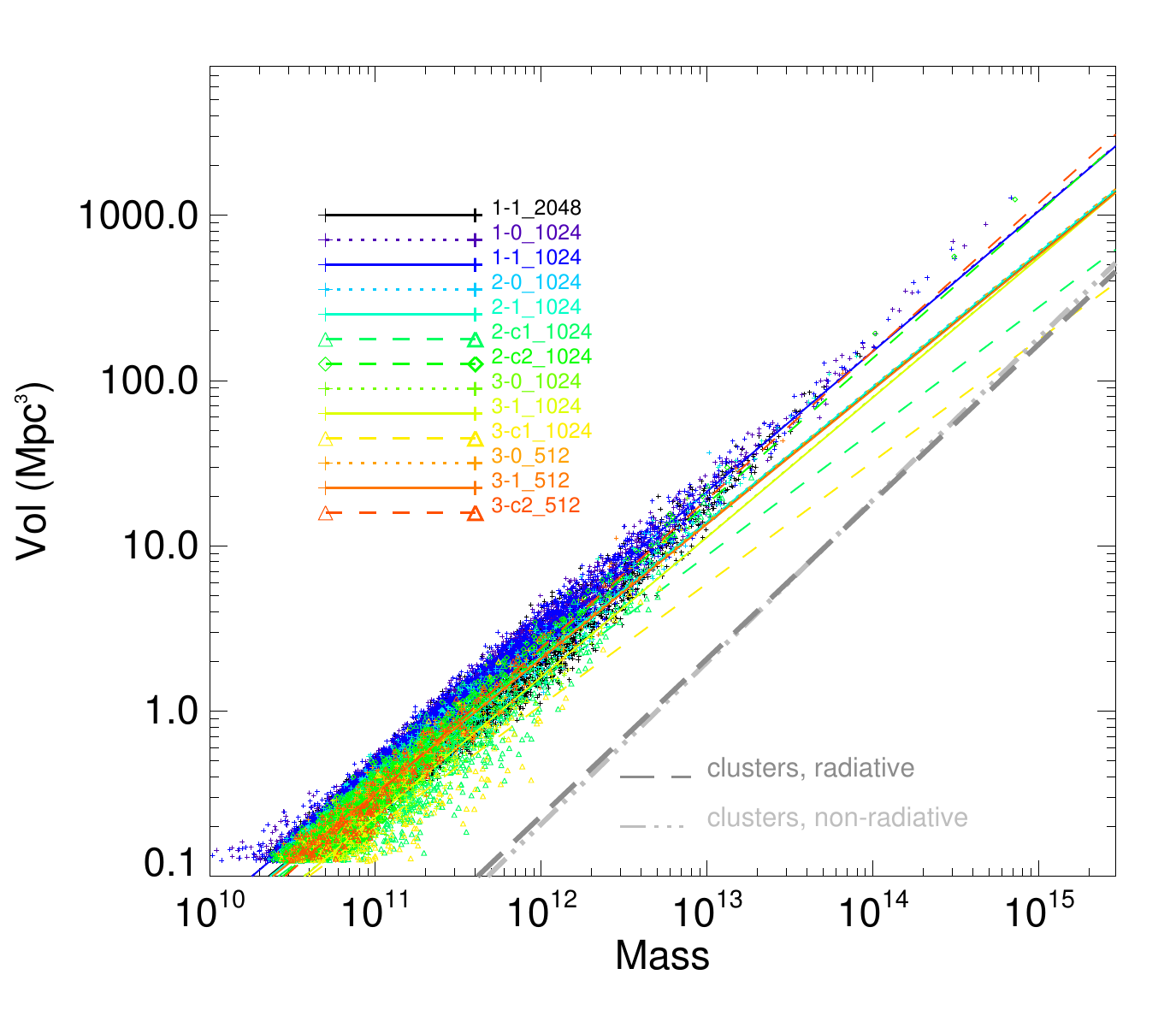}
  \caption{Relation between the enclosed gas mass and volume for the filaments in our sample (only 1/20 of filaments are shown for each model, for a clearer view). The additional lines in colour show the best-fit relation within each sample, while the two grey lines show the best-fit for the population of galaxy
clusters extracted from run 2-1\_1024 and 2-c1\_1024.}
  \label{fig:mv}
\end{figure*}

\begin{figure*}
  \includegraphics[width=0.45\textwidth]{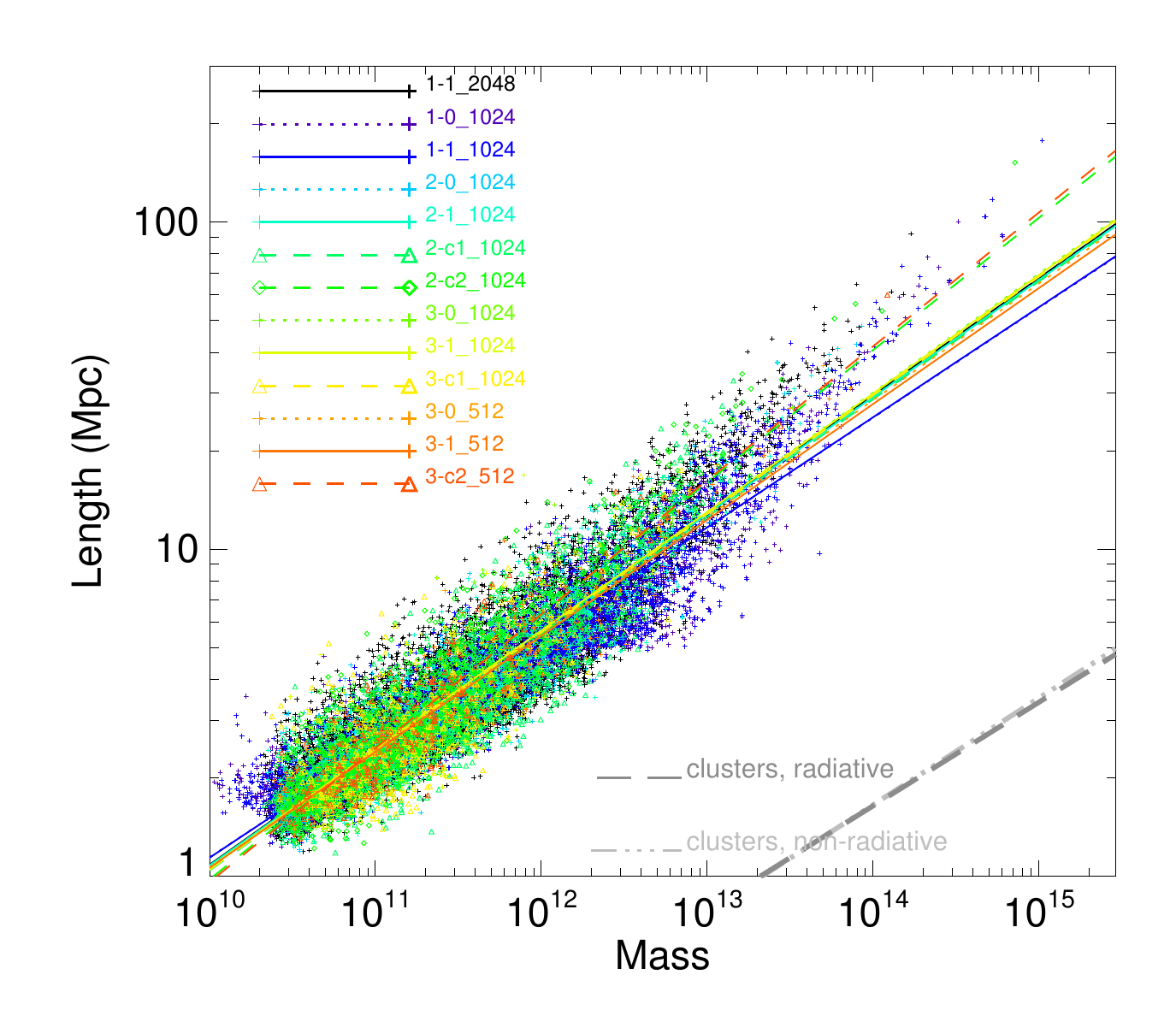}
  \includegraphics[width=0.45\textwidth]{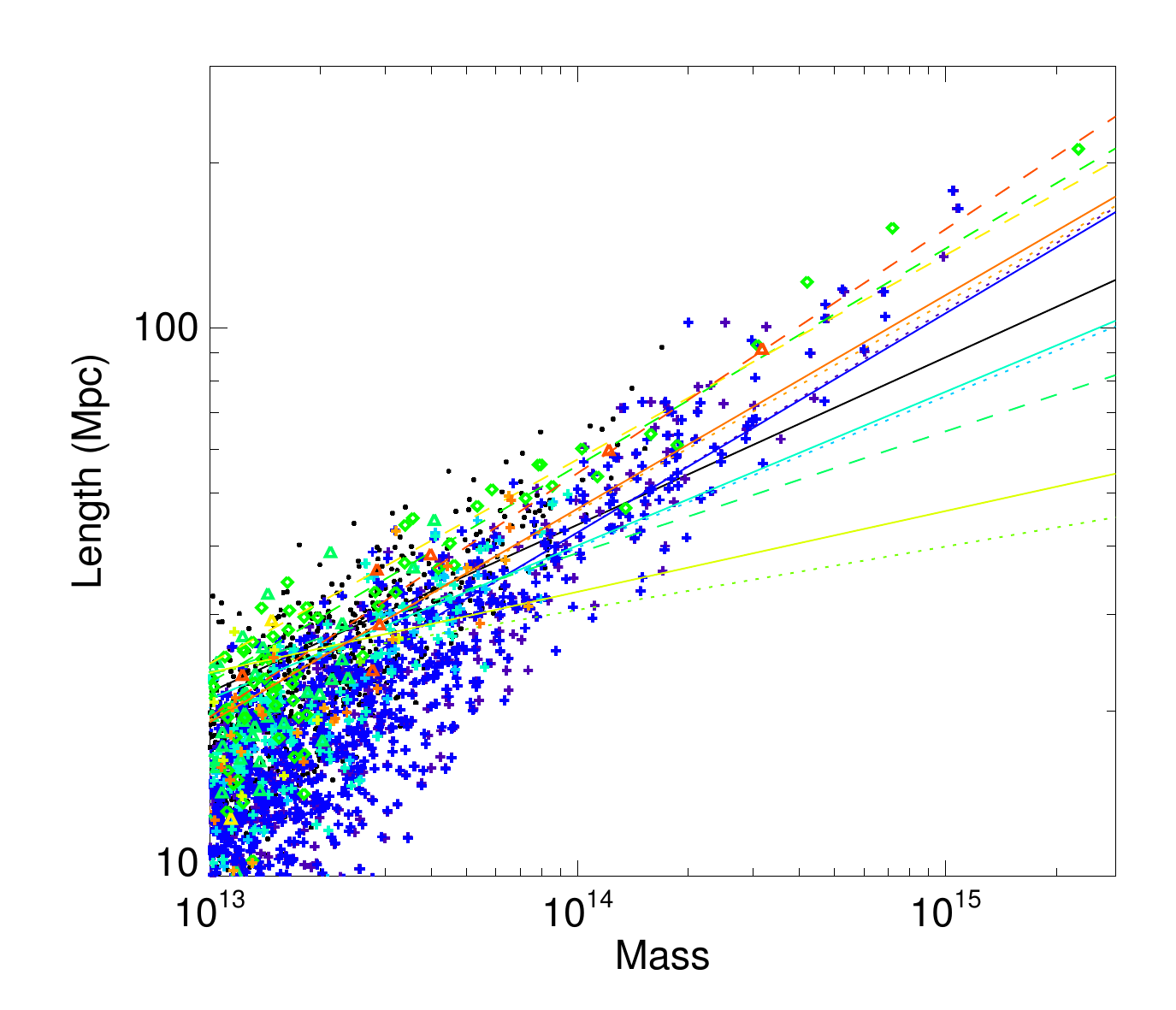}
  \caption{Relation between the enclosed gas mass and the estimated length of filaments in our sample. The left panel show the full range of values in the dataset (only 1/20 of filaments are shown for each model, for a clearer view). The additional lines in colour show the best-fit relation within each sample. The right panel only focuses on the objects with mass $M \geq 10^{13} \rm M_{\odot}$ (in this case all objects are plotted, same colour coding as in the left panel).
The length of each object is an estimate based on the bounding box of each filament, as described in Section \ref{sec:filaments}.}
  \label{fig:ml}
\end{figure*}

\begin{figure*}
  \includegraphics[width=0.45\textwidth]{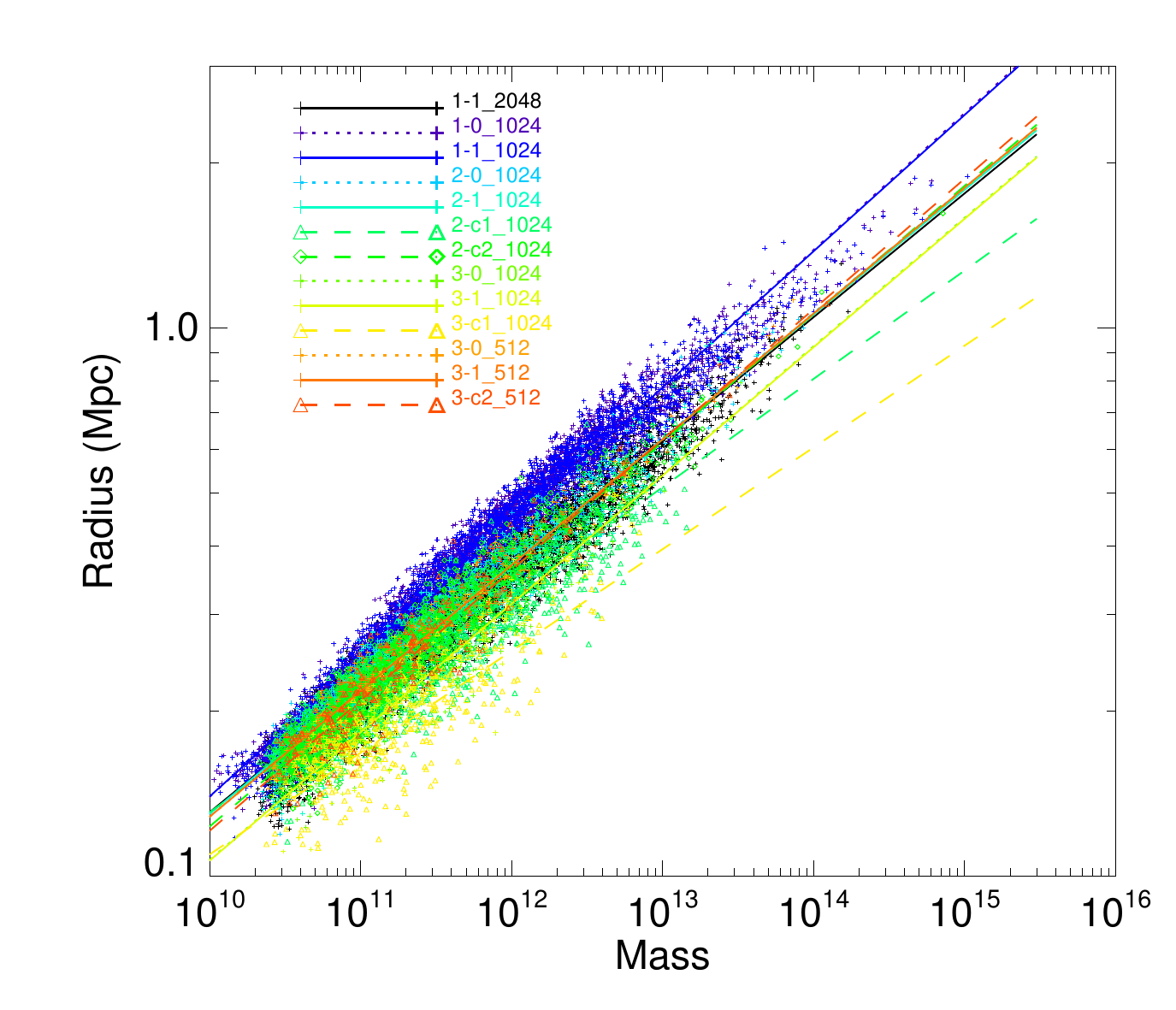}
  \includegraphics[width=0.45\textwidth]{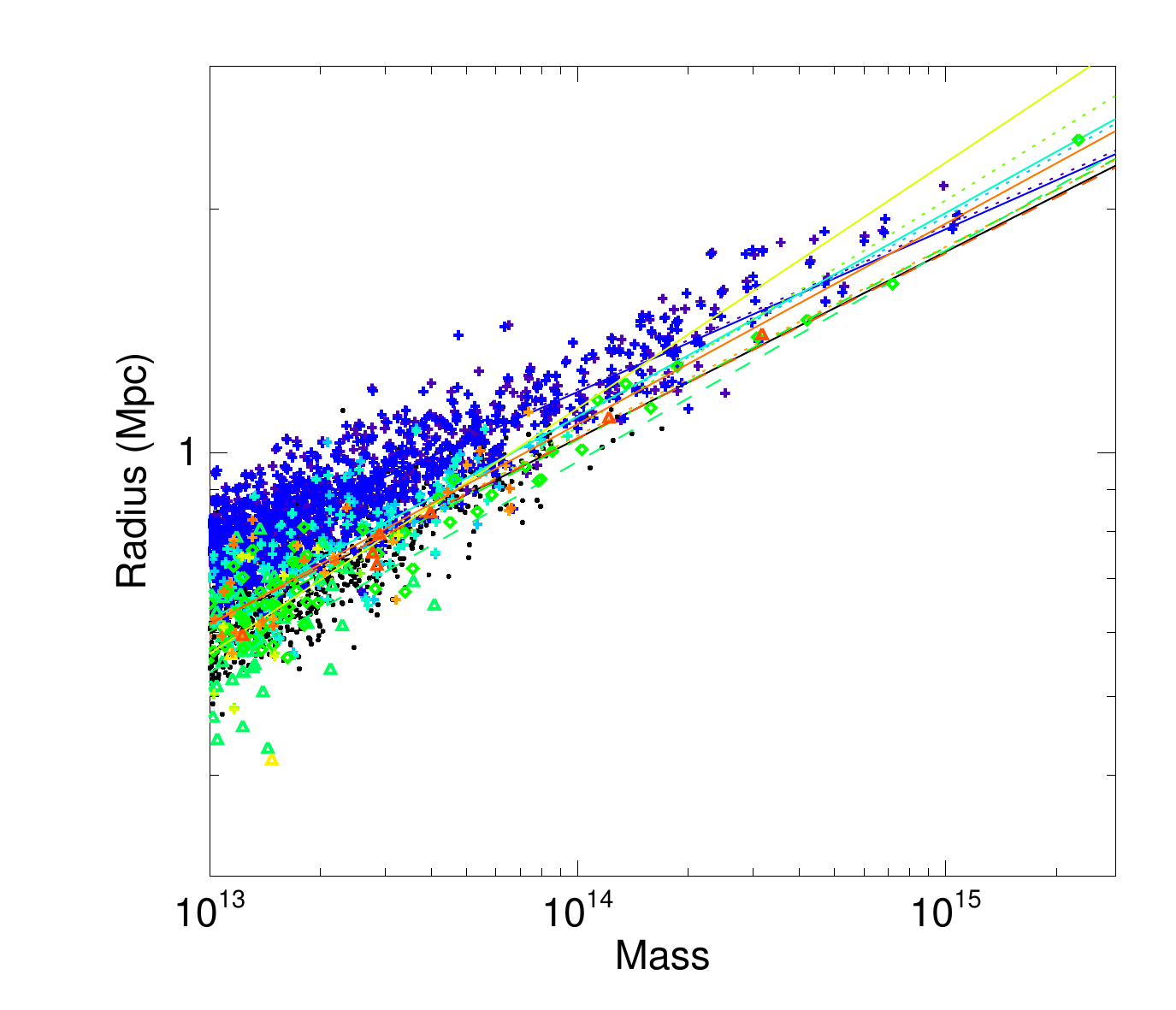}
  \caption{Relation between the enclosed gas mass and the estimated redius of filaments in our sample. The left panel show the full range of values in the dataset (only 1/20 of filaments are shown for each model, for a clearer view). The additional lines in colour show the best-fit relation within each sample. The right panel only focuses on the objects with mass $M \geq 10^{13} \rm M_{\odot}$ (in this case all objects are plotted, same colour coding as in the left panel). The fit for the 3-c1\_1024 model is omitted since statistically not meaningful (too few points available in the selected mass range). 
The radius of each object is estimated assuming cylindrical symmetry, as $R_{fil} = (V/\pi L)^{1/2}$, where $V$ and $L$ are the volume and the length of the filament respectively.}
  \label{fig:mr}
\end{figure*}

\begin{figure*}
  \includegraphics[width=0.7\textwidth]{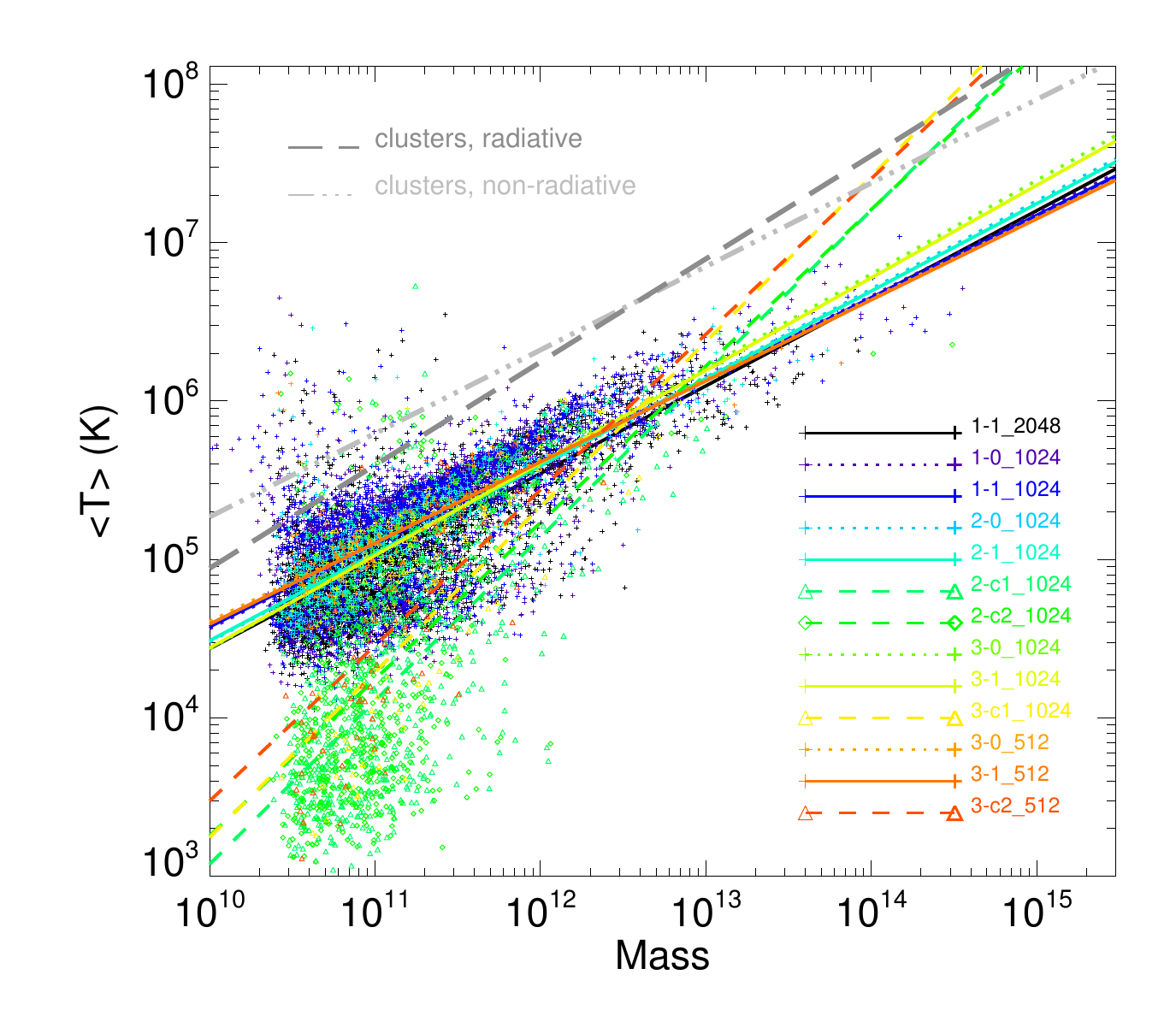}
  \caption{Relation between the enclosed gas mass and the average temperature for the filaments in our sample (only 1/20 of filaments are shown for each model, for a clearer view). The additional lines in colour show the best-fit relation within each sample, while the two grey lines show the best-fit for the population of galaxy clusters extracted from run 2-1\_1024 and 2-c1\_1024.}
  \label{fig:mt}
\end{figure*}

\section{Properties of the filaments}
\label{sec:prop}

By keeping our fiducial density threshold of $a_{\rm fil} = 1.0$ we identified all filaments in the simulated datasets 
and computed their average properties, as a function of resolution
and of the adopted physics. About 30000 filaments were identified in our largest run (1-1\_2048) and about 1000 objects in our highest resolution runs.

\subsection{Visual analysis}
\label{subsec:visual}

In Figure \ref{fig:filaments}, we show eight objects extracted from one of our highest resolution runs (3-1\_1024),
as a representative sample of the filaments identified by our procedure. 
The eight objects are between 10 and 28 Mpc long, the longest being that with ID=7887
(27.9 Mpc). In the top left
panel the selected filaments are presented together in order 
to show their location and to compare them to each other. 
The other panels are close-ups on each single object. Colours represent the temperature of the density isovolumes.
The images show how filaments can have heterogeneous geometries, depending on the environment in 
which they lie. In some cases, as for filaments 678, 6397, and 2957, they have elongated and rather
regular shapes, with comparable thickness and surface temperature. They have one or two main blob-like structures and they typically bridge pairs of massive galaxy 
clusters. We indicate such structures as ``giant bridges''. 
On the other hand, filaments 7887, 3153 and 14599 have a more branchy structure,
with multiple blob-like components and very thin segments, which present a systematically colder surface, and
several branches. 
Another peculiar type of filament is represented by filament 2131, which is  
more than $16$ Mpc long and is much thinner and colder than all the other objects.  
The nature of these objects require specific investigation to be fully understood. However,
Their features point towards filaments in a early evolutionary stage, with ongoing mass accretion, heated up
by a first generation of weak shocks propagating in a rather under-dense environment, and 
a total enclosed over-density which is just above our detection threshold. This class of filaments can be called ``long-thin''.
Finally, as in the case of filament 8975, we find objects for which
it is difficult to recognise the typical characteristics expected for filaments.  
They are found in rich regions and connect multiple larger halos. As for the other classes, they contain large 
almost-spherical matter clumps. Objects with these features are classified as ``irregular''. 

Irrespective of the classification, most of our filaments contains spherical clumps that have low average over-densities (around 20-50) and temperatures (around 0.1 keV).
Furthermore, they have radii smaller than 1 Mpc and masses around 10$^{12} M_{\odot}$ or lower. They can thereby be
considered as proper sub-structures of filaments, at least at the resolution we can achieve.

According to the above simple classification, we have estimated that about 30\% of the filaments in our simulation box
can be classified as ``giant bridges'', while approximately the 36\% are ``branchy''. ``Long-thin'' objects are less frequent, but still
account for the 12\% of the filaments. ``Irregular'' objects represent the 16\% of the population. It must
be stressed that this classification is performed visually on a limited sample of large objects (about 100 objects longer than $7$ Mpc), while smaller objects tend to be more difficult to classify, as resolution effects coarsen their substructures and smooth out their shape.

The remaining objects comprise isolated 
clumps emerging from the gravitational collapse process and spurious leftovers of the 
filament reconstruction, in particular at the edges of our boxes (see also Section \ref{subsec:statprop}). This also affects the low-mass end of the M-T relation as will be discussed in the following sections.

\subsection{Statistical properties of the filaments}
\label{subsec:statprop}

Figure \ref{fig:mv} shows the distribution of the gas mass of filaments  versus volume. 
Most objects fall along a narrow scaling relation, fitted by $V \approx 1000$ Mpc$^3 \cdot M/(2 \cdot 10^{14} M_{\odot})$.  For comparison, we also show the best fit relations for the case of galaxy clusters, extracted 
from the 2-1\_1024 and 2-c1\_1024 runs (for a total of $\sim 300$ objects in the gas-mass range $10^{12} \leq M/M_{\odot} \leq 10^{15}$). 
Independent of resolution and physics,  a self-similar scaling relation for filaments is found, as for galaxy clusters, yet with a higher normalisation.  
However, the normalisation of the best fit is smaller for the runs with efficient cooling (``c1" models), implying that cooling produces significantly more compact filaments, and that this removes more volume from the WHIM phase and locks it into overcooling halos, in turn reducing the volume of the filaments.

\begin{figure*}
  \includegraphics[width=0.7\textwidth]{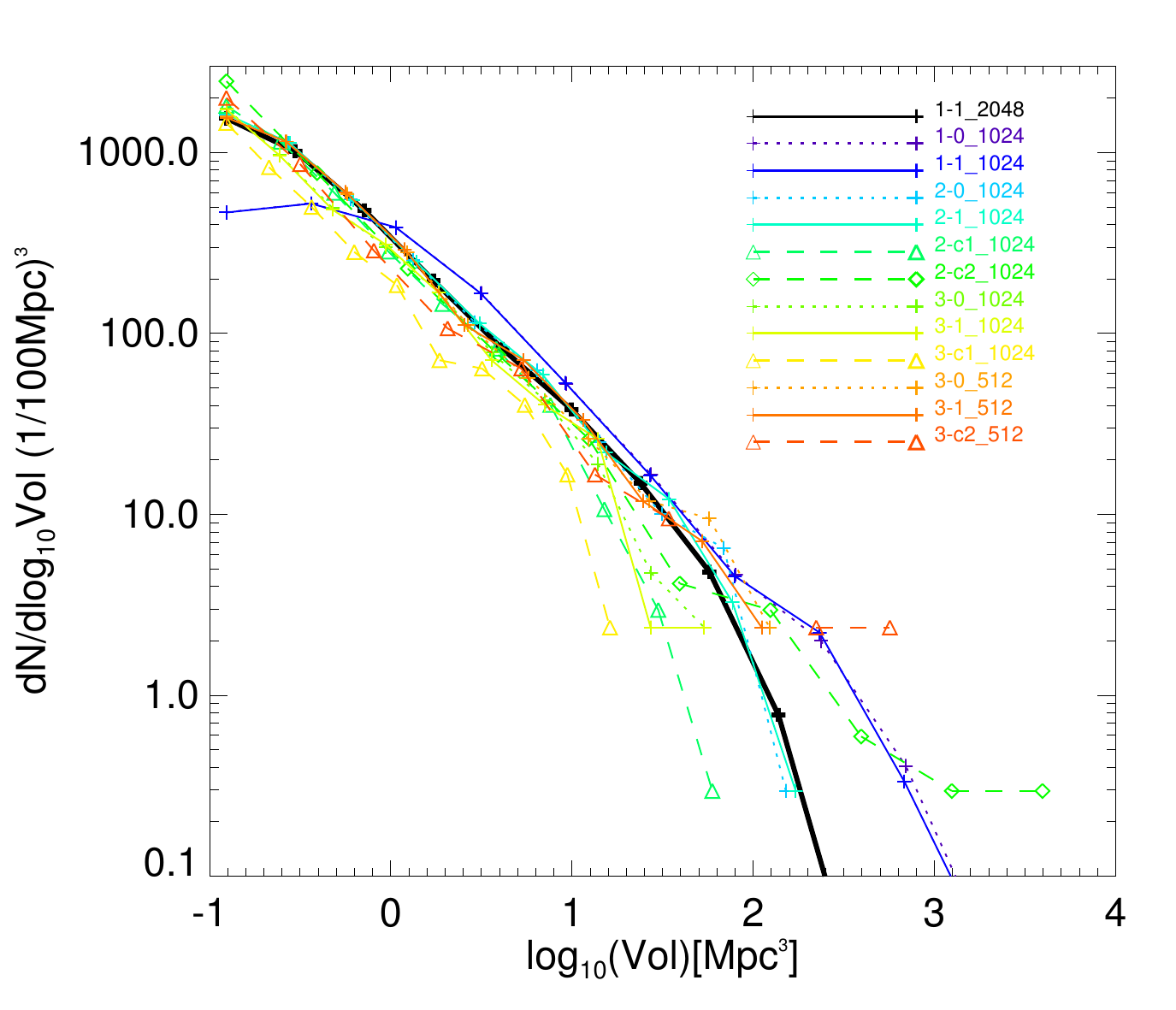}
  \caption{Number of filaments normalised to 100 cubic Mpc as a function of their volume.}
  \label{fig:voldist}
\end{figure*}

\begin{figure*}
  \includegraphics[width=0.7\textwidth]{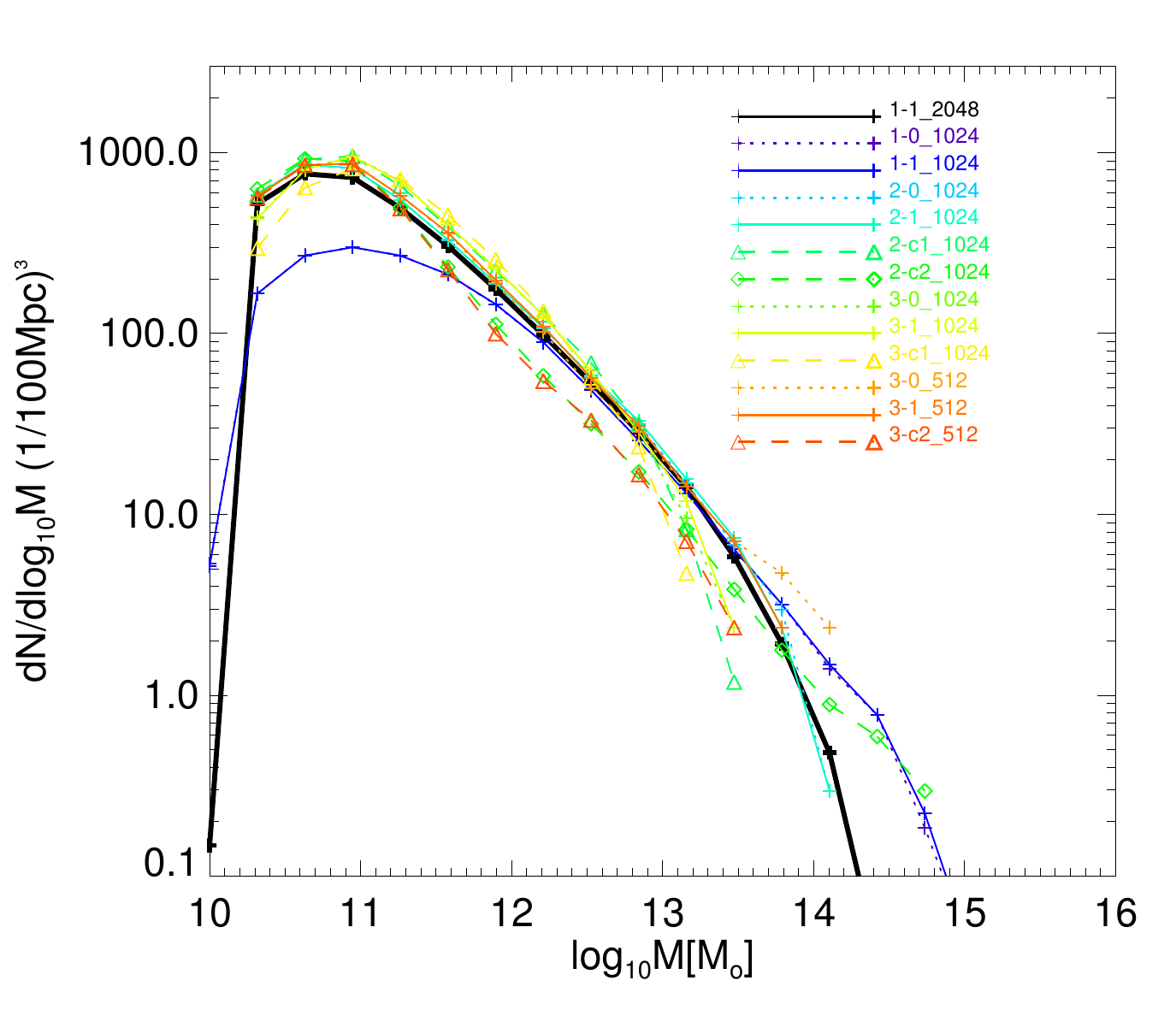}
  \caption{Number of filaments normalised to 100 cubic Mpc as a function of their gas mass}
  \label{fig:massdist}
\end{figure*}

\begin{figure*}
  \includegraphics[width=0.7\textwidth]{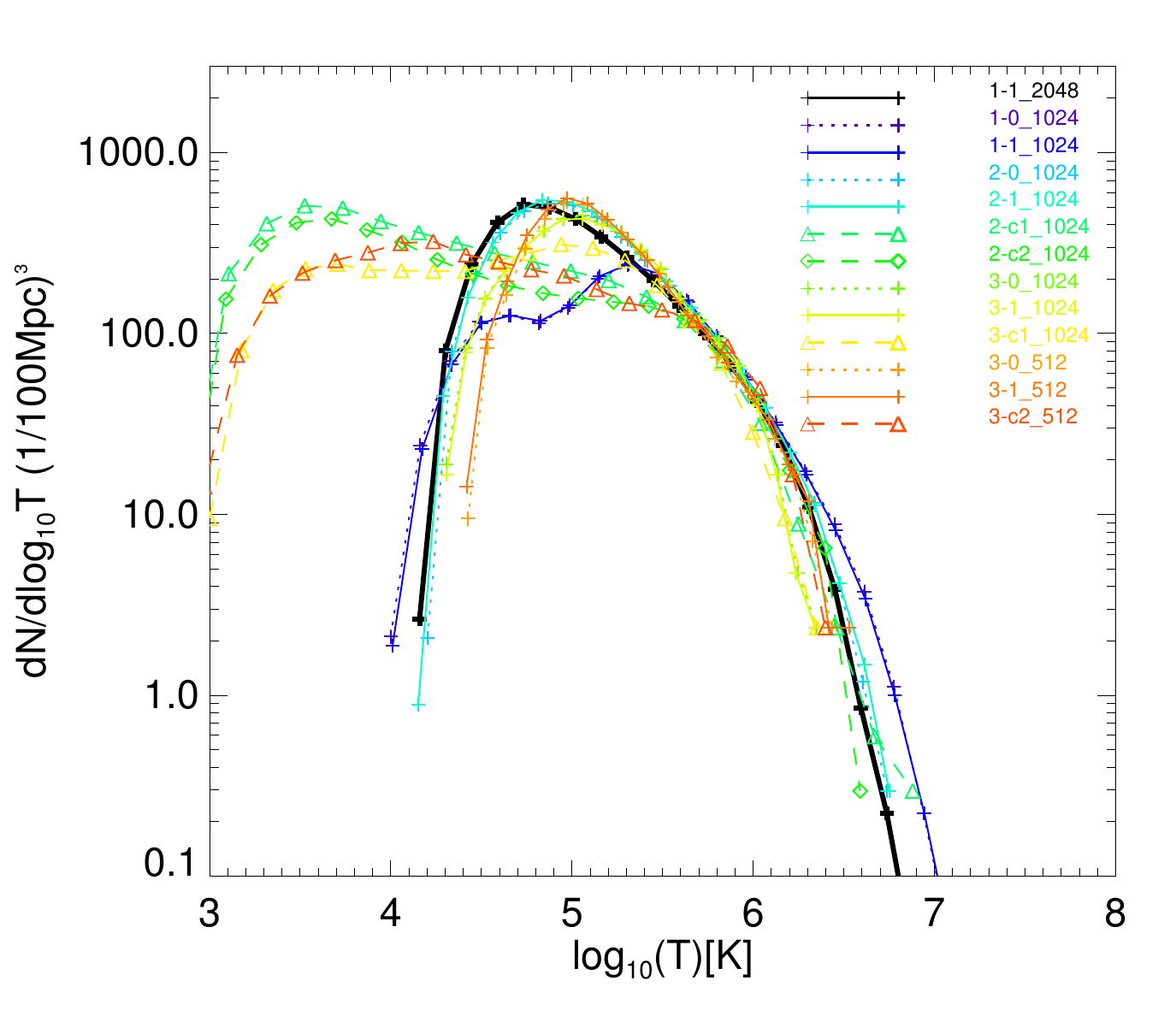}
  \caption{Number of filaments normalised to 100 cubic Mpc as a function of average temperature}
  \label{fig:tempdist}
\end{figure*}

Figure \ref{fig:ml} gives the distribution of gas mass versus the length of each filament, which is 
approximated using the diagonal of the bounding box defined by the reconstruction algorithm (Section \ref{sec:filaments}). 
The distribution for the full sample of each run is also described by a well-defined power-law with a rather narrow dispersion
of values around the mean, and with a self-similar scaling with mass, $L \propto M^{1/3}$. 
The presence of unresolved small-size halos embedded within irregular filaments can artificially steepen the relation by increasing the total
mass of gas in filaments, while for higher masses their contribution becomes negligible. Therefore, we also show the 
distribution and best fit (which become just a bit steeper, but still consistent with the $\propto M^{1/3}$ scaling) for the distribution
of filaments limited to $\geq 10^{13} M_\odot$ objects, which are characterised by $L \geq 10 ~\rm Mpc$. 
However, notice that our method systematically underestimates lengths, hence this trend represent a lower 
bound for the mass-length relation.  The runs with cooling show a rather steeper scaling, suggesting that filaments of the same length are 
less massive than in non-radiative runs. This is easily explained by the mass drop-out of cooling gas onto
clumps within filaments, which can significantly remove a fraction of the WHIM from the most diffuse phase, and at the same time
promote the full collapse of small-size halos and their detection by our algorithm. 

The radii of the filaments, presented in Figure \ref{fig:mr}, are estimated assuming cylindrical symmetry, such that
$R = (V/\pi L)^{1/2}$, where $R$, $V$ and $L$ are the radius, the volume and the length of the filament, respectively.
This is a rough estimate since many objects are far from a simple cylindrical geometry. However, it provides an indication
of the characteristic transversal size of a filament and its distribution with mass, with radii
between 0.1 and 0.3 Mpc for objects up to $10^{12}  M_{\odot}$, growing to more than 1 Mpc for the largest filaments,
with masses between $10^{14}$ and $10^{15} M_{\odot}$. 
If resolution is increased, the radii of filaments shrink, as indicated
by comparing the 1-0\_1024 and 1-1\_1024 to the 1-1\_2048 interpolation fits.
This affects in particular smaller objects, as shown by 
the right panel of Figure \ref{fig:mr}, where we present the radius-mass relation for the sub-set of filaments
with masses above $10^{13} M_{\odot}$. The transverse size of these large filaments is less affected
by resolution. This indicates that the properties of those objects whose radius is comparable to the cell size
(low mass filaments) are not fully converged with spatial resolution.
Also radiative cooling
(run c1) leads to narrower filaments, with the smallest radii associated with 3-c1\_1024 and 1-c1\_1024.
This effect is compensated by the efficient AGN feedback in the c2 models.

\begin{figure*}
  \includegraphics[width=0.7\textwidth]{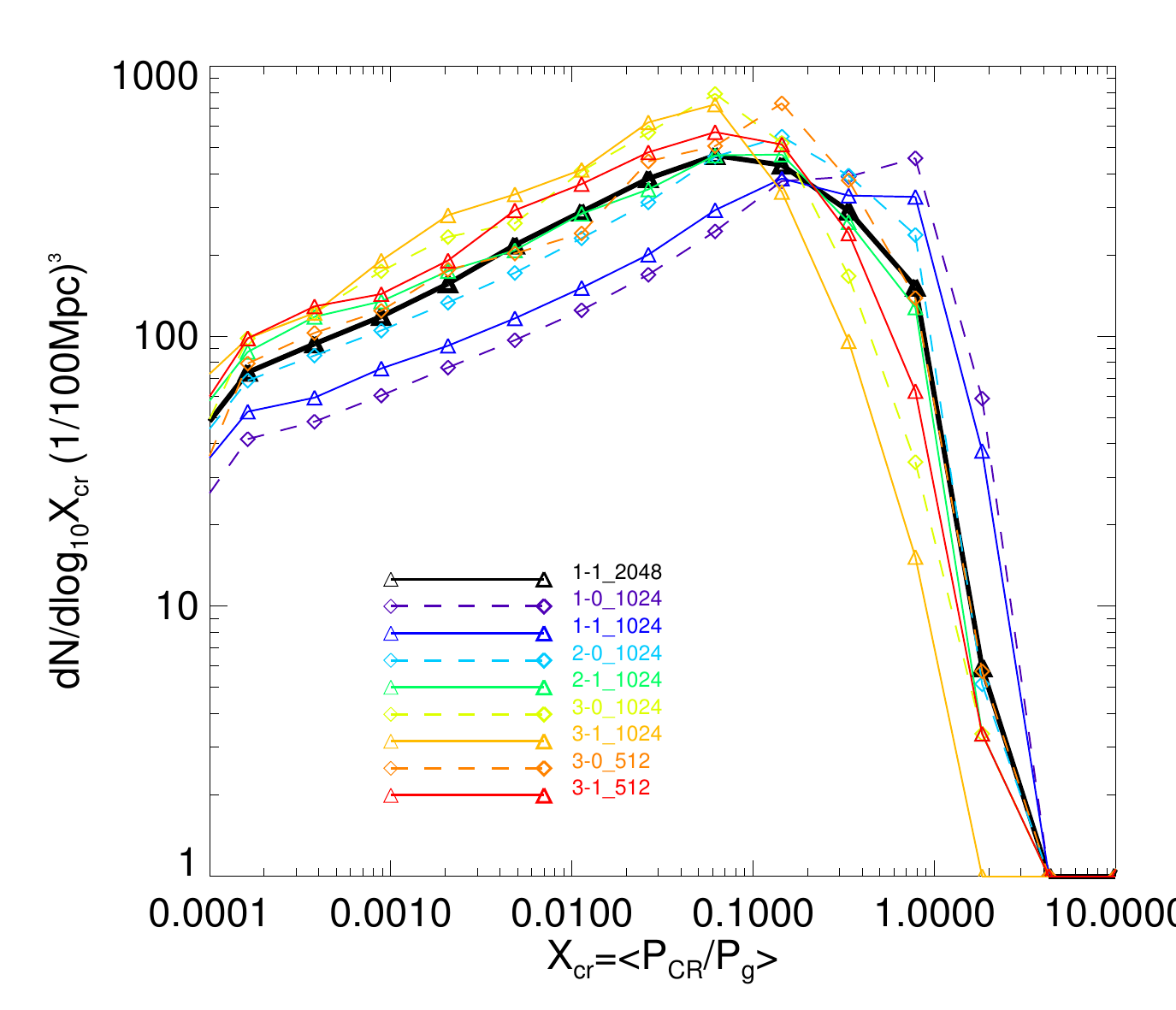}
  \caption{Number of filaments normalised to 100 cubic Mpc as a function of the pressure ratio between CR energy
and gas energy. }
  \label{fig:ecrpdist}
\end{figure*}

For the same objects,  we  show in Figure \ref{fig:mt} the relation between the enclosed gas mass 
and the volume-averaged gas temperature. 
Here a less defined scaling relation is seen.
The overall distribution follows a scaling law similar to galaxy clusters in the 
non-radiative case ($T \sim M^{2/3}$), with a  $\sim 1$ order of magnitude lower normalisation. 
However, a small fraction of the identified objects has unexpected properties, with low masses (below 10$^{12} M_\odot$) 
and high temperatures 
above 10$^5$K). These outliers are due to our reconstruction procedure, which does not support periodic 
boundary conditions. They correspond to small leftovers of larger (and hotter) objects crossing one of the 
sides of the computational box and there cut in two distinct parts. 

Variations of CR physics produce extremely small effects, and the scaling relations are essentially the same for the ``1" and ``0" models. 
Instead, the role played by gas cooling and feedback from AGN becomes much more relevant compared to the previous scaling 
relations since temperature is directly affected.
As for galaxy clusters, runs including cooling and AGN feedback produce 
a steepening of the scaling relation in the low-mass end of the distribution. 
Due to cooling, the average temperature at all masses is reduced significantly, 
and also the best-fit relations for the runs including cooling are steeper than in the non-radiative case. 
The effect is even bigger than in the case of galaxy clusters, 
as shown by Figure \ref{fig:mt}, comparing the best fit for clusters and filaments in the same volume. 
This can be explained considering that first, below $T \leq 10^{5}$ K the radiative cooling is strongly affected also by  
line cooling and second that although AGN feedback is effective
within clusters and can reach out to the WHIM in filaments, it is obviously less efficient there given the 
distance (several $\sim$ Mpc) from its release.  
As also pointed out in the next Section, the cooling efficiency in this regime is probably overestimated in our case,  
since we assumed a (constant) metallicity of $0.3 Z_{\odot}$ everywhere in the simulated volume. 
Furthermore, the mass dropout into forming stars is not included. \\

The volume/mass/temperature number distributions for all objects are given in Figures \ref{fig:voldist}-\ref{fig:tempdist}. 
Apart from a deficit of objects with $V \leq 1$ Mpc$^3$ in the 1-0\_1024 and 1-1\_1024 runs (due to
coarse resolution effects, Section \ref{sec:tuning}),  the volume distribution of objects is very similar in all
runs and is described by a simple $\rm log_{\rm 10}(N_{\rm fil}) \propto V^{-1}$ relation across $\sim 3$ decades in volume (Figure \ref{fig:voldist}). The various curves show a good numerical convergency down to gas masses of $\sim 10^{11} \rm M_{\odot}$ indicating that our reconstructed distribution of objects is complete down to this mass.
The total volume occupied by filaments ranges from $3.8$ to $4.5\%$ of the computational volume in all our runs.\\

\begin{figure*}
  \includegraphics[width=0.95\textwidth]{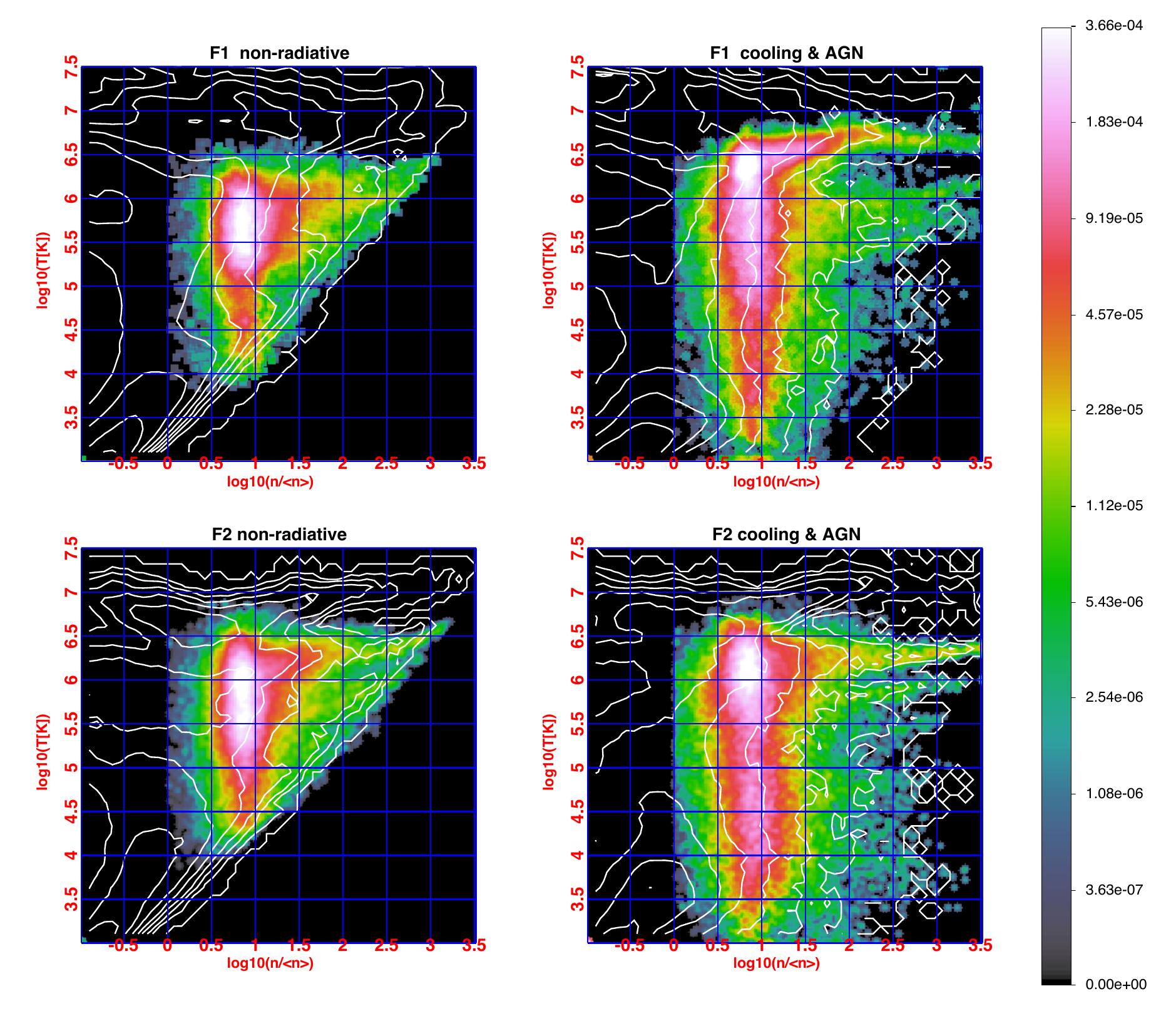}
  \caption{Phase diagram ($\rho$,$T$) for the filaments F1 (top panels) and F2 (bottom panels), comparing
 the non-radiative run and in the cooling and feedback run (c2). The colour coding gives the fraction of volume characterised by a given phase. For each object we also show the contours of the phase diagram for a larger $\approx (35 \rm ~ Mpc)^3$ box in order to compare with the entire distribution of cosmic baryons.}
  \label{fig:obj_phase}
\end{figure*}

When physical resimulations of the same volume are compared, we we find a significant excess of large-volume filaments when the more efficient ``c2" feedback model
is used, for $\geq 100$ Mpc$^3$ as well as a significant deficit of halos in the range 10 Mpc$^3\leq V \leq 100$ Mpc$^3$.
This is due to the powerful AGN feedback at high redshifts which leads to the expulsion of gas from
the centre of dense proto-clusters and affects the properties of a few tens of massive filaments around the most massive halos in the box. The efficient feedback releases hot gas beyond the radius achieved by non-radiative runs, with a strong impact on the densest regions, where the AGN feedback mostly takes place. 
This enriches even rarefied environments with additional gas, and leads to the junction of otherwise disconnected structures through the extra gas. 
A more detailed analysis will be given in the next section (e.g. Figure \ref{fig:obj_map}). \\
On the other hand, the effect of radiative cooling is enough to produce more compact objects within filaments than in the non-radiative case, thereby causing an increase of the low-volume objects in the sample. Both effects are relevant
for any filament detection scheme based on gas. Very similar trends are found for the distribution of filament mass (Figure \ref{fig:massdist}).
We conclude that on average only one filament with $M\geq 10^{14} M_{\odot}$ and $V \geq 1000$ Mpc$^3$ is found within a volume of (100 Mpc)$^3$, while within the same volume we can find $\sim 10$ objects with $M\geq 10^{13} M_{\odot}$ and $\sim 10^2$ objects with $M\geq 10^{12} M_{\odot}$. \\

The distribution of temperatures (Figure \ref{fig:tempdist}) clearly shows the imprint of radiative cooling
on the WHIM. Cooling removes a significant part of the cosmic population of filaments
from the soft X-ray bands and pushes it much below $10^{5}-10^{6}$ K (still above the minimum level set by our assumed UV heating background, which had dropped off at a level of a few $\sim 10^2 \rm K$ at $z=0$), as already seen in the average temperatures in Figure \ref{fig:denstemp}. 
The coarse resolution in the 1-0\_1024 and 1-1\_1024 runs is responsible for an excess of large temperatures compared to the other runs with a higher resolution.\\
Finally, we calculate the ratio between the thermal gas pressure and the pressure of CR-protons accelerated by shock waves, as in \citet{va14curie}. Figure \ref{fig:ecrpdist} gives the number distribution of filaments and compares all our ``1" and ``0" models for the non-radiative case. For the bulk of the population the pressure ratio within
the volume of filament is $P_{\rm CR}/P_{\rm g} \sim 0.1-0.2$, with a tendency of larger filaments
to have lower ratios. As expected, the lower efficiency ``1" model \citep[][]{kr13}
produces a smaller CR-pressure compared to the higher efficiency ``0" model \citep[][]{kj07},
However, the differences are small. At the same resolution, the two distributions differ by $\sim 20-30$ percent. 
This is expected because in the different simulations the enrichment of CRs into filaments is equally strong, owing to the fact that
 the saturated efficiency of the two acceleration models is only different by a factor $\sim 30$ percent in our treatment \citep[][]{va14curie}. 
The effect of increasing resolution is to lower the average pressure ratio due to the overall weakening of shocks, which is made manifest by the shift
 of the peak of this relation to lower values, going from the coarse 1-0\_1024 and 1-1\_1024 runs (where the number of objects is also found to be smaller than in the 
 other runs) to the 3-0\_1024, 3-1\_1024 and 3-c1\_1024 runs.
This suggests
that very large values of the pressure ratio are mostly driven by resolution effects on the shock population \citep[][]{ry03,va11comparison}.
However, thorough investigations on resolution effects in the distribution of CRs in these runs  have been already presented \citet[][, Figure 11]{va14curie} and showed that $P_{\rm CR}/P_{\rm g}$ is a rather well converged quantity (within a factor $\sim 2-3$) for all across the full range of cosmic environment, for spatial resolution equal or better than $\sim 200-300 \rm kpc$, i.e. well in the range of what explored by the runs we use in this work.
 Overall, these findings are in 
agreement with \citet[][]{va14curie}, and
indicate that, provided diffusive shock acceleration (DSA) can take place at low overdensities, cosmic filaments can store a large amount of CRs. 

\subsection{Properties of individual filaments}
\label{subsec:singleprop}

\begin{figure}
  \includegraphics[width=0.495\textwidth]{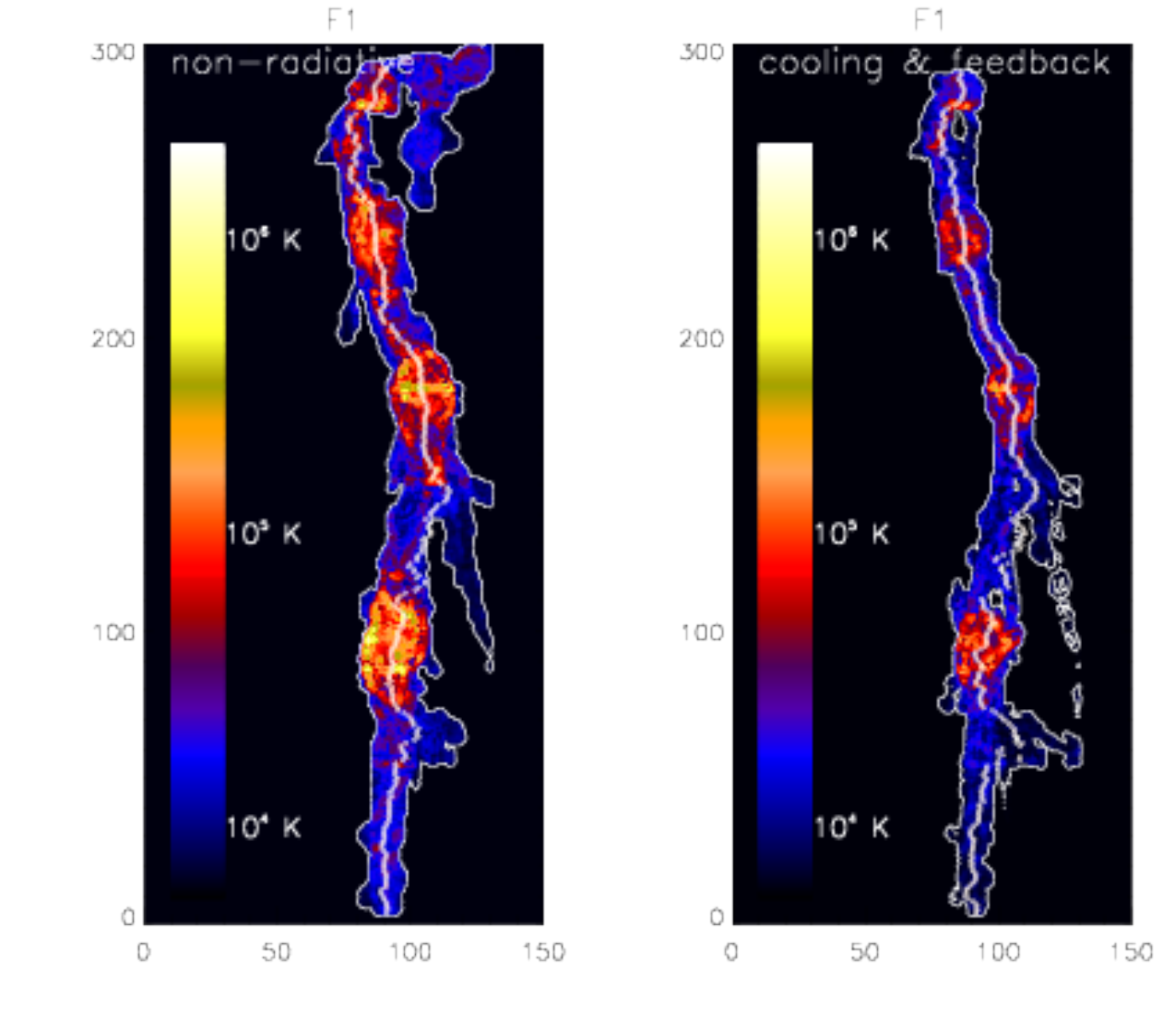}
  \includegraphics[width=0.495\textwidth]{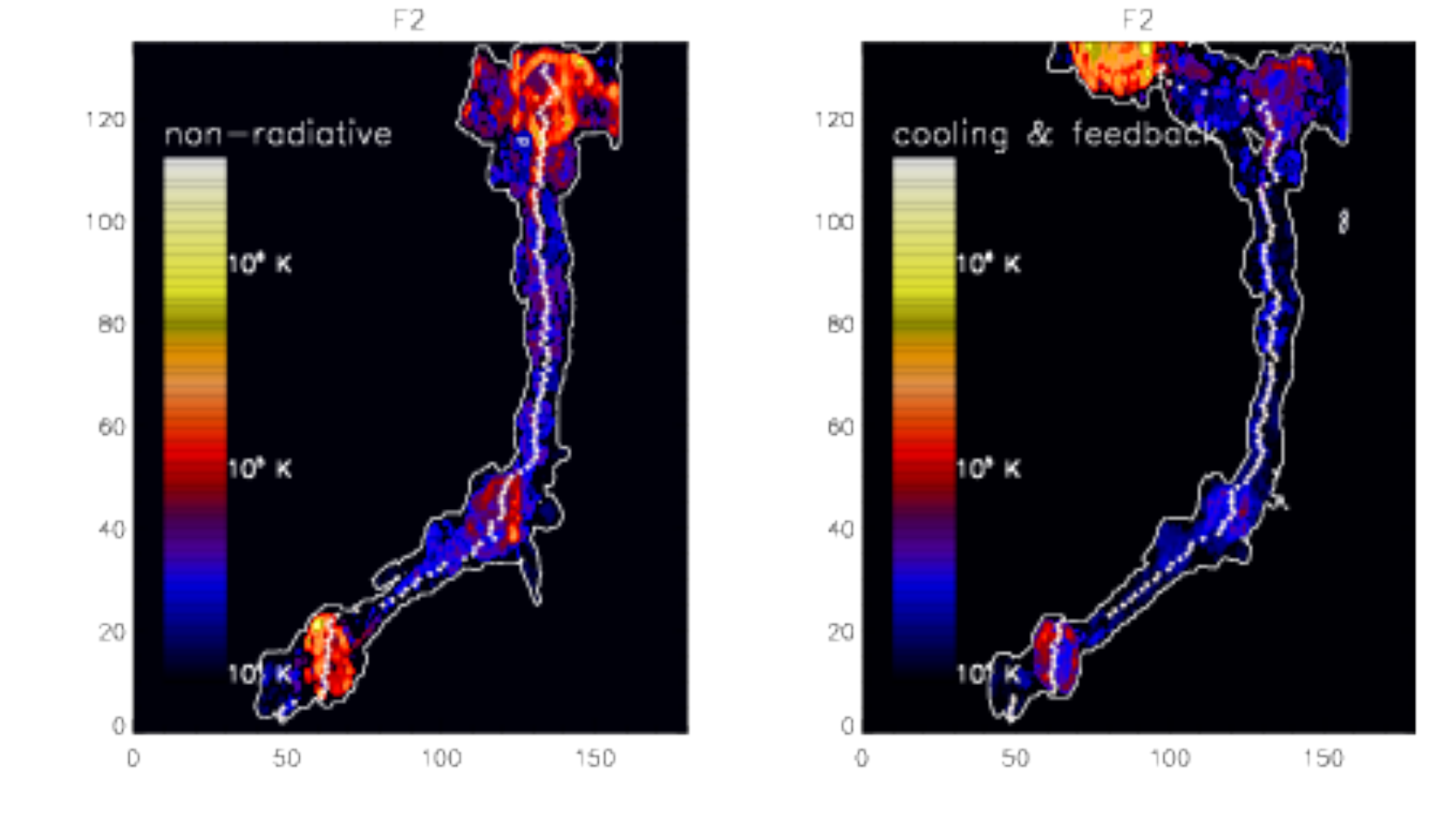}
  \caption{Maps of projected temperature for filament F1 and F2, for both non-radiative (left) and cooling \& feedback (right) runs. The white contours give the boundaries of the filaments as reconstructed by our algorithm, while the white points show the spine of each filament, that we used to compute the profiles in Figure \ref{fig:prof_fila1}-\ref{fig:prof_fila2}. Each panel has axes is cell units (here $\Delta x=74$ kpc).}
    \label{fig:obj_map}
\end{figure}

In order to monitor the impact of non-gravitational physics (cooling and feedback from AGN) onto
filaments, we analysed the properties of individual objects. 
For this purpose, we extracted from the 3-1\_1024 and 3-c1\_1024 runs all cells belonging to two objects: filament ``F1"  with a length of   $\sim 15.5$ Mpc and  filament ``F2" $\sim 10.2$ Mpc long. The maps of projected gas temperatures for these objects are shown in Figure \ref{fig:obj_map}.
For the two different physical models, the filaments' morphologies are quite similar. However, the combined effect of radiative cooling and feedback produces more substructures within each filament which are narrower compared to the non-radiative runs. Also the presence of more compact sub-halos causes a higher fragmentation in several portions of the filaments in the radiative runs.

{Other thermodynamical differences are more conspicuous in phase diagrams, as in the 
 ($T$,$\rho$) phase diagrams of Figure \ref{fig:obj_phase}.  The properties of the cells in filaments can be compared to the global phase diagram of the larger $\approx (35 \rm ~ Mpc)^3$ boxes containing the two filaments, in order to compare with the phase diagram of cosmic baryons in the volume. The two objects show a very similar thermodynamic pattern and contain most of the dense and warm-hot intergalactic medium ($T \leq 10^7$ K) outside galaxy clusters. Most of the volume in these two objects is taken up by gas with densities $\rho \sim 10 \langle \rho\rangle$ and temperature $T \sim 10^{6} \rm K$. 
The combined action of cooling and feedback is relevant in both cases.
Cooling promotes the formation of denser clumps in both filaments (visible as horizontal stripes on the phase diagrams), and it is found to affect the filaments even at $T \leq 10^{7} \rm K$, i.e.
in a regime where metal cooling can be dominant and that future X-ray telescopes \citep[e.g. ATHENA,][]{2013arXiv1306.2307N} will probe.}
Although recent observations of metallicity at the edge of the Perseus cluster found a mean metallicity of the order of $\sim 0.2-0.3 ~\rm Z_{\odot}$ \citep[][]{2014MNRAS.437.3939U}, our assumption is likely to overestimate the presence of metals at early times, and translates into an upper limit on the amount of radiative cooling of the WHIM \citep[e.g.][]{2006ApJ...650..560C,2011ApJ...731....6S,2011MNRAS.414.1145M}. \\
At the same time, the effect of feedback from the nearby clusters is enough to increase the temperature by a factor of 2-3 in most of the volume compared to the non-radiative run. The most peripheral parts of some clusters expand due to the AGN outbursts and, by getting more rarefied, are identified as part of filaments by our algorithm (this is particularly evident in the upper part of filament F2). This leads to a slight increase of the size of the structure due to percolation of expanding shells of matter, which in turn enhances the maximum length of several filaments in our radiative runs (as also pointed out in Section \ref{subsec:statprop}).

\begin{figure*}
  \includegraphics[width=0.9\textwidth]{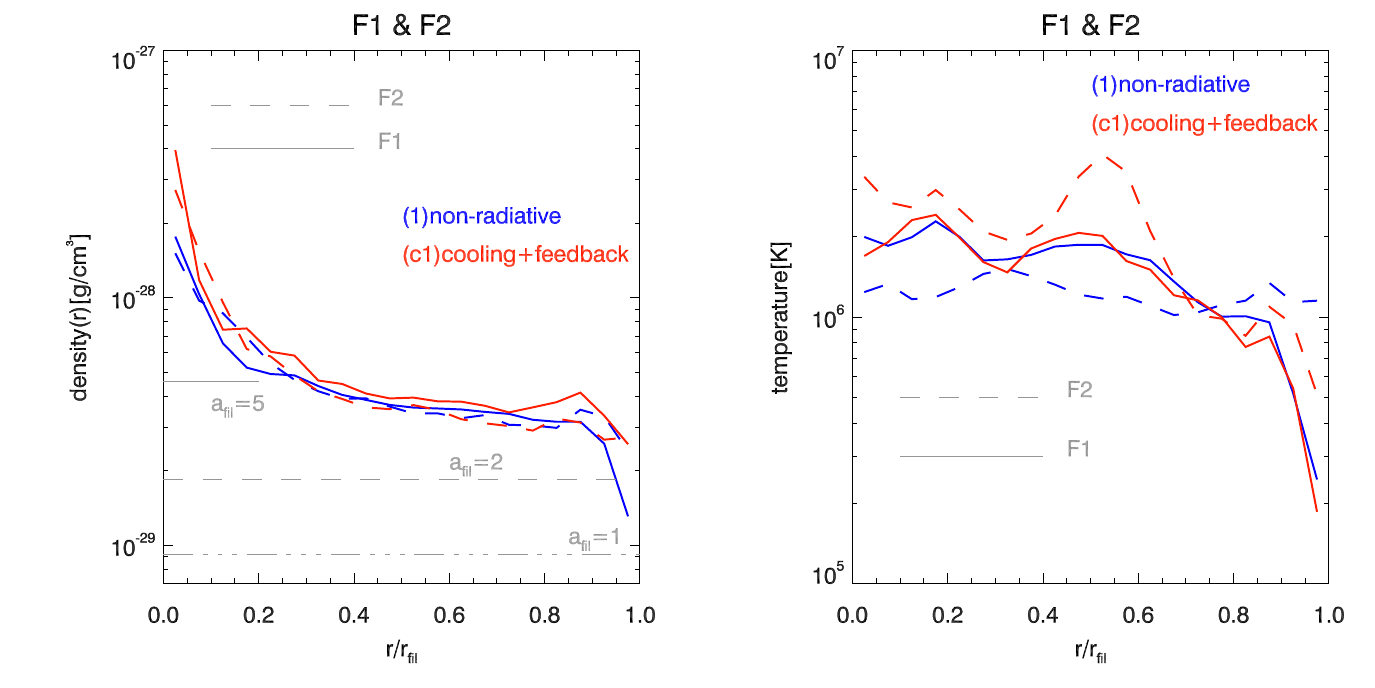}
  \caption{Transverse profile of gas density (left) and gas temperature (right) for the filaments F1 (solid) and F2 (dashed), in the non-radiative run and in the cooling and feedback run (c2). The horizontal lines in the first panel show the density threshold corresponding to different values of $a_{\rm fil}$ (Section \ref{sec:tuning}). The radial coordinate is normalised to the transversal size of each filament along the spine (see Section \ref{subsec:singleprop}).}
  \label{fig:prof_fila1}
\end{figure*}

\begin{figure}
  \includegraphics[width=0.45\textwidth]{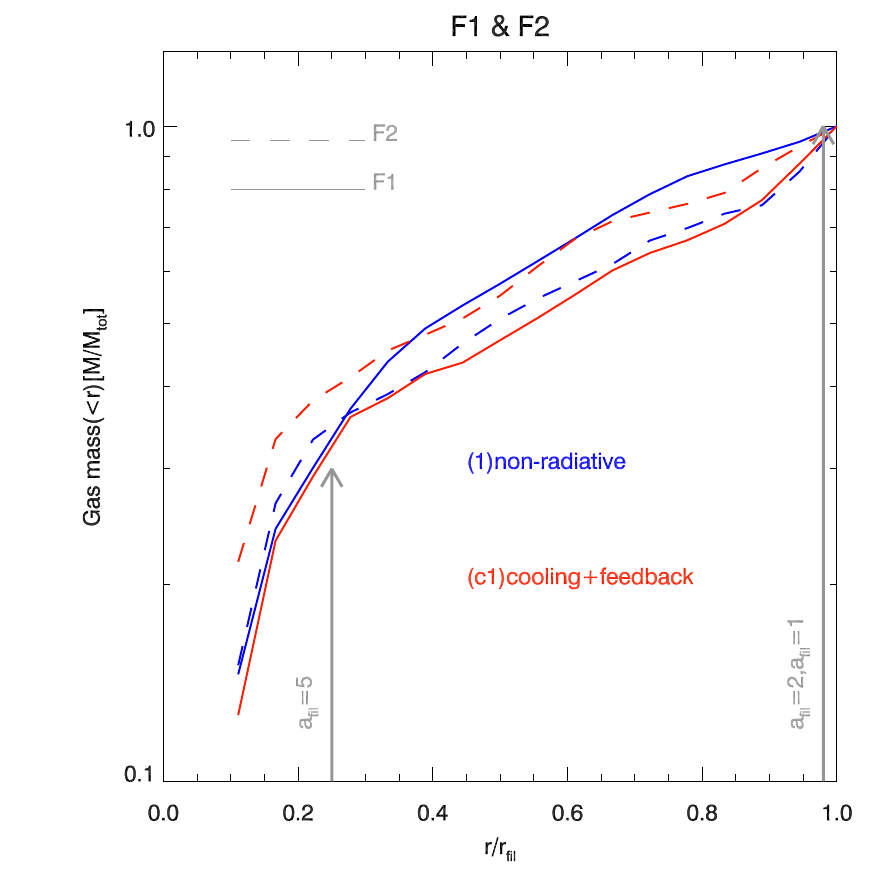}
  \caption{Transverse profile of the enclosed gas mass  for the filaments F1 (solid) and F2 (dashed), in the non-radiative run and in the cooling and feedback run (c2). The grey arrows show the approximate location where the mean density profile of filaments is equal to the different values of the density thresholds $a_{\rm fil}$. The profile of each filament has been normalised for the total enclosed mass to the last radial bin.}
  \label{fig:prof_fila0}
\end{figure}

\begin{figure*}
  \includegraphics[width=0.9\textwidth]{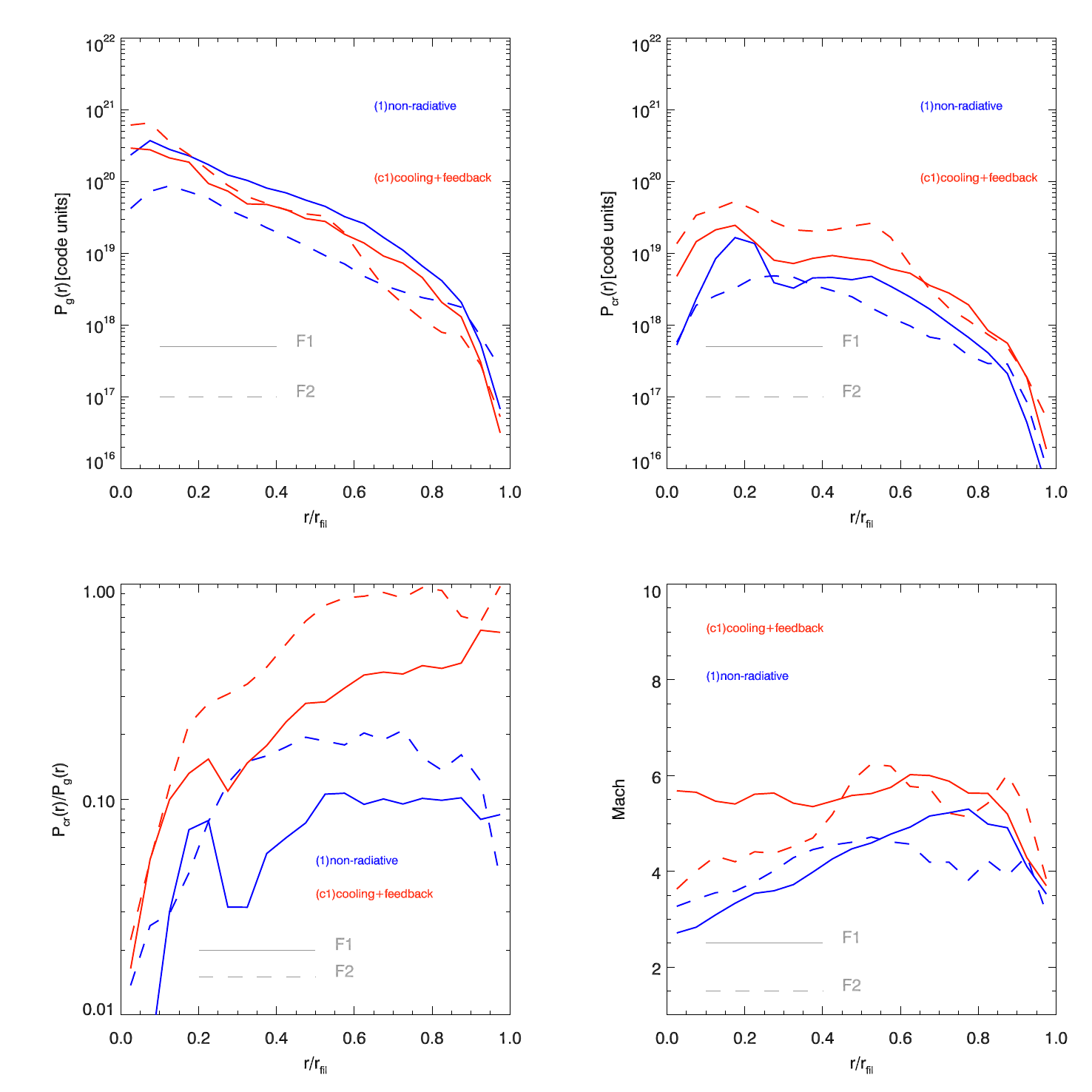}
  \caption{Transverse profiles of gas pressure (top left), CR pressure (top right), pressure ratio (bottom left) and shock Mach number (bottom right) for filaments F1 and F2, for the
   non-radiative run and in the cooling and feedback runs.}
  \label{fig:prof_fila2}
\end{figure*}

The effect of our  prescription for the thermal feedback and the fact it
reaches out to the outer part of filaments is similar to what is reported by other authors, who investigated
similar feedback schemes \citep[][]{ka07,mcc2010}.

In each of the two objects, we computed the location of the spine along the major axis of the filament and computed the profiles transverse to it.
The spine is computed by first identifying the major axis of each structure (based on the bounding box of each object, as in Section \ref{sec:identification}) and then computing the centre of gas mass in consecutive slices (each one cell thick) perpendicular to the main axis, as shown in Figure \ref{fig:obj_map}. Figures \ref{fig:prof_fila1}-\ref{fig:prof_fila0} show the profiles of gas density, temperature and enclosed gas mass transverse to the main
axis of the filaments, out to $r_{\rm fil} \sim 3.3 ~\rm Mpc$ (F1) and $2 ~\rm Mpc$ (F2). Since the transverse size of each object might vary along the spine (Figure \ref{fig:obj_map})
our radial bins are normalised to the transverse size of the filament along the spine, i.e. we bin our profiles according to $r/r_{\rm fil}$, where $r_{\rm fil}$ is
defined as the minimum distance of the filament edge from the spine in the corresponding transversal plane.
The density profiles for the same physical model are rather similar. 
In the non-radiative case, the central mass density is several $\sim 10^{-28}$ g/cm$^3$
becoming $\sim 2-3$ times larger in the radiative case, due to the higher compression. 
The mass density smoothly drops by a factor
$\sim 5-10$ at $0.5 r_{\rm fil}$ from the spine of each filament, then slowly declining outwards. Notice that the density at the last radial bin is 
larger than our fiducial threshold density ($a_{\rm fil}$) due to the averaging at $r=r_{\rm fil}$ which can include cells with mass density above
the threshold resulting from the compression produced by strong external shocks.
In all cases the central temperature is around $\sim 10^6 \rm K$. However, the corresponding radial profiles are
remarkably different. The F1 non-radiative filament shows a gently increasing temperature up to its external
boundaries where most of shock heating takes place. The same F1 object is highly affected by AGN feedback and cooling,
which leads to a higher central over-density and to a higher temperature peak $\sim 0.5 r_{\rm fil}$ from the spine of the filament. The effect is 
less evident in filament F2. 
The horizontal lines show the mass cut-off induced by different settings of the $a_{\rm fil}$ parameter.
It is clear how setting $a_{\rm fil} > 2$ significantly limits the fraction of gas mass assigned to filaments.
This is confirmed by Figure \ref{fig:prof_fila0} that shows
the cumulative gas mass for the various filaments and models, showing that most of the mass 
is contained at large distances from the spine, $\geq 0.3-0.5 r_{\rm fil}$ (i.e. $\geq 700-1500$ kpc) and 
how only using $a_{\rm fil} \leq 2$ this mass can be fully captured by our algorithm. 
Furthermore, as evident in the temperature profiles, we thus trace the filament  out to the position of the main shock fronts including most of the cell that have undergone compression and heating. 

In the panels of  Figure \ref{fig:prof_fila2} we present the profiles of the 
gas pressure, the CR pressure, the pressure ratio and the shock Mach number for the same objects and runs.
In both objects, the thermal gas pressure is dominant at all radii compared to the CR pressure. However, the CR pressure becomes $\sim 10-20$\% of the thermal pressure at large radii in the non-radiative runs, and {\bf $\sim 40-80$ \%} of the thermal pressure in the cooling and feedback runs. The latter is a combined effect of the decreased gas temperature and of the stronger shocks (as an effect of additional feedback bursts) in most of the filament volume, which yield a more efficient injection of CRs from DSA. However, inside most of the filament the gas motions are supersonic \citep[][]{ry08,va14mhd} and that, therefore, both gas and CR pressure are subdominant compared to the ram pressure of the gas. \\
Future radio telescopes (such as SKA or, in the nearest future, LOFAR) will be able to investigate this regime and might provide new clues about the degree of plasma collisionality and cosmic rays in this environment, by directly
imaging the morphology of strong accretion shocks in radio.  One significant emission channel for CRs is represented by direct
synchrotron radiation by accelerated CR-electrons at strong shocks \citep[][]{2011JApA...32..577B,2012MNRAS.423.2325A}. In this case, also the level of large-scale magnetisation will affect the chance of detection, and present uncertainties about the WHIM leave room for
several scenarios \citep[][]{2008Sci...320..909R,donn09,va14mhd}. \\

\section{Discussion}
\label{sec:discussion}

Compared to previous work (see Section \ref{sec:intro}) that studied the distribution of DM in the cosmic web (mostly using DM-only simulations), our 
work focused on the thermodynamic characterisation of the baryon component of the cosmic web. \\
The methods that we have used are particularly sensitive to gas densities close to the average cosmic value and have no artefacts caused by a particle-based DM distribution that especially affect poorly sampled low-density regions. 
As a consequence, the average over-density of our population of filaments is smaller than what has been quoted in the literature \citep[e.g.][]{2005MNRAS.359..272C}.  \\
This procedure enables us to extract the cosmic web down to 
an over-density very close to the critical density and to segment it into its filamentary components. Thus, we can analyse a large dynamic range in the mass/volume distribution of filaments, using the recent $2048^3$ and
$1024^3$ grids produced by our group \citep[][]{va14curie}.  This led to the extraction of a very large
number of filaments at high resolution, e.g. $\sim$ 30000 objects in our $300^3 \rm Mpc^3$ volume, significantly higher than what is typically found in DM-only simulations.  \\

Our method is simpler than other algorithms discussed in the literature  \citep[e.g.][]
{2005A&A...434..423S,2007MNRAS.381...41H,2007A&A...474..315A, 
2008MNRAS.383.1655S,2014MNRAS.441.2923C,2010MNRAS.408.2163A,2010MNRAS.407.1449G}
and is likely prone to possible misidentification in several pathological cases as, for instance, for irregular filaments (see Section~4.1) that in some cases may actually be parts of sheets with density comparable to that of filaments. 
It can also be affected by resolution effects, e.g. in small clumps whose typical size is close
to the resolution limit.
However, a recent comparison has shown that the volume occupied by filaments (and, to a lesser extent, the mass locked into filaments) can in principle be measured equally well by both density-based and
more sophisticated methods \citep[][]{2014MNRAS.441.2923C}. We conclude that, while additional checks might be necessary to assess the nature of specific low mass/size objects (e.g. also accessing the DM properties of small objects), in general our method gives robust results.\\

We find that the total volume occupied by filaments ranges from $3.8$ to $4.5\%$ of the computational volume, which is consistent with the range provided by 
 \citet[][]{2014MNRAS.441.2923C}, and significantly smaller than what is found by previous N-body simulations  \citep[e.g.][]{2007MNRAS.381...41H,2010MNRAS.408.2163A}
The largest objects identified in our dataset have a mass and estimated length 
of the order of what is found by large N-body simulations
\citep[e.g.][]{2005MNRAS.359..272C, 2014MNRAS.441.2923C}, provided that we rescale our gas masses by the cosmic baryon fraction in order to get the total mass (which is legitimate here since at the typical filaments' over-density the baryon fraction is close to the cosmic baryon fraction). However,  the mass distribution of filaments in the cosmic volume (Figure \ref{fig:massdist}) shows a more extended power-law behaviour down to the smallest masses compared to the literature \citep[e.g.][Figure 53]{2014MNRAS.441.2923C}. Here we can measure an unbroken power-law distribution in the range $10^{11} -10^{14} \rm M_{\odot}$, suggesting that our sample of filaments is complete down to total masses of $6 \cdot 10^{11} M_{\odot}$, i.e. 1-2 orders of magnitude deeper than what is typically achieved with N-body simulations.\\

We also performed an analysis of the properties of 
individual filaments, focusing on two massive objects as representative cases.
The profiles of gas density across the filaments is similar to what is
reported elsewhere \citep[][]{2005MNRAS.359..272C, 2006MNRAS.370..656D, 2010MNRAS.408.2163A}. 
The description of sharp discontinuities of the temperature at the outer accretion shocks is a feature in which our simulations are particularly effective.  Qualitatively similar results for the density and temperature profiles were reported by \citep[][]{2012MNRAS.423..304K}, but these were based on idealised simulations of a forming filament that were not set in a cosmological framework.
The pattern of strong accretion shocks surrounding filaments are consistent with previous works in the literature \citep[][]{ry03,sk08,va09shocks,va11comparison,2012MNRAS.423.2325A}, which 
also predict high CR acceleration efficiency at the scale of filaments \citep[][]{pf06}.  Based on our run-time modelling of CR physics \citep[][]{va14curie}, the pressure ratios within filaments show that CRs can contribute  
up to $\sim 10-20$ percent of the total pressure. However, the present uncertainties CR-acceleration efficiencies, especially in (low-density) environments \citep[e.g.][]{kr13} force us to treat this number with caution.
The impact of radiative cooling and feedback from AGN on the WHIM has been explored by several authors
\citep[e.g][]{2006ApJ...650..560C,2006MNRAS.368...74R,2007ApJ...671...27H}. Still, as far as we know, our work represents the first study of how cosmic filaments change with feedback. However, our modelling of radiative cooling over-estimates the gas metallicity at high redshift, and possibly
also in the outer parts of filaments at low redshifts, even if the observational constraints are still weak \citep[e.g.][]{2014MNRAS.437.3939U}.

\section{Conclusions}
\label{sec:conclusions}

In this work we investigated the main properties of the baryonic matter  in cosmic filaments by using a 
large set of cosmological grid simulations recently obtained with the \enzo code \citep[][]{va14curie}.
We built a filament identification procedure upon the \visit data analysis and visualisation 
software, exploiting a combination of its {\it Isovolume} and {\it Connected Components} algorithms.
The method separates over-dense from under-dense regions, identifying a filament as a connected set of cells with 
mass density above a given threshold, $a_{\rm fil}$. The filament identification is further refined by eliminating clusters and small clumps and accepting only elongated 
objects. 
The resulting methodology depends on six parameters, the most important being the mass density threshold 
$a_{\rm fil}$, whose value lies in the range $0.5 - 2.0$. The exact value can influence some 
statistical properties of the identified objects (e.g. the maximum size of filaments extracted from a simulation). 
Other properties (e.g. the average density within filaments) are unaffected. The remaining parameters
are either more tightly constrained or they have only a minor impact on the results (see Appendix A for further discussion).

Our most significant findings can be summarised as follows:

\begin{itemize}

\item Morphology and thermodynamic properties: filaments, especially long objects (longer than $\sim 7$ Mpc), 
show a broad variety of shapes and thermodynamic properties, that depend on the environment in which 
they lie and on the evolutionary stage. A first rough visual classification as been attempted,
but further investigation is needed to reach comprehensive and robust conclusions.

\item Average properties of the WHIM: the population of filament extracted setting $\varrho \geq \varrho_0$ and removing collapsed halos, is of the
order of $\sim 3500$ objects within (100 Mpc)$^3$. The enclosed gas density averaged over the 
whole population is $\sim 3-5 ~\rho_0$, the average temperature is a few $\sim 10^5 \rm K$.

\item Scaling relations: filaments follow well-defined scaling relations in the  $(T, M)$ and $(V, M)$ plane, with slopes similar to those of galaxy clusters but different normalisation (i.e. lower temperature and larger masses). The observed scaling shows that also in filaments gravity sets a clear dependence between the enclosed mass
and the gas temperature, ($T \propto M^{2/3}$) even if at low masses the effect of radiative cooling can steepen the relation significantly, similar to the case of radiative galaxy clusters.

\item Massive filaments: the most massive objects found in our suite of simulations have gas masses in excess of $\sim 10^{15} M_{\odot}$, an average (volume-weighted) temperatures of $\sim 10^{7} \rm K$ and a total volume of a few $\sim 10^3 \rm Mpc^3$, reaching a total length of the order of $100 ~\rm Mpc$.  Only about 1 object of such size can be found on average within $100^3 \rm Mpc^3$.

\item Smallest filaments: the smallest filaments that our algorithm can reliably extract have a length of about 1-2 Mpc, gas mass of a few $\sim 10^{10} M_{\odot}$  and temperature of $\sim 10^{4}-10^{5} \rm K$ (or down to $\sim 10^{3} ~\rm K$ in the radiative case).

\item Resolution effects: spatial resolution can affect the population of filaments at all scales.
The most evident effect is the high diffusion in low resolution simulations, that leads to the percolation 
of otherwise disconnected structures. This leads to a drop of the number of small objects per unit of volume 
and to an increase in the number of large objects. As a consequence, the same trend can be found in the mass distributions. 
Temperature distributions show a similar behaviour due also to the largest distance travelled by shock waves in low resolution
simulations, heating up larger volumes. 
In summary, our analysis on the thermodynamical and statistical properties of the WHIM suggest that a spatial 
resolution equal or better than $\sim 100 ~\rm kpc/h$ is necessary to have converging results on the simulated filaments properties, considering objects of masses larger than $10^{10}M_{\odot}$.

\item Effects of radiative cooling: cooling (mostly dominated by line cooling here, given the assumed fixed large metallicity) decreases the average temperature of the WHIM by a factor $\sim 3-5$ compared to the non-radiative case. In the presence of small-scale clumps contained in filaments, the minimum temperature inside filaments is 
reduced by more than one order of magnitude. The gas density profile
is also increased, by a factor $\sim 3$ close the axis of filaments in the presence of cooling.  

\item Effects of AGN feedback: efficient feedback from AGN can affect most of the volume of filaments, even out to large distances from the nearby halos. The thermal feedback implemented in our model causes the expulsion of gas from halos and the expansion of filaments, that can percolate and
produce significantly larger objects compared to the non-radiative case.

\item Effects of cosmic rays: the dynamical impact of CRs is small since where the gas density is higher, close to the axis of filaments, the ratio between CRs and the thermal gas is small ($\sim 10-20$ percent in the non-radiative case). However, the combination of cooling and AGN feedback is found to increase the budget of CRs significantly outwards, where strong accretion shocks dominate.

\end{itemize}

As a final remark, we stress that a comprehensive description of the observable properties of the cosmic web requires further important physical
mechanisms to be taken into account, such as star formation and chemical enrichment of the diffuse medium by galactic activity (which affects the distribution of elements in the WHIM, and determines
the emission/absorption properties of the gas through lines, \citealt[e.g.][]{2011ApJ...731...11C,2013ApJ...769...90N}), partial thermal equilibration of the WHIM due
to proton-electron equilibration \citep[e.g.][]{2009ApJ...701L..16R}, thermal conduction \citep[][]{2012MNRAS.423..304K} 
and large-scale magnetic fields \citep[e.g.][]{br05,do08,2008Sci...320..909R,va14mhd}. 
Such extensive studies will be the subject of forthcoming work. 
However, the methodology presented in this paper does not depend on the physics of the problem and can be applied to many kind cosmological simulations as is based on simple assumptions. It adapts very well to even larger datasets, and its parallel
strategy will allow us to exploit it effectively for the identification and 
characterisation of filamentary structures in any future simulations programme.

\section*{acknowledgements}

Computations described in this work were performed using the {\enzo} code (http://enzo-project.org), which is the product of a collaborative effort of scientists at many universities and national laboratories. We gratefully acknowledge the {\enzo} development group for providing helpful and well-maintained on-line documentation and tutorials.\\
We acknowledge PRACE for awarding us access to CURIE-Genci based in France at Bruyeres-le-Chatel. The support of the TGC Hotline from the Centre CEA-DAM Ile de France to the technical work is gratefully acknowledged.
We also acknowledge CSCS-ETHZ{\footnote{http://www.cscs.ch}} for the use of the Cray XC30 Piz Daint in order to complete the $2048^3$ run and of the Pilatus system for data processing and visualization.
F.V. and M.B. acknowledge support from Forschergruppe FOR1254 from the Deutsche Forschungsgemeinschaft. 
F.V. and M.B. acknowledge the usage of computational resources on the JUROPA cluster at the at the J\"ulich Supercomputing Centre (JSC), under project no. 5018, 5056, 6981 and 7006. 
We would like to thank M.G. Giuffreda for her valuable technical assistance at CSCS.

\bigskip

\bibliographystyle{mnras}
\bibliography{filaments}

\begin{thebibliography}{83}
\expandafter\ifx\csname natexlab\endcsname\relax\def\natexlab#1{#1}\fi

\bibitem[{Anderson {et~al}\mbox{.}(2010)Anderson, Garth, Duchaineau, \&
  Joy}]{10.1109/TVCG.2010.17}
Anderson J.~C., Garth C., Duchaineau M.~A., Joy K.~I., 2010, IEEE Transactions
  on Visualization and Computer Graphics, 16, 802

\bibitem[{{Arag{\'o}n-Calvo} {et~al}\mbox{.}(2007){Arag{\'o}n-Calvo}, {Jones},
  {van de Weygaert}, \& {van der Hulst}}]{2007A&A...474..315A}
{Arag{\'o}n-Calvo} M.~A., {Jones} B.~J.~T., {van de Weygaert} R., {van der
  Hulst} J.~M., 2007, \aap, 474, 315

\bibitem[{{Arag{\'o}n-Calvo} {et~al}\mbox{.}(2010){Arag{\'o}n-Calvo}, {van de
  Weygaert}, \& {Jones}}]{2010MNRAS.408.2163A}
{Arag{\'o}n-Calvo} M.~A., {van de Weygaert} R., {Jones} B.~J.~T., 2010, \mnras,
  408, 2163

\bibitem[{{Araya-Melo} {et~al}\mbox{.}(2012){Araya-Melo}, {Arag{\'o}n-Calvo},
  {Br{\"u}ggen}, \& {Hoeft}}]{2012MNRAS.423.2325A}
{Araya-Melo} P.~A., {Arag{\'o}n-Calvo} M.~A., {Br{\"u}ggen} M., {Hoeft} M.,
  2012, \mnras, 423, 2325

\bibitem[{{Bagchi} {et~al}\mbox{.}(2002){Bagchi}, {En{\ss}lin}, {Miniati},
  {Stalin}, {Singh}, {Raychaudhury}, \& {Humeshkar}}]{2002NewA....7..249B}
{Bagchi} J., {En{\ss}lin} T.~A., {Miniati} F., {Stalin} C.~S., {Singh} M.,
  {Raychaudhury} S., {Humeshkar} N.~B., 2002, \na, 7, 249

\bibitem[{{Bleuler} {et~al}\mbox{.}(2014){Bleuler}, {Teyssier}, {Carassou}, \&
  {Martizzi}}]{2014arXiv1412.0510B}
{Bleuler} A., {Teyssier} R., {Carassou} S., {Martizzi} D., 2014, ArXiv e-prints

\bibitem[{{Bolton} {et~al}\mbox{.}(2014){Bolton}, {Becker}, {Haehnelt}, \&
  {Viel}}]{2014MNRAS.438.2499B}
{Bolton} J.~S., {Becker} G.~D., {Haehnelt} M.~G., {Viel} M., 2014, \mnras, 438,
  2499

\bibitem[{{Bond} {et~al}\mbox{.}(1996){Bond}, {Kofman}, \&
  {Pogosyan}}]{1996Natur.380..603B}
{Bond} J.~R., {Kofman} L., {Pogosyan} D., 1996, \nat, 380, 603

\bibitem[{{Brown}(2011)}]{2011JApA...32..577B}
{Brown} S.~D., 2011, Journal of Astrophysics and Astronomy, 32, 577

\bibitem[{{Br{\"u}ggen} {et~al}\mbox{.}(2005){Br{\"u}ggen}, {Ruszkowski},
  {Simionescu}, {Hoeft}, \& {Dalla Vecchia}}]{br05}
{Br{\"u}ggen} M., {Ruszkowski} M., {Simionescu} A., {Hoeft} M., {Dalla Vecchia}
  C., 2005, \apjl, 631, L21

\bibitem[{{Bryan} {et~al}\mbox{.}(2014){Bryan}, {Norman}, {O'Shea}, {Abel},
  {Wise}, {Turk}, {Reynolds}, {Collins}, {Wang}, {Skillman}, {Smith},
  {Harkness}, {Bordner}, {Kim}, {Kuhlen}, {Xu}, {Goldbaum}, {Hummels},
  {Kritsuk}, {Tasker}, {Skory}, {Simpson}, {Hahn}, {Oishi}, {So}, {Zhao},
  {Cen}, {Li}, \& {Enzo Collaboration}}]{enzo13}
{Bryan} G.~L. {et~al.}, 2014, \apjs, 211, 19

\bibitem[{{Cautun} {et~al}\mbox{.}(2014){Cautun}, {van de Weygaert}, {Jones},
  \& {Frenk}}]{2014MNRAS.441.2923C}
{Cautun} M., {van de Weygaert} R., {Jones} B.~J.~T., {Frenk} C.~S., 2014,
  \mnras, 441, 2923

\bibitem[{{Cen} \& {Chisari}(2011)}]{2011ApJ...731...11C}
{Cen} R., {Chisari} N.~E., 2011, \apj, 731, 11

\bibitem[{{Cen} \& {Ostriker}(1999)}]{1999ApJ...514....1C}
{Cen} R., {Ostriker} J.~P., 1999, \apj, 514, 1

\bibitem[{{Cen} \& {Ostriker}(2006)}]{2006ApJ...650..560C}
{Cen} R., {Ostriker} J.~P., 2006, \apj, 650, 560

\bibitem[{{Chen} {et~al}\mbox{.}(2015){Chen}, {Ho}, {Freeman}, {Genovese}, \&
  {Wasserman}}]{2015arXiv150105303C}
{Chen} Y.-C., {Ho} S., {Freeman} P.~E., {Genovese} C.~R., {Wasserman} L., 2015,
  ArXiv e-prints

\bibitem[{Childs {et~al}\mbox{.}(2011)Childs, Brugger, Whitlock, Meredith,
  Ahern, Bonnell, Miller, Weber, Harrison, Fogal, Garth, S, Bethel, Durant,
  Camp, Favre, Rübel, Navrátil, Α, Α, \& Vivodtzev}]{Childs11visit:an}
Childs H. {et~al.}, 2011, in In Proceedings of SciDAC

\bibitem[{{Colberg} {et~al}\mbox{.}(2005){Colberg}, {Krughoff}, \&
  {Connolly}}]{2005MNRAS.359..272C}
{Colberg} J.~M., {Krughoff} K.~S., {Connolly} A.~J., 2005, \mnras, 359, 272

\bibitem[{{Colella} \& {Woodward}(1984)}]{cw84}
{Colella} P., {Woodward} P.~R., 1984, Journal of Computational Physics, 54, 174

\bibitem[{{Colless} {et~al}\mbox{.}(2003){Colless}, {Dalton}, {Maddox},
  {Sutherland}, {Norberg}, {Cole}, {Bland-Hawthorn}, {Bridges}, {Cannon},
  {Collins}, {Couch}, {Cross}, {Deeley}, {de Propris}, {Driver}, {Efstathiou},
  {Ellis}, {Frenk}, {Glazebrook}, {Jackson}, {Lahav}, {Lewis}, {Lumsden},
  {Madgwick}, {Peacock}, {Peterson}, {Price}, {Seaborne}, \&
  {Taylor}}]{2003yCat.7226....0C}
{Colless} M. {et~al.}, 2003, VizieR Online Data Catalog, 7226, 0

\bibitem[{{Courtois} {et~al}\mbox{.}(2013){Courtois}, {Pomar{\`e}de}, {Tully},
  {Hoffman}, \& {Courtois}}]{2013AJ....146...69C}
{Courtois} H.~M., {Pomar{\`e}de} D., {Tully} R.~B., {Hoffman} Y., {Courtois}
  D., 2013, \aj, 146, 69

\bibitem[{{Dav{\'e}} {et~al}\mbox{.}(2001){Dav{\'e}}, {Cen}, {Ostriker},
  {Bryan}, {Hernquist}, {Katz}, {Weinberg}, {Norman}, \&
  {O'Shea}}]{2001ApJ...552..473D}
{Dav{\'e}} R. {et~al.}, 2001, \apj, 552, 473

\bibitem[{{de Lapparent} {et~al}\mbox{.}(1986){de Lapparent}, {Geller}, \&
  {Huchra}}]{1986ApJ...302L...1D}
{de Lapparent} V., {Geller} M.~J., {Huchra} J.~P., 1986, \apjl, 302, L1

\bibitem[{{Dolag} {et~al}\mbox{.}(2008){Dolag}, {Bykov}, \& {Diaferio}}]{do08}
{Dolag} K., {Bykov} A.~M., {Diaferio} A., 2008, \ssr, 134, 311

\bibitem[{{Dolag} {et~al}\mbox{.}(2006){Dolag}, {Meneghetti}, {Moscardini},
  {Rasia}, \& {Bonaldi}}]{2006MNRAS.370..656D}
{Dolag} K., {Meneghetti} M., {Moscardini} L., {Rasia} E., {Bonaldi} A., 2006,
  \mnras, 370, 656

\bibitem[{{Donnert} {et~al}\mbox{.}(2009){Donnert}, {Dolag}, {Lesch}, \&
  {M{\"u}ller}}]{donn09}
{Donnert} J., {Dolag} K., {Lesch} H., {M{\"u}ller} E., 2009, \mnras, 392, 1008

\bibitem[{{Einasto} {et~al}\mbox{.}(2011){Einasto}, {H{\"u}tsi}, {Saar},
  {Suhhonenko}, {Liivam{\"a}gi}, {Einasto}, {M{\"u}ller}, {Starobinsky},
  {Tago}, \& {Tempel}}]{2011A&A...531A..75E}
{Einasto} J. {et~al.}, 2011, \aap, 531, A75

\bibitem[{{Einasto} {et~al}\mbox{.}(1984){Einasto}, {Klypin}, {Saar}, \&
  {Shandarin}}]{1984MNRAS.206..529E}
{Einasto} J., {Klypin} A.~A., {Saar} E., {Shandarin} S.~F., 1984, \mnras, 206,
  529

\bibitem[{{Farnsworth} {et~al}\mbox{.}(2013){Farnsworth}, {Rudnick}, {Brown},
  \& {Brunetti}}]{2013ApJ...779..189F}
{Farnsworth} D., {Rudnick} L., {Brown} S., {Brunetti} G., 2013, \apj, 779, 189

\bibitem[{{Ferland} {et~al}\mbox{.}(1998){Ferland}, {Korista}, {Verner},
  {Ferguson}, {Kingdon}, \& {Verner}}]{1998PASP..110..761F}
{Ferland} G.~J., {Korista} K.~T., {Verner} D.~A., {Ferguson} J.~W., {Kingdon}
  J.~B., {Verner} E.~M., 1998, \pasp, 110, 761

\bibitem[{{Finoguenov} {et~al}\mbox{.}(2003){Finoguenov}, {Briel}, \&
  {Henry}}]{2003A&A...410..777F}
{Finoguenov} A., {Briel} U.~G., {Henry} J.~P., 2003, \aap, 410, 777

\bibitem[{Fujishiro {et~al}\mbox{.}(1995)Fujishiro, Maeda, \&
  Sato}]{conf/visualization/FujishiroMS95}
Fujishiro I., Maeda Y., Sato H., 1995, in IEEE Visualization, pp. 151--158

\bibitem[{{Geller} \& {Huchra}(1989)}]{1989Sci...246..897G}
{Geller} M.~J., {Huchra} J.~P., 1989, Science, 246, 897

\bibitem[{{Giovannini} {et~al}\mbox{.}(2010){Giovannini}, {Bonafede},
  {Feretti}, {Govoni}, \& {Murgia}}]{2010A&A...511L...5G}
{Giovannini} G., {Bonafede} A., {Feretti} L., {Govoni} F., {Murgia} M., 2010,
  \aap, 511, L5

\bibitem[{{Gonz{\'a}lez} \& {Padilla}(2010)}]{2010MNRAS.407.1449G}
{Gonz{\'a}lez} R.~E., {Padilla} N.~D., 2010, \mnras, 407, 1449

\bibitem[{{Gott} {et~al}\mbox{.}(2005){Gott}, {Juri{\'c}}, {Schlegel}, {Hoyle},
  {Vogeley}, {Tegmark}, {Bahcall}, \& {Brinkmann}}]{2005ApJ...624..463G}
{Gott}, III J.~R., {Juri{\'c}} M., {Schlegel} D., {Hoyle} F., {Vogeley} M.,
  {Tegmark} M., {Bahcall} N., {Brinkmann} J., 2005, \apj, 624, 463

\bibitem[{{Guo} \& {Oh}(2008)}]{guo08}
{Guo} F., {Oh} S.~P., 2008, \mnras, 384, 251

\bibitem[{{Haardt} \& {Madau}(1996)}]{hm96}
{Haardt} F., {Madau} P., 1996, \apj, 461, 20

\bibitem[{{Hahn} {et~al}\mbox{.}(2007){Hahn}, {Carollo}, {Porciani}, \&
  {Dekel}}]{2007MNRAS.381...41H}
{Hahn} O., {Carollo} C.~M., {Porciani} C., {Dekel} A., 2007, \mnras, 381, 41

\bibitem[{{Hallman} {et~al}\mbox{.}(2007){Hallman}, {O'Shea}, {Burns},
  {Norman}, {Harkness}, \& {Wagner}}]{2007ApJ...671...27H}
{Hallman} E.~J., {O'Shea} B.~W., {Burns} J.~O., {Norman} M.~L., {Harkness} R.,
  {Wagner} R., 2007, \apj, 671, 27

\bibitem[{Harrison {et~al}\mbox{.}(2011)Harrison, Childs, \&
  Gaither}]{conf/egpgv/HarrisonCG11}
Harrison C., Childs H., Gaither K.~P., 2011, in EGPGV, Kuhlen T., Pajarola R.,
  Zhou K., eds., Eurographics Association, pp. 131--140

\bibitem[{{Iapichino} {et~al}\mbox{.}(2011){Iapichino}, {Schmidt}, {Niemeyer},
  \& {Merklein}}]{iapichino11}
{Iapichino} L., {Schmidt} W., {Niemeyer} J.~C., {Merklein} J., 2011, \mnras,
  414, 2297

\bibitem[{{Kang} \& {Jones}(2007)}]{kj07}
{Kang} H., {Jones} T.~W., 2007, Astroparticle Physics, 28, 232

\bibitem[{{Kang} \& {Ryu}(2013)}]{kr13}
{Kang} H., {Ryu} D., 2013, \apj, 764, 95

\bibitem[{{Kang} {et~al}\mbox{.}(2007){Kang}, {Ryu}, {Cen}, \&
  {Ostriker}}]{ka07}
{Kang} H., {Ryu} D., {Cen} R., {Ostriker} J.~P., 2007, \apj, 669, 729

\bibitem[{{Klar} \& {M{\"u}cket}(2012)}]{2012MNRAS.423..304K}
{Klar} J.~S., {M{\"u}cket} J.~P., 2012, \mnras, 423, 304

\bibitem[{{Komatsu} {et~al}\mbox{.}(2011){Komatsu}, {Smith}, {Dunkley},
  {Bennett}, {Gold}, {Hinshaw}, {Jarosik}, {Larson}, {Nolta}, {Page},
  {Spergel}, {Halpern}, {Hill}, {Kogut}, {Limon}, {Meyer}, {Odegard}, {Tucker},
  {Weiland}, {Wollack}, \& {Wright}}]{2011ApJS..192...18K}
{Komatsu} E. {et~al.}, 2011, \apjs, 192, 18

\bibitem[{{Kronberg} {et~al}\mbox{.}(2007){Kronberg}, {Kothes}, {Salter}, \&
  {Perillat}}]{2007ApJ...659..267K}
{Kronberg} P.~P., {Kothes} R., {Salter} C.~J., {Perillat} P., 2007, \apj, 659,
  267

\bibitem[{{Maio} {et~al}\mbox{.}(2011){Maio}, {Khochfar}, {Johnson}, \&
  {Ciardi}}]{2011MNRAS.414.1145M}
{Maio} U., {Khochfar} S., {Johnson} J.~L., {Ciardi} B., 2011, \mnras, 414, 1145

\bibitem[{{McCarthy} {et~al}\mbox{.}(2010){McCarthy}, {Schaye}, {Ponman},
  {Bower}, {Booth}, {Dalla Vecchia}, {Crain}, {Springel}, {Theuns}, \&
  {Wiersma}}]{mcc2010}
{McCarthy} I.~G. {et~al.}, 2010, \mnras, 406, 822

\bibitem[{Meredith(2004)}]{Meredith2004}
Meredith J.~S., 2004, in Nuclear Explosives Code Developers Conference (NECDC)

\bibitem[{Meredith \& Childs(2010)}]{Meredith:2010:VAR:2421836.2421898}
Meredith J.~S., Childs H., 2010, in Proceedings of the 12th Eurographics / IEEE
  - VGTC Conference on Visualization, EuroVis'10, Eurographics Association,
  Aire-la-Ville, Switzerland, Switzerland, pp. 1241--1250

\bibitem[{{Miniati} {et~al}\mbox{.}(2000){Miniati}, {Ryu}, {Kang}, {Jones},
  {Cen}, \& {Ostriker}}]{mi00}
{Miniati} F., {Ryu} D., {Kang} H., {Jones} T.~W., {Cen} R., {Ostriker} J.~P.,
  2000, \apj, 542, 608

\bibitem[{{Nandra} {et~al}\mbox{.}(2013){Nandra}, {Barret}, {Barcons},
  {Fabian}, {den Herder}, {Piro}, {Watson}, {Adami}, {Aird}, {Afonso}, \&
  et~al.}]{2013arXiv1306.2307N}
{Nandra} K. {et~al.}, 2013, ArXiv e-prints

\bibitem[{{Nicastro} {et~al}\mbox{.}(2013){Nicastro}, {Elvis}, {Krongold},
  {Mathur}, {Gupta}, {Danforth}, {Barcons}, {Borgani}, {Branchini}, {Cen},
  {Dav{\'e}}, {Kaastra}, {Paerels}, {Piro}, {Shull}, {Takei}, \&
  {Zappacosta}}]{2013ApJ...769...90N}
{Nicastro} F. {et~al.}, 2013, \apj, 769, 90

\bibitem[{{Nicastro} {et~al}\mbox{.}(2010){Nicastro}, {Krongold}, {Fields},
  {Conciatore}, {Zappacosta}, {Elvis}, {Mathur}, \&
  {Papadakis}}]{2010ApJ...715..854N}
{Nicastro} F., {Krongold} Y., {Fields} D., {Conciatore} M.~L., {Zappacosta} L.,
  {Elvis} M., {Mathur} S., {Papadakis} I., 2010, \apj, 715, 854

\bibitem[{{Obreschkow} {et~al}\mbox{.}(2013){Obreschkow}, {Power}, {Bruderer},
  \& {Bonvin}}]{2013ApJ...762..115O}
{Obreschkow} D., {Power} C., {Bruderer} M., {Bonvin} C., 2013, \apj, 762, 115

\bibitem[{{Pfrommer} {et~al}\mbox{.}(2006){Pfrommer}, {Springel}, {En{\ss}lin},
  \& {Jubelgas}}]{pf06}
{Pfrommer} C., {Springel} V., {En{\ss}lin} T.~A., {Jubelgas} M., 2006, \mnras,
  367, 113

\bibitem[{{Planck Collaboration} {et~al}\mbox{.}(2013){Planck Collaboration},
  {Ade}, {Aghanim}, {Arnaud}, {Ashdown}, {Atrio-Barandela}, {Aumont},
  {Baccigalupi}, {Balbi}, {Banday}, \& et~al.}]{2013A&A...550A.134P}
{Planck Collaboration} {et~al.}, 2013, \aap, 550, A134

\bibitem[{{Roncarelli} {et~al}\mbox{.}(2012){Roncarelli}, {Cappelluti},
  {Borgani}, {Branchini}, \& {Moscardini}}]{2012MNRAS.424.1012R}
{Roncarelli} M., {Cappelluti} N., {Borgani} S., {Branchini} E., {Moscardini}
  L., 2012, \mnras, 424, 1012

\bibitem[{{Roncarelli} {et~al}\mbox{.}(2006){Roncarelli}, {Moscardini},
  {Tozzi}, {Borgani}, {Cheng}, {Diaferio}, {Dolag}, \&
  {Murante}}]{2006MNRAS.368...74R}
{Roncarelli} M., {Moscardini} L., {Tozzi} P., {Borgani} S., {Cheng} L.~M.,
  {Diaferio} A., {Dolag} K., {Murante} G., 2006, \mnras, 368, 74

\bibitem[{{Rudd} \& {Nagai}(2009)}]{2009ApJ...701L..16R}
{Rudd} D.~H., {Nagai} D., 2009, \apjl, 701, L16

\bibitem[{{Ryu} {et~al}\mbox{.}(2008{\natexlab{a}}){Ryu}, {Kang}, {Cho}, \&
  {Das}}]{2008Sci...320..909R}
{Ryu} D., {Kang} H., {Cho} J., {Das} S., 2008{\natexlab{a}}, Science, 320, 909

\bibitem[{{Ryu} {et~al}\mbox{.}(2008{\natexlab{b}}){Ryu}, {Kang}, {Cho}, \&
  {Das}}]{ry08}
{Ryu} D., {Kang} H., {Cho} J., {Das} S., 2008{\natexlab{b}}, Science, 320, 909

\bibitem[{{Ryu} {et~al}\mbox{.}(2003){Ryu}, {Kang}, {Hallman}, \&
  {Jones}}]{ry03}
{Ryu} D., {Kang} H., {Hallman} E., {Jones} T.~W., 2003, \apj, 593, 599

\bibitem[{{Skillman} {et~al}\mbox{.}(2008){Skillman}, {O'Shea}, {Hallman},
  {Burns}, \& {Norman}}]{sk08}
{Skillman} S.~W., {O'Shea} B.~W., {Hallman} E.~J., {Burns} J.~O., {Norman}
  M.~L., 2008, \apj, 689, 1063

\bibitem[{{Smith} {et~al}\mbox{.}(2011){Smith}, {Hallman}, {Shull}, \&
  {O'Shea}}]{2011ApJ...731....6S}
{Smith} B.~D., {Hallman} E.~J., {Shull} J.~M., {O'Shea} B.~W., 2011, \apj, 731,
  6

\bibitem[{{Smith} {et~al}\mbox{.}(2001){Smith}, {Brickhouse}, {Liedahl}, \&
  {Raymond}}]{2001ApJ...556L..91S}
{Smith} R.~K., {Brickhouse} N.~S., {Liedahl} D.~A., {Raymond} J.~C., 2001,
  \apjl, 556, L91

\bibitem[{{Sousbie} {et~al}\mbox{.}(2008){Sousbie}, {Pichon}, {Colombi},
  {Novikov}, \& {Pogosyan}}]{2008MNRAS.383.1655S}
{Sousbie} T., {Pichon} C., {Colombi} S., {Novikov} D., {Pogosyan} D., 2008,
  \mnras, 383, 1655

\bibitem[{{Stoica} {et~al}\mbox{.}(2005){Stoica}, {Mart{\'{\i}}nez}, {Mateu},
  \& {Saar}}]{2005A&A...434..423S}
{Stoica} R.~S., {Mart{\'{\i}}nez} V.~J., {Mateu} J., {Saar} E., 2005, \aap,
  434, 423

\bibitem[{{Tegmark} {et~al}\mbox{.}(2004){Tegmark}, {Blanton}, {Strauss},
  {Hoyle}, {Schlegel}, {Scoccimarro}, {Vogeley}, {Weinberg}, {Zehavi},
  {Berlind}, {Budavari}, {Connolly}, {Eisenstein}, {Finkbeiner}, {Frieman},
  {Gunn}, {Hamilton}, {Hui}, {Jain}, {Johnston}, {Kent}, {Lin}, {Nakajima},
  {Nichol}, {Ostriker}, {Pope}, {Scranton}, {Seljak}, {Sheth}, {Stebbins},
  {Szalay}, {Szapudi}, {Verde}, {Xu}, {Annis}, {Bahcall}, {Brinkmann},
  {Burles}, {Castander}, {Csabai}, {Loveday}, {Doi}, {Fukugita}, {Gott},
  {Hennessy}, {Hogg}, {Ivezi{\'c}}, {Knapp}, {Lamb}, {Lee}, {Lupton}, {McKay},
  {Kunszt}, {Munn}, {O'Connell}, {Peoples}, {Pier}, {Richmond}, {Rockosi},
  {Schneider}, {Stoughton}, {Tucker}, {Vanden Berk}, {Yanny}, {York}, \& {SDSS
  Collaboration}}]{2004ApJ...606..702T}
{Tegmark} M. {et~al.}, 2004, \apj, 606, 702

\bibitem[{{Urban} {et~al}\mbox{.}(2014){Urban}, {Simionescu}, {Werner},
  {Allen}, {Ehlert}, {Zhuravleva}, {Morris}, {Fabian}, {Mantz}, {Nulsen},
  {Sanders}, \& {Takei}}]{2014MNRAS.437.3939U}
{Urban} O. {et~al.}, 2014, \mnras, 437, 3939

\bibitem[{{Vazza} {et~al}\mbox{.}(2013){Vazza}, {Br{\"u}ggen}, \&
  {Gheller}}]{va13feedback}
{Vazza} F., {Br{\"u}ggen} M., {Gheller} C., 2013, \mnras, 428, 2366

\bibitem[{{Vazza} {et~al}\mbox{.}(2012){Vazza}, {Br{\"u}ggen}, {Gheller}, \&
  {Brunetti}}]{scienzo}
{Vazza} F., {Br{\"u}ggen} M., {Gheller} C., {Brunetti} G., 2012, \mnras, 2518

\bibitem[{{Vazza} {et~al}\mbox{.}(2014{\natexlab{a}}){Vazza}, {Br{\"u}ggen},
  {Gheller}, \& {Wang}}]{va14mhd}
{Vazza} F., {Br{\"u}ggen} M., {Gheller} C., {Wang} P., 2014{\natexlab{a}},
  \mnras, 445, 3706

\bibitem[{{Vazza} {et~al}\mbox{.}(2009){Vazza}, {Brunetti}, \&
  {Gheller}}]{va09shocks}
{Vazza} F., {Brunetti} G., {Gheller} C., 2009, \mnras, 395, 1333

\bibitem[{{Vazza} {et~al}\mbox{.}(2010){Vazza}, {Brunetti}, {Gheller}, \&
  {Brunino}}]{va10kp}
{Vazza} F., {Brunetti} G., {Gheller} C., {Brunino} R., 2010, \na, 15, 695

\bibitem[{{Vazza} {et~al}\mbox{.}(2011){Vazza}, {Dolag}, {Ryu}, {Brunetti},
  {Gheller}, {Kang}, \& {Pfrommer}}]{va11comparison}
{Vazza} F., {Dolag} K., {Ryu} D., {Brunetti} G., {Gheller} C., {Kang} H.,
  {Pfrommer} C., 2011, \mnras, 418, 960

\bibitem[{{Vazza} {et~al}\mbox{.}(2014{\natexlab{b}}){Vazza}, {Gheller}, \&
  {Br{\"u}ggen}}]{va14curie}
{Vazza} F., {Gheller} C., {Br{\"u}ggen} M., 2014{\natexlab{b}}, \mnras, 439,
  2662

\bibitem[{{Viel} {et~al}\mbox{.}(2005){Viel}, {Branchini}, {Cen}, {Ostriker},
  {Matarrese}, {Mazzotta}, \& {Tully}}]{2005MNRAS.360.1110V}
{Viel} M., {Branchini} E., {Cen} R., {Ostriker} J.~P., {Matarrese} S.,
  {Mazzotta} P., {Tully} B., 2005, \mnras, 360, 1110

\bibitem[{{Vogelsberger} {et~al}\mbox{.}(2014){Vogelsberger}, {Genel},
  {Springel}, {Torrey}, {Sijacki}, {Xu}, {Snyder}, {Bird}, {Nelson}, \&
  {Hernquist}}]{2014Natur.509..177V}
{Vogelsberger} M. {et~al.}, 2014, \nat, 509, 177

\bibitem[{{Werner} {et~al}\mbox{.}(2008){Werner}, {Finoguenov}, {Kaastra},
  {Simionescu}, {Dietrich}, {Vink}, \& {B{\"o}hringer}}]{2008A&A...482L..29W}
{Werner} N., {Finoguenov} A., {Kaastra} J.~S., {Simionescu} A., {Dietrich}
  J.~P., {Vink} J., {B{\"o}hringer} H., 2008, \aap, 482, L29

\bibitem[{{Zeldovich} {et~al}\mbox{.}(1982){Zeldovich}, {Einasto}, \&
  {Shandarin}}]{1982Natur.300..407Z}
{Zeldovich} I.~B., {Einasto} J., {Shandarin} S.~F., 1982, \nat, 300, 407

\end{thebibliography}

\begin{figure*}
  \includegraphics[width=0.7\textwidth]{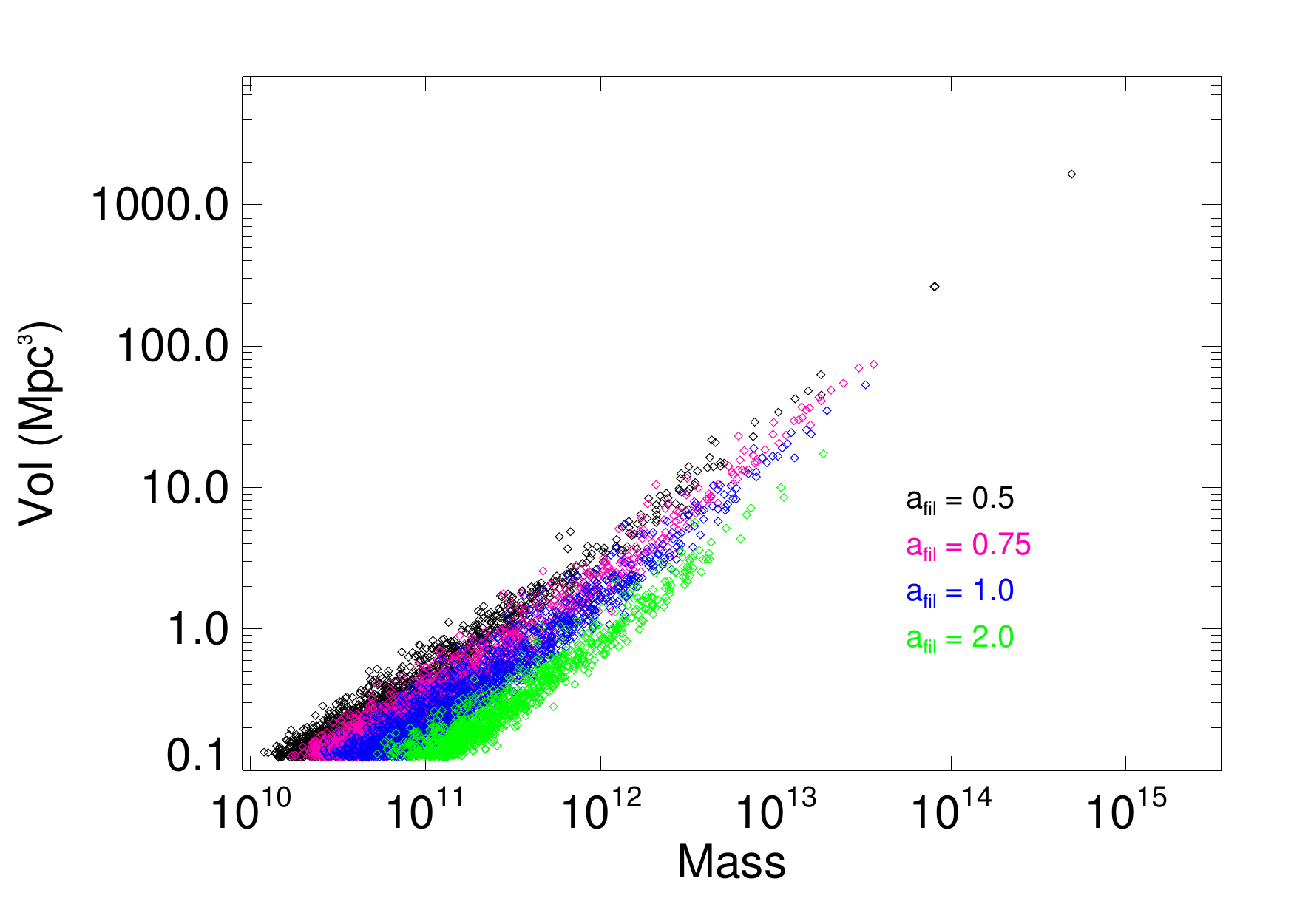}
  \caption{Mass-Volume relation for simulation 1-1\_1024 for different values of $a_{\rm fil}$.}
  \label{fig:M-V-A}
\end{figure*}

\section*{Appendix A: Parameters and Performance Analysis}

\subsection*{Parameters analysis}

Our filaments identification procedure is characterized by six parameters, the most relevant
being the mass density threshold $a_{\rm fil}$. Its value has been set according to various qualitative
and statistical indications, as described in Section \ref{sec:tuning}. 
Although $a_{\rm fil} = 1$ has been set, alternative values in the range $0.5-2.0$ are acceptable.
In order to verify the influence of $a_{\rm fil}$ on the results,
we have calculated for simulation 1-1\_1024 the Mass-Volume 
relation obtained adopting $a_{\rm fil} = 0.5, 0.75, 1.0, 2.0$. Figure \ref{fig:M-V-A} compares the distributions 
obtained for the different values of $a_{\rm fil}$,
showing that the trends and the dispersions obtained for the different thresholds are similar, hence the 
obtained scaling relations are reliable. 
On the other hand, the  mass of the largest objects increases with decreasing $a_{\rm fil}$, due to percolation at low thresholds,
and the normalizations change. This indicates that the methodology is robust 
for comparative studies of different models or multiple realisations of the same model. Absolute estimates (as, for instance, the number 
of objects with a given mass) have to be taken with some care. 

\begin{figure*}
  \includegraphics[width=0.7\textwidth]{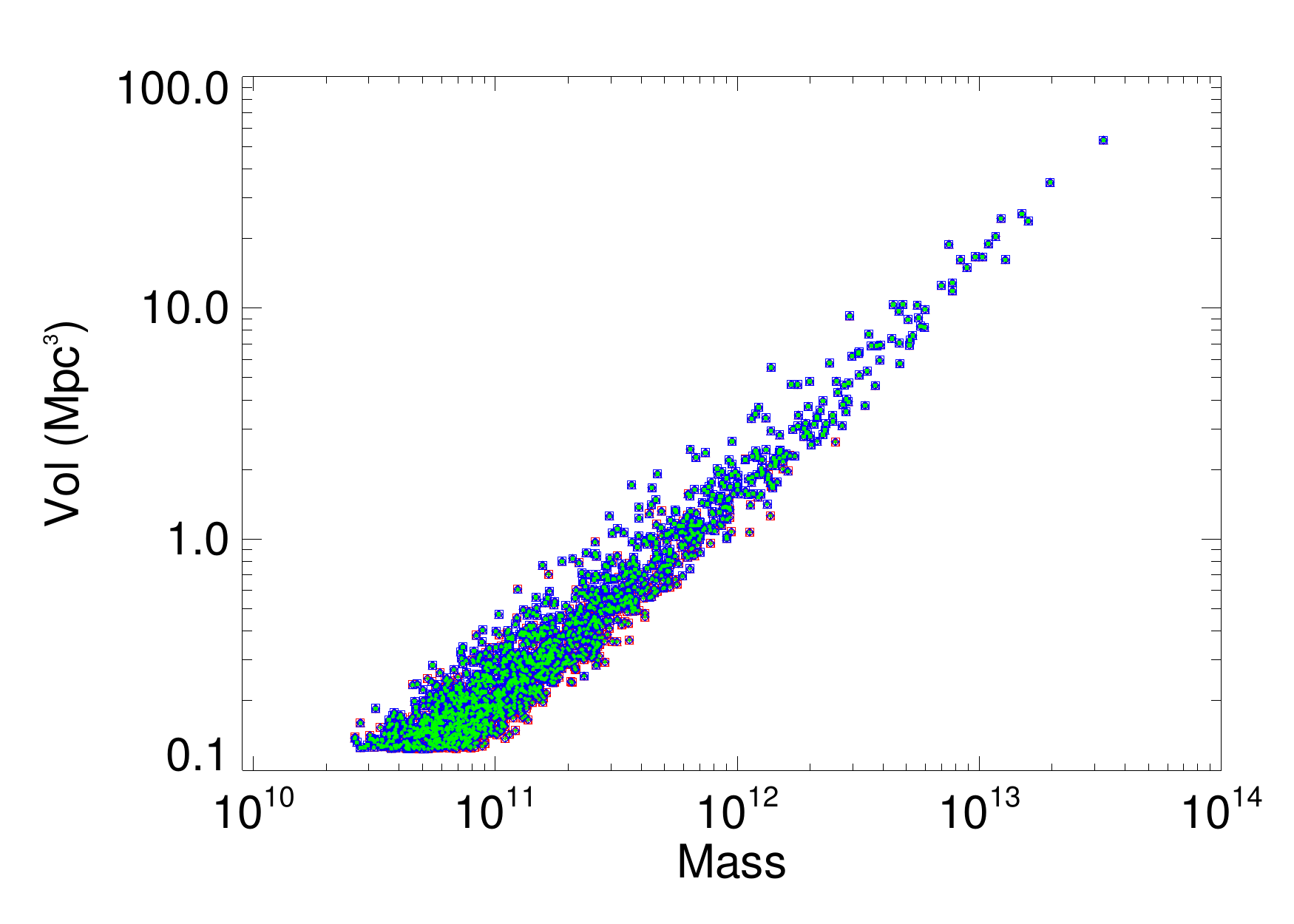}
  \caption{Mass-Volume relation for simulation 1-1\_1024 for $\alpha = 1.0$ (red), $\alpha = 2.0$ (blue), $\alpha = 4.0$ (green) and
           $L_{\varphi} = 0.1$ (crosses), $L_{\varphi} = 2.5$ (diamonds), $L_{\varphi} = 5.0$ (triangles), $L_{\varphi} = 10.0$ (squares). 
           The overall influence of the different settings on the Mass-Volume statistics is shown.}
  \label{fig:M-V-alpha-phi}
\end{figure*}

\begin{figure*}
\centering
\begin{minipage}[c]{0.48\linewidth}
  \includegraphics[width=1.1\textwidth]{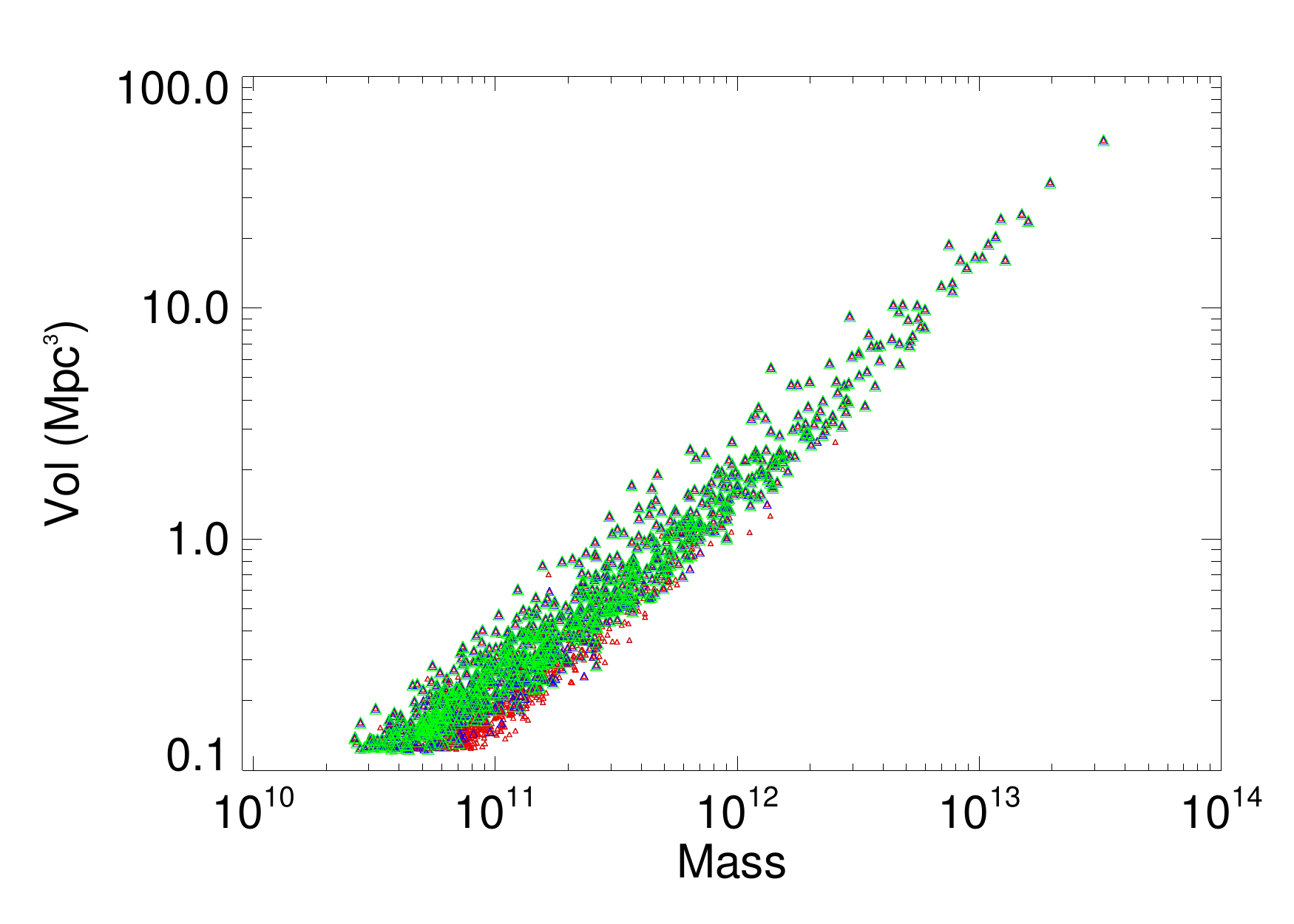}
\end{minipage}
\quad
\begin{minipage}[c]{0.48\linewidth}
  \includegraphics[width=1.1\textwidth]{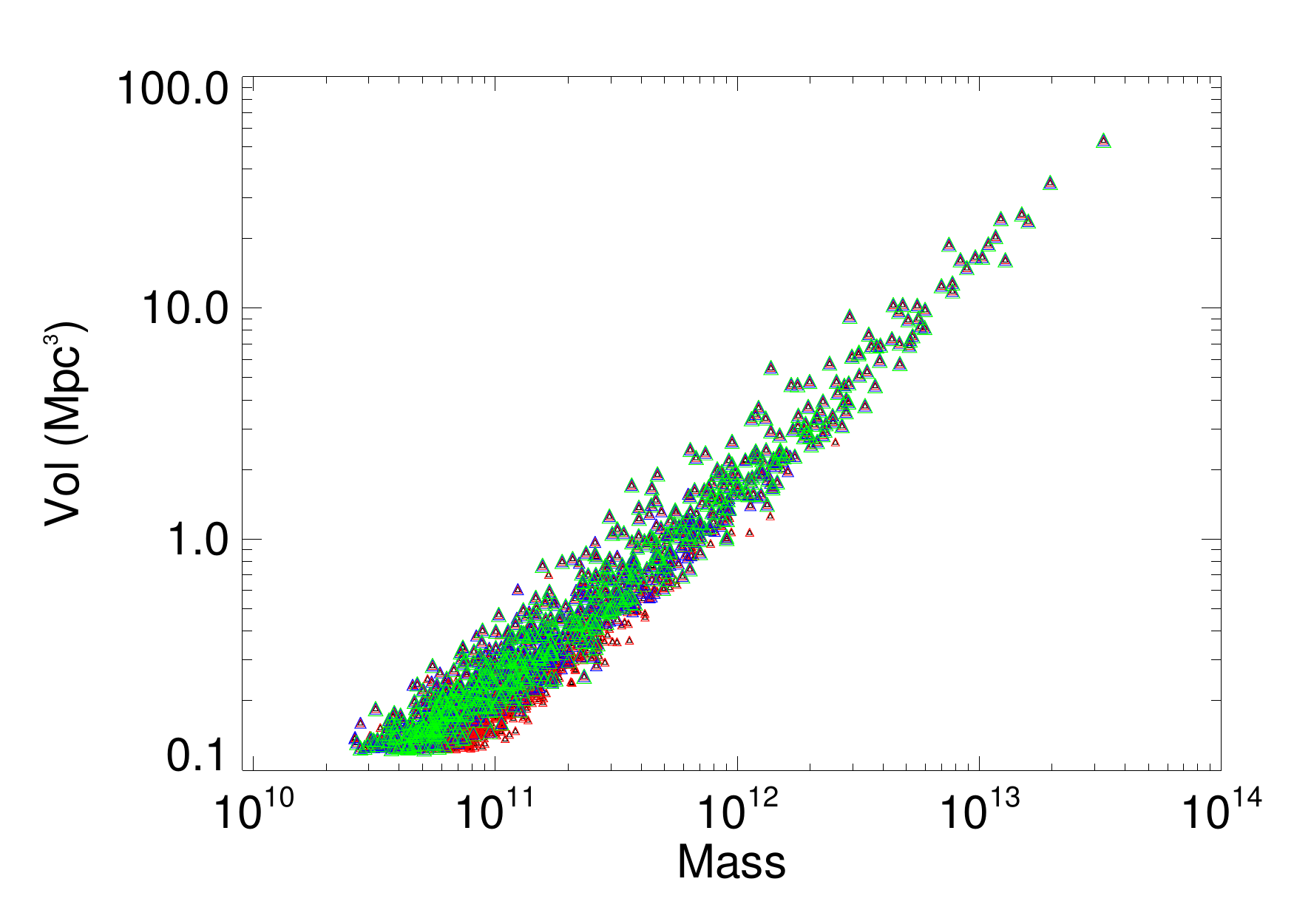}
\end{minipage}
\caption{Mass-Volume relation for simulation 1-1\_1024 for $\alpha = 1.0$ (red), $\alpha = 2.0$ (blue), $\alpha = 4.0$ (green) and
           $L_{\varphi} = 5.0$ (left panel) and $L_{\varphi} = 0.1$ (black), $L_{\varphi} = 2.5$ (red), $L_{\varphi} = 5.0$ (blue),
           $L_{\varphi} = 10.0$ (green) and $\alpha = 2.0$ (right panel). The impact of the various settings for each of the
           two parameters individually on the Mass-Volume statistics is shown.
           }
\label{fig:M-V-2}
\end{figure*}

The remaining five parameters are:
\begin{enumerate}
\item $a_{\rm cl}$: mass density threshold to select the higest peaks of the mass distribution;
\item $\beta$: scale factor to get clusters' radii of the order of 1 Mpc;
\item $V_{\rm res}$: objects whose volume is below this value are discarded;
\item $\alpha$: maximum ratio between the axes of the bounding box containing a filament;
\item $\varphi$: filling factor; filaments with ratio between their volume and the associated bounding box volume is above
this value are discarded.
\end{enumerate}

The first three parameters can be considered as fixed, being related to the physical properties of galaxy clusters
(the first two) and to the spatial resolution of the simulations (the third). For the remaining two parameters, 
we have performed the identification procedure several times, changing their values in order to investigate the influence
on the results. Again, we have used simulation 1-1\_1024 as a representative case. 

\begin{figure*}
  \includegraphics[width=0.7\textwidth]{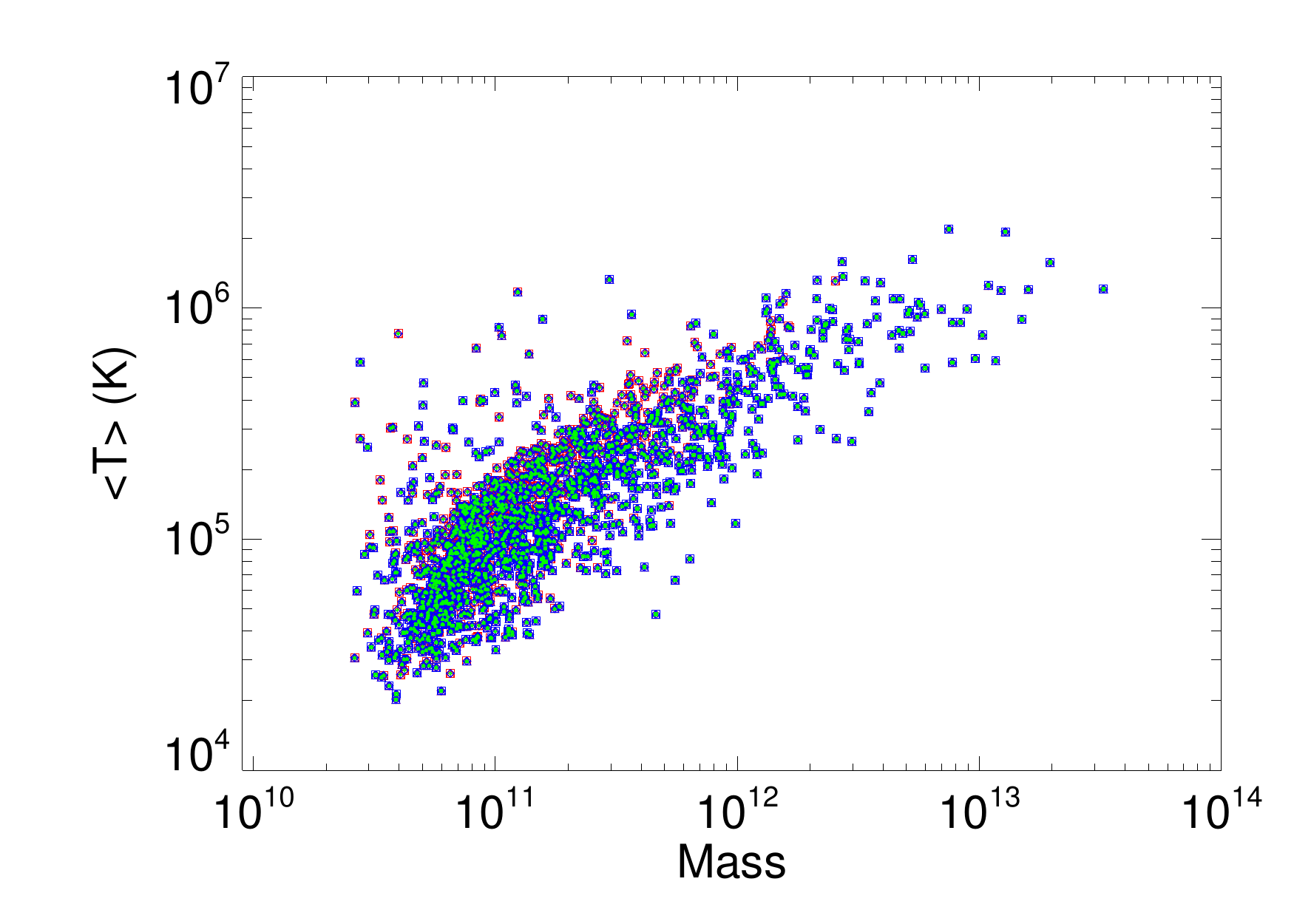}
  \caption{Mass-Temperature relation for simulation 1-1\_1024 for $\alpha = 1.0$ (red), $\alpha = 2.0$ (blue), $\alpha = 4.0$ (green) and
           $L_{\varphi} = 0.1$ (crosses), $L_{\varphi} = 2.5$ (diamonds), $L_{\varphi} = 5.0$ (triangles), $L_{\varphi} = 10.0$ (squares).
           The overall influence of the different settings on the Mass-Temperature statistics is shown.
           }
  \label{fig:M-T-alpha-phi}
\end{figure*}

\begin{figure*}
\centering
\begin{minipage}[c]{0.48\linewidth}
  \includegraphics[width=1.1\textwidth]{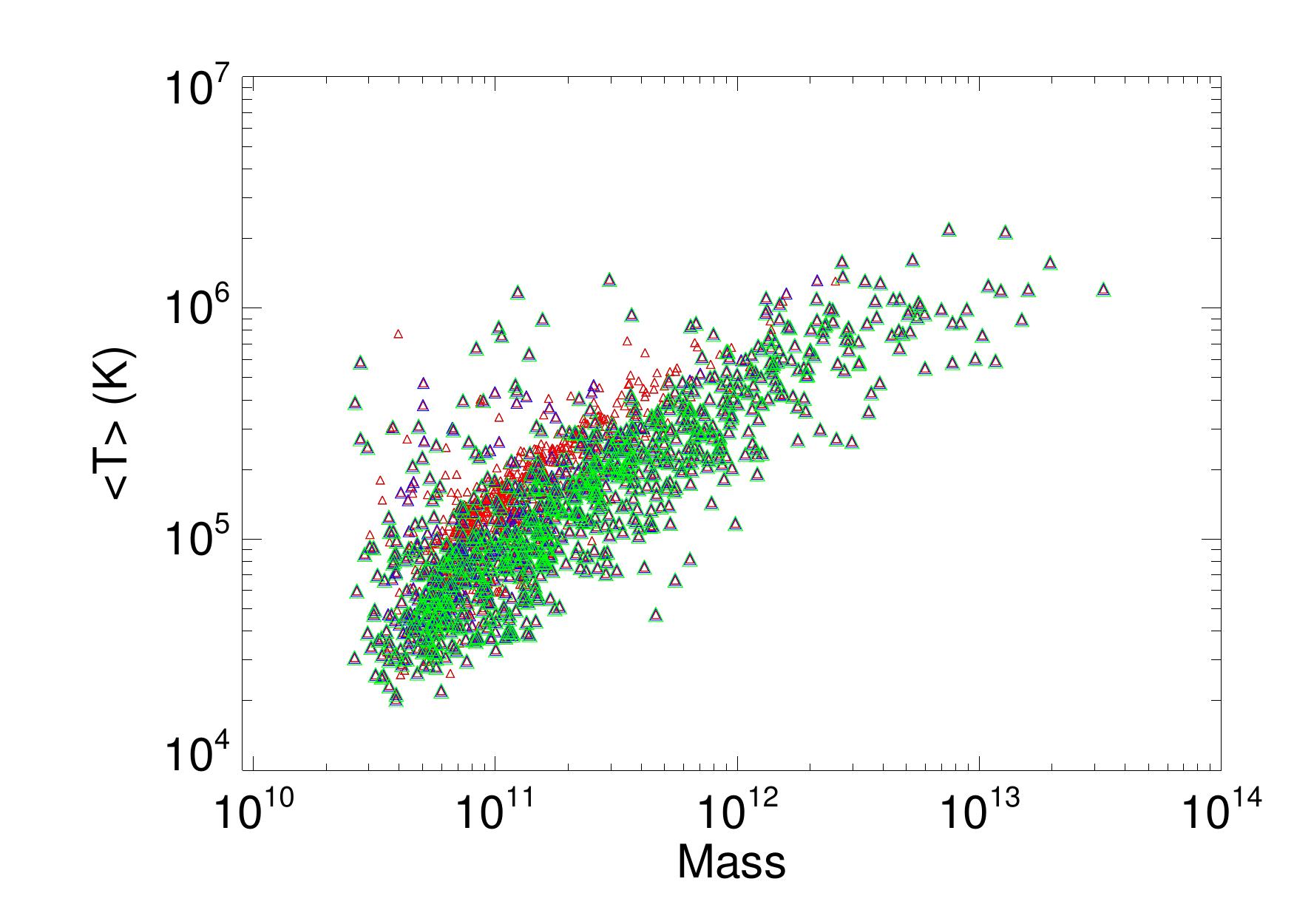}
\end{minipage}
\quad
\begin{minipage}[c]{0.48\linewidth}
  \includegraphics[width=1.1\textwidth]{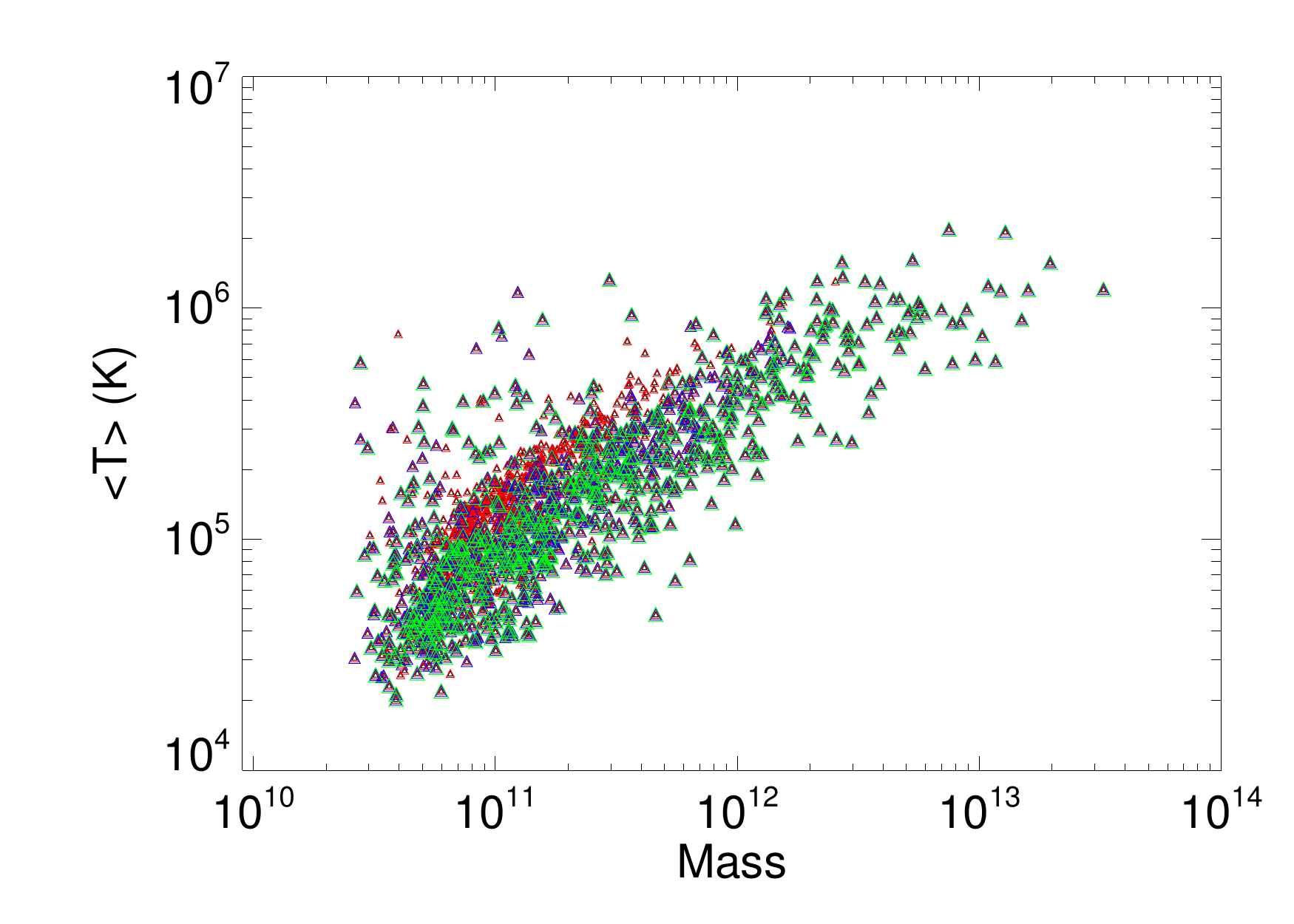}
\end{minipage}
\caption{Mass-Temperature relation for simulation 1-1\_1024 for $\alpha = 1.0$ (red), $\alpha = 2.0$ (blue), $\alpha = 4.0$ (green) and
           $L_{\varphi} = 5.0$ (left panel) and $L_{\varphi} = 0.1$ (black), $L_{\varphi} = 2.5$ (red), $L_{\varphi} = 5.0$ (blue),            
           $L_{\varphi} = 10.0$ (green) and $\alpha = 2.0$ (right panel). The impact of the various settings for each of the
           two parameters individually on the Mass-Temperature statistics is shown.}
\label{fig:M-T-2}
\end{figure*}

Figure \ref{fig:M-V-alpha-phi} shows the Mass-Volume relation for all twelve cases obtained setting $\alpha = 1, 2, 4$ and $L_{\varphi} = 0.1, 2.5, 5, 10$,
this last parameter being the ratio between the bounding box diagonal length and the radius of the inscribed cylinder (see Section \ref{sec:tuning}
for more details): the larger the value of $L_{\varphi}$, the smaller the threshold filling factor. Figure \ref{fig:M-V-2} show a subset 
of the mass-volume data, obtained selecting only those values obtained for $L_{\varphi} = 5$ and $\alpha = 2$ respectively (the values 
adopted in this work). Figures \ref{fig:M-T-alpha-phi} and \ref{fig:M-T-2} show the Mass-Temperature relation for the same parameter settings.

It is clear how changing the value of these two parameters has a minor impact on the results, all the main characteristics of the
presented statistics being preserved. However, a proper selection of their values allows eliminating outliers and objects 
whose geometry is not compatible with that expected for typical filaments (e.g. small, round shaped clumps), fine tuning the results.
The values $L_{\varphi} = 5$ and $\alpha = 2$ have proved to be effective without being too restrictive, leading to 
the loss of significant objects.  

\subsection*{Performance analysis}

The data analysis has been performed on the 
{\it Pilatus}\footnote{http://user.cscs.ch/computing\_resources/pilatus} HPC data processing system
of ETHZ-CSCS, using the client-server 2.8 MPI-parallel version of VisIt. 
Pilatus is a cluster composed of 44 nodes
each with 2 Intel Xeon CPUs E5-2670 
(2.60GHz, 16 cores, 64GB RAM). 

In a first series of tests, strong scalability (i.e. the change in computing
time with increasing the number of MPI tasks) is investigated. 
The 2-1\_1024 model is adopted as reference. Table~\ref{tab:strong} presents the CPU time
as a function of the number of MPI tasks. Acceptable scalability (though not linear) is achieved
up to 16 tasks, although at 32 and 64 tasks the computing time is still decreasing. In the same table, the memory usage is shown. Due to the data replica needed for parallel processing,
the memory required grows with the number of tasks, reaching, at 64 tasks, more than
three times that required for a single task, which is about 17GB. 
Two variables, mass density and temperature, are used
each represented by a $1024^3$ cells floating point single precision variable, accounting for
8 GB. The remaining 50\% of memory is required to handle isosurfaces and connected components.  

\begin{table}
\caption{Strong scalability for the 2-1\_1024 model. Computing time (column 2) and
memory usage (column 3) are shown as a function of the number of cores (column 1).}
\centering \tabcolsep 5pt
\begin{tabular}{c|c|c}
  MPI tasks & CPU time (sec) & Memory (GB) \\  \hline
  1   & 633 &  17.309\\
  2   & 369 & 18.381 \\
  4   & 207 & 19.296 \\
  8   & 123 & 21.229 \\
  16  & 83  & 25.256 \\
  32  & 63  & 34.592 \\
  64  & 50  & 53.158 \\
\end{tabular}
\label{tab:strong}
\end{table}

For the entire data analysis, we have set the number of MPI tasks to 32. This represents a good trade-off between 
computing time and memory usage. In Table~\ref{tab:cputime} we present the CPU time needed to
run the filament reconstruction procedure for the different simulations. Only 
the 1-1\_2048 case required 64 MPI tasks, due to memory constraints. 
The ``0'', ``c1'' and ``c2'' models have similar requirements.

\begin{table}
\caption{Computing time (column 2) and memory usage (column 3) for selected models using 32 cores.)}
\centering \tabcolsep 5pt
\begin{tabular}{c|c|c}
  Model & CPU time (sec) & Memory (GB) \\  \hline
  3-1\_512   & 20  & 18.166\\
  3-1\_1024  & 52  & 32.526 \\
  2-1\_1024  & 63  & 34.592 \\
  1-1\_1024  & 67  & 36.785\\
  1-1\_2048  & 771 & 185.541 \\
\end{tabular}
\label{tab:cputime}
\end{table}

The usage of 32 tasks keeps the CPU time reasonably low. This is particularly important
during the set-up of the different model parameters when the filament identification 
procedure must be repeated many times, with slightly different values of each parameter.
Note that the analysis of the 1-1\_2048 case requires a huge memory allocation (more than 185 GB).
Furthermore, in the $1024^3$ models, the memory usage grows from 3-1\_512 to 1-1\_2048 runs because of the increasing
simulation volume and hence the larger number of identified objects.

Finally, we analysed the load balancing among MPI tasks, to check if the workload
is evenly distributed among the cores. Again, we adopt as reference
the 2-1\_1024 model. As a measure of the load balancing, in Table~\ref{tab:balancing}
we show the minimum and maximum memory used by the various tasks. This can be assumed
as an effective 
measure for the workload, due to the homogeneous spatial distribution of the filaments.
In all tests, the difference between the maximum and minimum values, so the work
imbalance, is between the 11 and 15 \%.  \\
The results of the performance analysis clearly show the feasibility of this approach for big datasets 
by exploiting HPC systems and parallel processing, and its potential scalability to the analysis 
of future larger simulations.

\begin{table}
\caption{Minimum (column 2) and maximum (column 3) memory usage on the different MPI tasks
for the 2-1\_1024 model as a function of the number of cores (column 1)}
\centering \tabcolsep 5pt
\begin{tabular}{c|c|c}
  MPI tasks & Min Mem (GB) & Max Mem (GB) \\  \hline
  4   & 4.553  &  5.140\\
  8   & 2.586  &  2.817\\
  16  & 1.443  &  1.703\\
  32  & 1.011  &  1.151\\
  64  & 0.770  &  0.868\\
\end{tabular}
\label{tab:balancing}
\end{table}

\end{document}